\documentclass[a4paper,useAMS,usenatbib]{mnras}
\usepackage{amsmath}
\usepackage{amssymb}
\usepackage{float}
\usepackage{graphicx}
\usepackage{subcaption}
\usepackage{mathptmx}
\usepackage{newtxtext,newtxmath}
\usepackage{hyperref} 

\newcommand{\dd}{\mathrm{d}}
\newcommand{\pdd}[2]{\frac{\partial #1}{\partial #2}}
\newcommand{\tensor}[1]{\mathbb{#1}}
\newcommand{\Msun}{$\mathrm{M}_{\odot}$}

\graphicspath{{./figures/}}

\def\equationautorefname~#1\null{Eq.~(#1)\null}

\title[The peak of the IMF]{On the origin of the peak of the stellar initial mass function: exploring the tidal screening theory}

\author[T. Colman et al.] {Tine Colman\thanks{tcolman@physik.uzh.ch}$$ and Romain Teyssier$$\\ \\
  {Center for Theoretical Astrophysics and Cosmology, Institute for Computational Science, University of Zurich,} \\ 
  {Winterthurerstrasse 190, 8057 Zurich, Switzerland} \\ 
}

\begin{document}
\maketitle

\begin{abstract}
Classical theories for the stellar initial mass function (IMF) predict a peak mass which scales with the properties of the molecular cloud.
In this work, we explore a new theory proposed by Lee \& Hennebelle (2018). The idea is that the tidal field around first Larson cores prevents the formation of other collapsing clumps within a certain radius.
The protostar can then freely accrete the gas within this radius. This leads to a peak mass of roughly $10 \, M_{\mathrm{1LC}}$, independent of the parent cloud properties.
Using simple analytical arguments, we derive a collapse condition for clumps located close to a protostar.
We then study the tidal field and the corresponding collapse condition using a series of numerical simulations.
We find that the tidal field around protostars is indeed strong enough to prevent clumps from collapsing unless they have high enough densities.
For each newly formed protostar, we determine the region in which tidal screening is dominant.
We call this the tidal bubble. The mass within this bubble is our estimate for the final mass of the star.
Using this formalism, we are able to construct a very good prediction for the final IMF in our simulations.
Not only do we correctly predict the peak, but we are also able to reproduced the high and low mass end of the IMF.
We conclude that tidal forces are important in determining the final mass of a star and might be the dominant effect in setting the peak mass of the IMF.
\end{abstract}


\section{Introduction}\label{sec:intro}
Star formation is a long standing problem in astrophysics.
While a general idea of how stars form is well established, see e.g. the reviews by \cite{McKee&Ostriker2007} and \cite{Krumholz2014}, it is unclear what processes determines their final mass.
A star's mass sets other important quantities, such as the luminosity and whether it will end its life with a supernova.
Recent work has established that supernova feedback is important for the formation and evolution of galaxies.
On smaller scales, the formation of planets is influenced by the luminosity of the parent star.
So knowing the distribution of stellar masses at birth, also called the stellar initial mass function or IMF, and understanding its origin is not only crucial for a complete theory of star formation but also has an impact on both planet and galaxy formation theories.

The IMF has been measured quite extensively inside the Milky and appears to be remarkably similar for different molecular clouds \citep{Bastian_Covey_Meyer2010, Offner_et_al2014, Dib_Schmeja_Hony2017, Hopkins2018}.
This hints that star formation is governed mainly by processes that are independent of the global cloud properties.
Recently, much effort is being put into measuring the IMF in other galaxies (e.g.  \cite{vanDokkum_Conroy2010, Gunawardhana_et_al2011}) though this is not an easy task due to the limited resolution.
It is still debated whether the results from these observations indicate an IMF that is universal across galaxies or not \citep{elmegreen2009, Kroupa_et_al2013, Hopkins2018}.

\begin{figure}
\center
\includegraphics[scale=0.65]{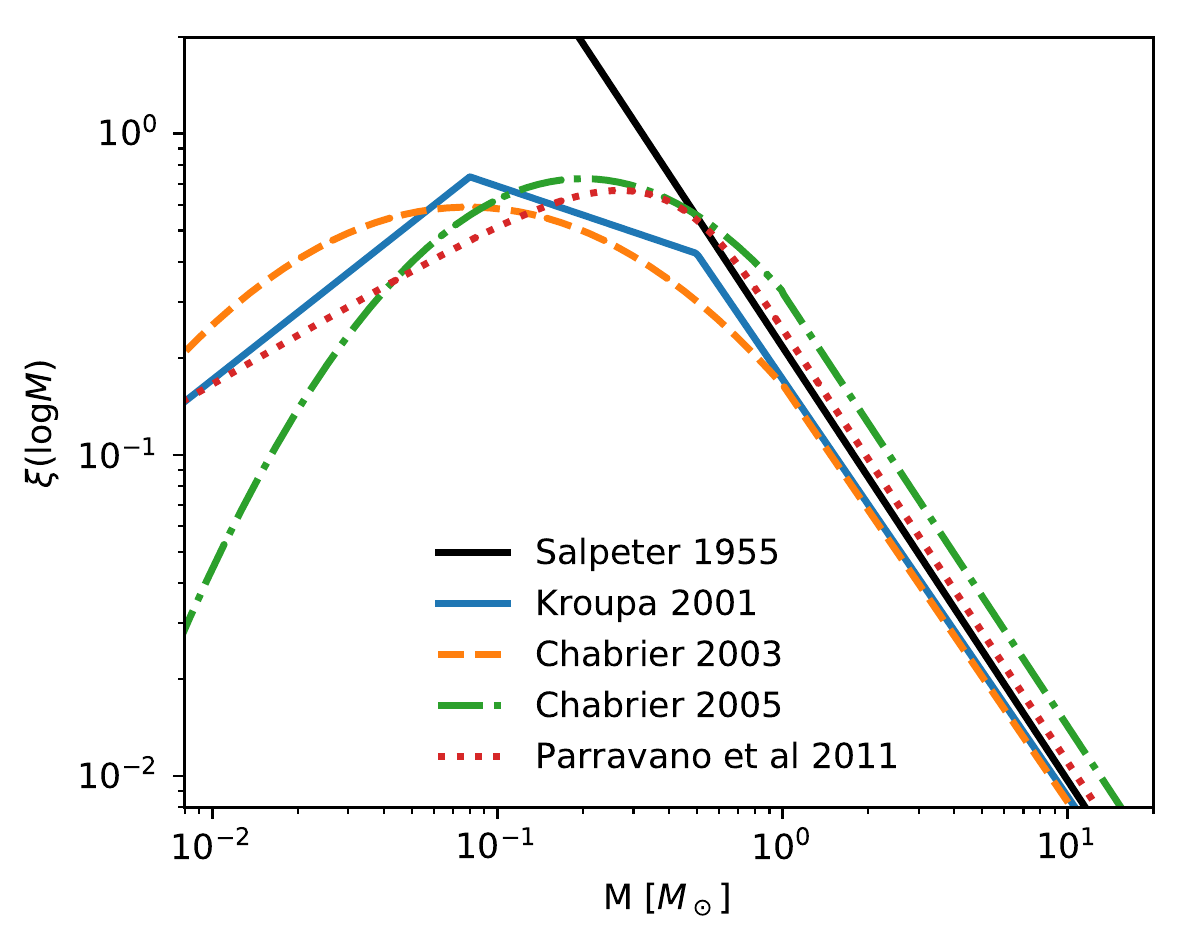}
\caption{Several functional forms for the IMF (individual stars).}
\label{fig:imf_forms}
\end{figure}

Since \cite{Salpeter1955} first introduced the concept, many different analytic forms have been proposed, defining the IMF $\xi$ as a probability function
\begin{align}
\dd N = \xi(\log M) \, \dd \log M
\end{align}
with several free parameters which are determined by fitting the observations.
The most commonly used is the Chabrier IMF, consisting of a lognormal and a powerlaw tail \citep{Chabrier2003, Chabrier2005}, but several others exist (e.g. \cite{Kroupa2001, Parravano_McKee_Hollenbach2011}).
These different forms are compared in Fig.~\ref{fig:imf_forms}.
While, it is commonly accepted that the high mass tail is a powerlaw with a slope close to the Salpeter value, there is a large uncertainty at the low mass end.
All of these forms, however, show a peak around 0.1 - 0.3 \Msun which indicates a characteristic stellar mass.
In this work, we focus on explaining this observed peak in the IMF.

This paper is structured as followed.
In section \ref{sec:theories_peak}, we explore some of the existing explanations for the IMF peak mass.
Typical for these is that they predict a scaling with global cloud properties, something which is in contradiction with observations and recent simulations.
However, such a scaling is not predicted by a recent theory proposed by \cite{Lee_Hennebelle2018b}, which states that the peak mass is set by tidal screening of the first hydrostatic core (also called the first Larson core).
This tidal theory is explored further in section \ref{sec:tidal_theory}, where we give a simple and intuitive analytical description.
In section \ref{sec:setup}, we describe the simulations used to investigate this theory.
In section \ref{sec:cloud_results}, we present the results from the simulations.
We study the tidal field around newly born stars and try to build a predicted IMF which is compared to the IMF measured at the end of the simulation.
In section \ref{sec:discussion}, we discuss what our results mean in the context of a general star formation theory, as well as the caveats associated with our approach. 
We conclude the paper in section \ref{sec:conclusion}.
\section{Theories for the IMF's peak}\label{sec:theories_peak}

Over the years, many people have come up with theories to explain the origin of a typical mass scale for stars.
In this section, we briefly review several of these with a focus on understanding why the peak of the IMF lies around 0.1 \Msun. Other interesting related questions
are: What is the smallest mass of a star? What is the largest possible mass? But this is beyond the scope of this paper.

\subsection{The Jeans mass at the opacity limit}
\label{sec:jeans_mass}
A strong property of the isothermal gas that defines the interstellar medium (ISM) is the absence of a characteristic mass scale owing to the scale free nature of gravity.
From the collapse condition of an isothermal sphere, one can derive the Jeans mass as
\begin{equation}
M_{\mathrm{J}} = \frac{4 \pi}{3} \rho \left( c_{\mathrm{s}} t_{\rm ff}\right)^3
\end{equation}
with $c_{\mathrm{s}}$ is the sound speed, given by
\begin{equation}
c_{\mathrm{s}} = \sqrt{\frac{k_{\mathrm{B}} T}{\mu \, m_{\mathrm{H}}}} 
\end{equation}
and the free-fall time $t_{\mathrm{ff}}$ is defined here as
\begin{equation}\label{eq:tff}
t_{\mathrm{ff}} = \sqrt{\frac{3\pi}{32G \rho}}.
\end{equation}
One sees immediately that for isothermally collapsing gas, we cannot choose a unique characteristic density that would result in a unique characteristic mass.
The larger the density of the gas during collapse and fragmentation, the smaller the Jeans mass.
At very large densities, though, owing to the increased dust absorption coefficient, the gas becomes opaque to its own radiation.
This is sometimes referred to as the opacity limit \citep{Low_lynden-Bell1976, Rees1976}.
At this point the fragmentation is halted, setting a scale below which no fragments form.
It is possible to estimate the characteristic density at which this process occurs by requiring the optical depth of one Jeans radius to be unity
\begin{equation}
\tau = \kappa_{\rm dust} \rho R_{\rm J} = 1~~~{\rm with}~~~R_{\rm J} = c_{\mathrm{s}} t_{\rm ff}
\end{equation}
For typical molecular clouds conditions in the Milky Way with $T=10$~K, $\mu=2.2$
 and $\kappa_{\rm dust}=0.1~{\rm cm}^2~{\rm g}^{-1}$, we find for the critical density
that defines the opacity limit
\begin{equation}
\rho_{\rm crit} = \frac{32 G}{3 \pi \kappa_{\rm dust}^2 c_s^2 } \simeq 5 \times 10^{-14} ~{\rm g}~{\rm cm}^{-3}
\end{equation}
At this critical density, we obtain a unique Jeans mass of the order
\begin{equation}
M_{\mathrm{J}} = 0.670 \frac{c_{\mathrm{s}}^3}{\sqrt{G^3 \rho_{\mathrm{crit}}}} \approx 6 \times 10^{-4} \mathrm{M_\odot}.
\end{equation}
Note that the critical density can also be derived using slightly more complicated arguments, leading to a very similar value for typical Milky Way conditions \citep{Krumholz_book2017}. 
This mass corresponds to the smallest gas clumps that can overcome the pressure gradients and collapse.
At this characteristic density, the gas evolution transitions from isothermal to adiabatic, which prevents the collapse of smaller fragments, even at higher densities.
The Jeans mass at the opacity limit is obviously much smaller than the observed characteristic mass of the IMF. It does not even corresponds to the minimal mass of a star, 
as the smallest of these fragments won't be able to collapse enough to reach stellar densities in their centres. 

\subsection{The mass of the first Larson core}\label{sec:1LC}

Once one of these small stable fragments form, it can accrete mass from the surroundings and build up a hydrostatic core,
a very important concept in the theory of star formation.
\cite{Larson1969} was the first to show that during the collapse of an isothermal cloud, a small core forms in hydrostatic equilibrium.
He called this the first hydrostatic core, now also known as the first Larson core.
Its formation is triggered when the density reaches the critical value defined above, changing the nature of the collapse from isothermal to adiabatic, 
with the formation of an accretion shock at the boundary of the core.
Accretion continues, with more and more mass accumulating in the hydrostatic envelope, compressing and heating the central core.
At some point, the centre reaches a temperature high enough for molecular hydrogen to dissociate.
Most of the energy now goes into dissociation and the collapse resumes, roughly isothermal again.
This second collapse leads to the formation of a second hydrostatic core, which is the actual protostar.
The remainder of the first core is then quickly absorbed by the protostar, which continues to accrete mass from the surrounding envelope.

Following this scenario, one computes the mass of the first Larson core by solving the Lane-Emden equation for an hydrostatic sphere, with a polytropic
index close to $5/3$, valid in the adiabatic, optically thick regime. The properties of the resulting polytrope will depend on two parameters, the internal density in the centre $\rho_{\rm int}$
and the density at the edge of the core $\rho_{\rm ext}$. Setting $\rho_{\rm ext}=\rho_{\rm crit}$ (corresponding to the transition from the isothermal to the adiabatic regime) 
and $\rho_{\rm int}=10^3 \rho_{\rm crit}$ (corresponding to molecular hydrogen dissociation) fully specifies the mass of the core around $M_{\mathrm{1LC}} \simeq 0.02$ \Msun,
a numerical value which Larson found to depend only weakly on the gas and dust temperatures.
Many authors have since repeated and improved Larson's numerical experiments, 
but the mass of the first Larson core $M_{\mathrm{1LC}}$ is still found to be of the order 
$0.02$ \Msun and has proven to be very robust against variations in the equation of state and the properties of the surrounding envelop \citep[e.g.][]{Vaytet_et_al2013, Krumholz_book2017, Bhandare_et_al2018, Lee_Hennebelle2018b}.
When the second Larson core forms, it accretes the leftovers of the first Larson core on a very short timescale.
All protostars will therefore have a mass of about $M_{\mathrm{1LC}} \approx 0.02$ \Msun shortly after their birth.
This is why the mass of the first Larson core is probably a good estimate of the mass of the {\it smallest} stars, but is still 
one order of magnitude lower than the observed peak of the IMF.
This indicates that the majority of protostars continue to accrete material from a surrounding envelope after they have been formed and that the typical mass reservoir in the envelope should be about 0.1 \Msun.

\subsection{The Bonnor-Ebert mass}

On scales larger than the first Larson core, the flow is fully isothermal and scale invariant. Extracting a characteristic mass is therefore quite difficult.
One strategy is to look for hydrostatic envelopes emerging from the parent molecular cloud. Hydrostatic spheres in this case are described using the
isothermal Lane-Emden equation, leading to a family of solutions, with here again the internal and external densities $\rho_{\rm int}$ and $\rho_{\rm ext}$ as the two parameters.
A particular solution, known as the Bonnor-Ebert sphere, corresponds to the largest stable isothermal self-gravitating sphere, with still one remaining free parameter $\rho_{\mathrm{ext}}$ \citep{Bonnor1956, Ebert1957}.
Its mass, the Bonnor-Ebert mass, is given by
\begin{equation}
M_{\mathrm{BE}} = 1.182 \frac{c_{\mathrm{s}}^3}{\sqrt{G^3 \rho_{\rm ext}}} 
\end{equation}
and looks very similar to the Jeans mass, except that now the density is the density at the external boundary.
We still have to find the value for $\rho_{\rm ext}$. One possibility is to use the typical density of molecular clouds.
With $n_{\rm H} \simeq 100~{\rm H}/{\rm cm}^{3}$ we obtain $M_{\mathrm{BE}} \simeq 21$ \Msun, a value too large to explain the characteristic mass of the IMF.
Moreover, molecular clouds are turbulent, strongly structured and have different sizes, resulting in a wide range of densities and thus a wide range of possible Bonnor-Ebert masses, 
in contrast to the requirement of having a universal characteristic mass.

A variation of this idea
is based on the concept of {\it turbulent} Bonnor-Ebert spheres \citep{Haugbolle_Padoan_Nordlund2018}. We know indeed that the ISM is turbulent and follows an observed universal scaling relation,
known as Larson's relation \citep{Larson1981, Heyer_et_al2009}, which related the velocity dispersion $\sigma$ to the diameter of a cloud $L$
\begin{equation}
\sigma_{\rm 1D}(L) = \sigma_0 \left( \frac{L}{L_0} \right)^{1/2}
\end{equation}
where $\sigma_0=1~{\rm km}~{\rm s}^{-1}$ and $L_0=1~{\rm pc}$ are constant parameters valid throughout the Milky Way. 
This scaling relation is typical for supersonic (or Burgers) turbulence, in which strong isothermal shocks compress the gas to much higher densities than the
initial cloud mean density.
Hydrostatic envelopes will emerge from these compressed regions and deliver the envelopes to be accreted by the first Larson cores. 
Using the Rankine-Hugoniot relations for isothermal shocks, we can derive the external density of the Bonnor-Ebert sphere as
\begin{equation}
\rho_{\mathrm{ext}} = \overline{\rho} \left(1 + \mathcal{M}^2 \right) \simeq  \overline{\rho} \mathcal{M}^2
\end{equation}
where $\overline{\rho}$ is the mean cloud density and $\mathcal{M}$ is the Mach number of the turbulence defined by
\begin{equation}
\mathcal{M} = \frac{\sigma_{\rm 1D}}{c_{\mathrm{s}}}
\end{equation}
The resulting Bonnor-Ebert mass still depends on the cloud density, but if one assumes that star forming molecular clouds are in rough
virial equilibrium,
one has
\begin{equation}
\alpha_{\rm vir} = \frac{15}{\pi} \frac{\sigma_{\rm 1D}^2}{G \overline{\rho}L^2} \simeq 1
\end{equation}
which results in a constant post-shock density with
\begin{equation}
\rho_{\rm ext} \simeq \overline{\rho} \mathcal{M}^2 = \frac{15}{\pi} \frac{\sigma_0^4}{G c_{\mathrm{s}}^2 L_0^2}
\end{equation}
Injecting this in the Bonnor-Ebert mass formula finally gives for the turbulent Bonnor-Ebert mass (adding superscript ``turb'')
\begin{equation}
M_{\mathrm{BE}}^{\rm turb} = 0.54 \, \frac{c_{\mathrm{s}}^4 L_0}{G \sigma_0^2} \simeq 0.2 \, \mathrm{M}_{\odot}
\end{equation}
whose numerical value spectacularly fits the observed characteristic mass of the IMF \citep{Haugbolle_Padoan_Nordlund2018} .
One problem of this approach is that other galaxies have different global turbulent properties with different values for $L_0$ and $\sigma_0$,
resulting in a different IMF characteristic mass. A fair question to ask also, is whether the formation of hydrostatic envelopes is at all possible
within a self-gravitating, turbulent isothermal fluid. A more common situation in simulations of isothermal turbulence is a free-falling envelope around 
first Larson cores. Moreover, the numerical experiments we perform in this paper do not seem to support this scenario.

\subsection{Turbulent fragmentation theory}

The turbulent fragmentation scenario, introduced by 
\cite{Padoan_et_al1997} and further developed by \cite{Hennebelle_Chabrier2008} and \cite{Hopkins2012},
provides a more realistic framework for the origin of individual star forming regions.
This theory builds on the result that turbulence creates overdensities which follow a lognormal density PDF
\begin{equation}
\mathcal{P}(\delta) = \frac{1}{\sqrt{2\pi \sigma_0^2}} \exp \left[- \frac{\left(\delta - \bar{\delta}\right)^2}{2 \sigma_0^2} \right]
\end{equation}
with $\delta = \log (\rho/\bar{\rho})$ and $\bar{\delta} = - \sigma_0^2/2$.
The width of the distribution depends on the Mach number
\begin{equation}
\sigma_0^2 = \ln(1 + b^2 \mathcal{M}(L)^2)
\end{equation}
where the Mach number is defined at the scale of the entire molecular cloud and $b$ is a parameter related to the nature of the turbulent forcing,
either compressive with $b=1$ or solenoidal with b=$1/3$.
The spectrum of turbulence is here again described using Burgers theory, with velocity fluctuations depending on each scale $\ell$ as
\begin{equation}
\sigma_{\rm 1D}(\ell) = \sigma_0 \left( \frac{\ell}{L_0} \right)^{1/2}~~~{\rm where}~~~ L > \ell > \ell_{\rm sonic}
\end{equation}
where $\ell_{\rm sonic}$ is the {\it sonic length}, defined by
\begin{equation}
\sigma_{\rm 1D}(\ell_{\rm sonic}) = c_{\mathrm{s}}~~~{or}~~~\ell_{\rm sonic} = {L_0}\left( \frac{c_{\mathrm{s}}}{\sigma_0} \right)^2
\end{equation}
corresponding to scales where turbulence becomes subsonic and density fluctuations become small.

Using the analytical Press-Schechter theory, \cite{Hennebelle_Chabrier2008} and \cite{Hopkins2012} computed the mass fraction 
in collapsed structures, using a scale-dependent collapse criterion based on the virial parameter, which writes
\begin{equation}
\alpha_{\rm vir}(\ell) = \frac{15}{\pi} \frac{\sigma_{\rm 1D}(\ell)^2+c_{\mathrm{s}}^2}{G \rho \ell^2} < 1
\end{equation}
The excursion set formalism allowed them to compute the molecular {\it cloud} mass function using the first crossing of the collapse barrier,
while the molecular {\it core} mass function was obtained using the last crossing of the collapse barrier. In other words, molecular clouds are the
largest bound objects in a galaxy, while molecular cores are the smallest bound objects, at the bottom of the gravitational fragmentation hierarchy.
Interestingly enough, this analytical formalism allowed these authors to compute the characteristic mass
of the molecular cores as the mass of the smallest gravitational bound objects whose sizes are equal to the sonic length and the density 
is equal to the minimum density for gravitational collapse.  Using Equation~(18) and (46) of \cite{Hennebelle_Chabrier2008}, we obtain  
\begin{equation}
M_{\rm peak} = \frac{c_{\mathrm{s}}^3}{\sqrt{G^3 \overline{\rho}}}  \frac{1}{\left( 1 + b {\cal M}^{2} \right)^{3/4}}\simeq  0.13 \, \mathrm{M}_\odot
\end{equation}
This peak mass, under Milky Way conditions, is close to the value of the turbulent Bonnor-Ebert mass, 
although it has a slightly different dependency on the turbulent flow parameters. The numerical value we quote here was obtained using Larson's relation, $b=0.3$ and ${\cal M} = 10$

\subsection{Competitive accretion theory}
\label{sec:competitive_accretion}

A different theory was proposed by \cite{Bonnell_et_al2001}, who argued that observationally most young stars are seen in star clusters, sometimes
still embedded in their parent gas cloud. The stellar density can be high, leading to dynamical interactions causing star ejections, mass segregation
and various scenarios of secular evolution. In this complex dynamical environment, young protostars, in particular the first Larson cores, interacts with one another
and compete for accreting gas. This scenario, called {\it competitive accretion}, is based on early first Larson core accreting gas at the core of their parent gas envelope, 
that are ejected after some typical interaction time by a more massive core, so that the accretion abruptly stops. As a result, smaller stars are the one ejected early,
while more massive stars are the ones that managed to remain in their envelope until the exhaustion of the full gas reservoir.

One can estimate the characteristic mass of the resulting IMF in the competitive accretion scenario, using
\begin{equation}
M_{\rm CA} = \dot{M}_{\rm acc} \, t_{\mathrm{dyn}}
\end{equation}
where $\dot{M}_{\rm acc}$ is a typical accretion rate and $t_{\mathrm{dyn}}$ is the dynamical interaction timescale. 
For the latter, one uses the local gas dynamical time scale in the gas envelope
\begin{equation}
t_{\mathrm{dyn}} = \sqrt{\frac{3\pi}{32 G \rho_{\rm env}}}
\end{equation}
while for the accretion rate, one can use the singular isothermal sphere model that  gives
\begin{equation}
\dot{M}_{\rm acc} = \frac{c_{\mathrm{s}}^3}{G}
\end{equation}
This leads to a characteristic mass
\begin{equation}
M_{\rm CA} \approx 0.54 \frac{c_{\mathrm{s}}^3}{\sqrt{G^3 \rho_{\rm env}}} 
\end{equation}
equal within a factor of two to the Bonnor-Ebert mass. Here again, the difficulty is to adopt the right value for $\rho_{\rm env}$,
using either the cloud mean density or the typical post-shock density emerging from the supersonic turbulence. 
For the latter, more realistic scenario, the resulting characteristic mass is half of the turbulent Bonnor-Ebert mass
\begin{equation}
M_{\rm CA} \approx 0.1 M_{\odot} 
\end{equation}
with exactly the same scaling than for the other previously discussed scenarios. 
Note that a  more complicated derivation based on Bondi-Hoyle accretion was performed by \cite{Bonnell_et_al2001},
leading to a similar result, in very good agreement with the observed IMF.

\subsection{Tidal screening theory}
\label{sec:tidal_theory_quick}

The previously discussed theories are based on supersonic self-gravitating isothermal turbulence, and always lead to a characteristic mass close to the turbulent Bonnor-Ebert mass.
As a consequence, one expects the IMF to vary widely with the two parameters controlling the turbulence, namely the injection scale $L_0$ and the total velocity dispersion $\sigma_0$,
and a third parameter, namely the molecular gas sound speed $c_{\mathrm{s}}$. It is quite possible that the Universe conspires to keep the turbulent Bonnor-Ebert mass constant across very different environments,
from quiescent early type galaxies to violent starbursts or high redshift galaxies.  On the other hand, the mass of the first Larson core depends only weakly on only one parameter, 
the coupled gas and dust temperature in the deepest and densest regions of molecular cores,
which is likely to remain very close to 10~K in a wide range of galactic environments  \citep{Juvela_Ysard2011}.
This makes the first Larson core a very robust candidate for explaining why the characteristic mass of the IMF would remain constant in different environments.

In a recent series of papers, \cite{Lee_Hennebelle2018b, Lee_Hennebelle2018c} proposed to explain the characteristic mass of the IMF by tidal forces caused by the first Larson core and its envelope,
preventing the formation of new cores in their surroundings. This allows the central core to gather all the mass available in this tidally screened reservoir, 
without sharing it with the other cores that would have collapsed otherwise. A rough estimate of the tidal screening process can be made using a simplified version of the virial theorem,
applied to a small spherical region in the vicinity of an existing first Larson core. A more rigorous derivation follows in the next section.

We compare the tidal field of the first Larson core to the self-gravity of a collapsing gas clump by writing
\begin{equation}
\frac{2 G M_{\rm 1LC}}{r^3}=\frac{4\pi}{3} G \rho_{\rm gas}
\end{equation}
Assuming that gas around the newly formed core is distributed in a envelope described by a singular isothermal sphere, with
\begin{equation}
\rho_{\rm gas}= \rho_0 \left( \frac{r}{r_0}\right)^{-2},
\end{equation}
we can deduce the radius within which the tidal field is strong enough to prevent the collapse as
\begin{equation}
r_{\rm tidal} = \frac{3 M_{\rm 1LC}}{2\pi \rho_0 r_0^2}
\end{equation}
Integrating the total mass enclosed within this tidally screened region, we get
\begin{equation}
M_{\rm tidal} = \int_0^{r_{\rm tidal}} \rho_{\rm gas}(r) 4 \pi r^2 {\rm d}r = 6 M_{\rm 1LC} \simeq 0.12 M_\odot
\end{equation}
independent on the envelope's properties (this is only valid when the exponent of the envelope profile is -2). Note that this very naive estimate is quite close to the numerical value found by \cite{Lee_Hennebelle2018b},
which was $M_{\mathrm{tidal}} \approx 10 \, M_{\mathrm{1LC}} \simeq 0.2 M_\odot$ and also quite close to the characteristic mass of the observed IMF.
Note that it is completely independent on any parent cloud physical parameters or turbulent galactic scaling relations. 

The goal of this paper is precisely to investigate this tidal screening theory with high-resolution simulations, using a complementary approach to \cite{Lee_Hennebelle2018b}.


\section{Tidal disruption theory}\label{sec:tidal_theory}

\begin{figure*}
\center
\includegraphics[scale=0.8]{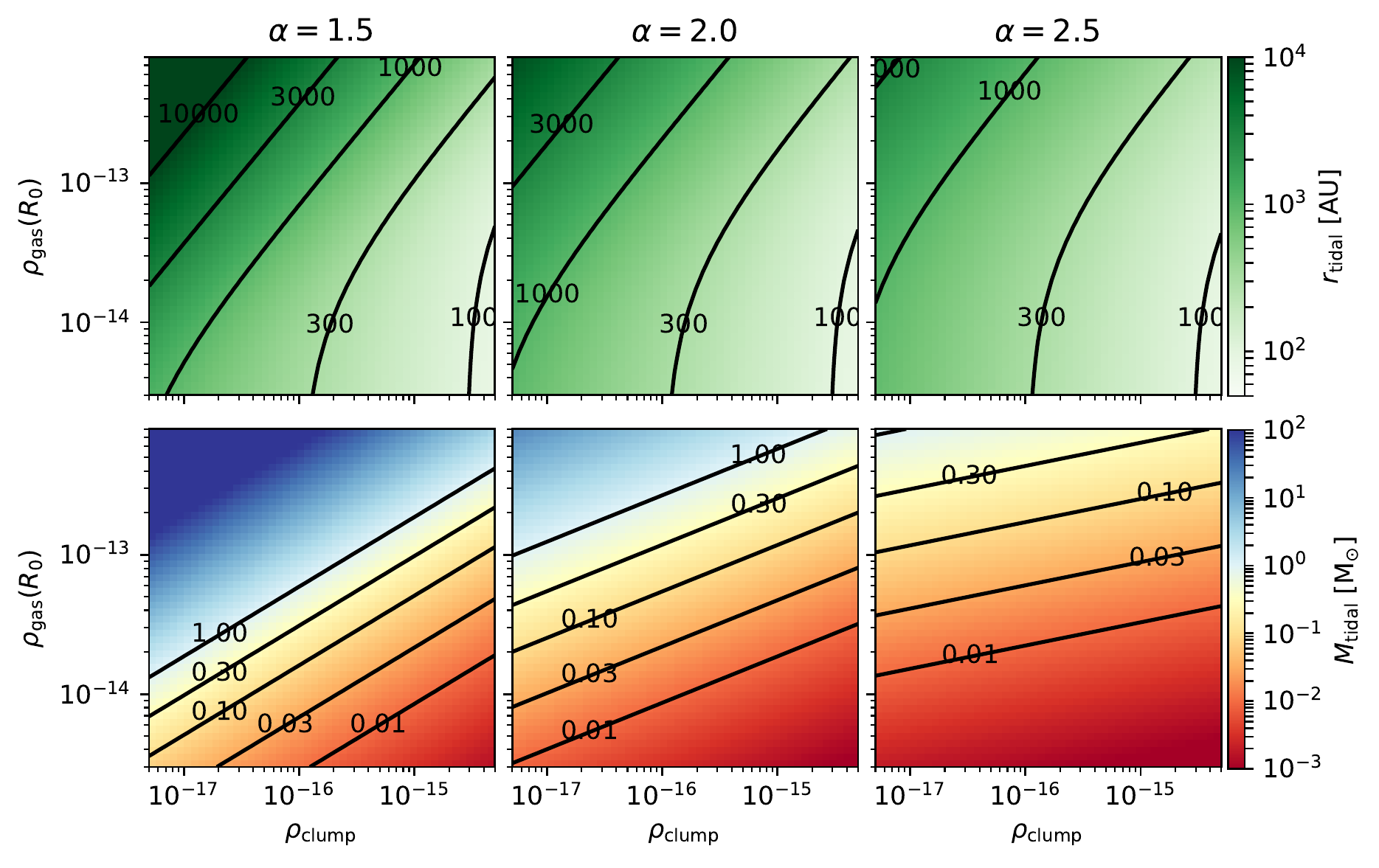}
\caption{The tidal radius and corresponding tidal mass for a large range of clump densities and envelope profiles. $M_{*}$ is fixed to 0.1 \Msun. $R_0$ is set to 12.5 AU.}
\label{fig:dependence}
\end{figure*}

In this paper, we study specifically tidal screening of the first Larson core as the origin of the peak of the IMF.
The mathematical aspects of the tidal screening theory are quite complex.
For the original calculations, we refer to \cite{Lee_Hennebelle2018b}.
Here, we adopt a slightly different point of view, exploiting directly our simulation data and highlighting the physical mechanism at work. 

\subsection{Model and assumptions}

We consider here a basic configuration with a protostar (or a first Larson core) embedded in its gaseous envelope described by a power law density profile
\begin{equation} \label{eq:rho_profile}
\rho(r) = \rho_0 \: \left( \frac{r}{r_0} \right) ^{-\alpha}
\end{equation}
In reality, star forming cores are not spherically symmetric.
Moreover, the parent clouds are highly turbulent and the envelopes are expected to contain filaments and voids.
Nevertheless, we consider that a small, spherical clump with uniform density $\rho_{\mathrm{clump}}$ is forming at a distance $r$ from the protostar in its spherically symmetric envelope.
Certain conditions are required for this clump to be able to collapse and form a star.
The fundamental idea is that due to the already existing protostar (and its power law envelope), a tidal force is exerted on the clump, stretching it in the radial direction.
If this stretching is too strong, the clump will be tidally stripped and will be prevented from forming a companion star.
In this case, the gas in the envelope has no other option than to be accreted onto the central protostar.
This process is called tidal screening and the gas mass in this tidally stripped region will determine the final mass of the star. 

\subsection{Tidal tensor}

In general, the tidal tensor is defined as
\begin{equation}
\label{eq:def_tidal}
T_{ij} \equiv \frac{\partial^2 \phi}{\partial x_i \partial x_j} = - \frac{\partial g_j}{\partial x_i}
\end{equation}
with $\phi$ the gravitational potential and $\mathbf{g}$ the gravitational acceleration.
We have used the relation
\begin{equation}
\mathbf{g} = - \nabla \phi \mathrm{~~~~or~~~~} g_i = - \frac{\partial \phi}{\partial x_i}
\end{equation}
in the second equality.
The derivatives commute so the tidal tensor is symmetric and can be diagonalised to the form
\begin{equation}
\mathbb{T} = 
\begin{bmatrix}
\lambda_1 & 0 & 0\\
0 & \lambda_2 & 0\\
0 & 0 & \lambda_3
\end{bmatrix}
\end{equation}
with $\lambda_i$ the three eigenvalues.
The sign of an eigenvalue indicates whether a local gas clump is expanding ($\lambda_i < 0$) or contracting ($\lambda_i > 0$) in the direction of the corresponding eigenvector. 
In the case of a spherically symmetric potential in spherical coordinates, the tidal tensor is given by
\begin{equation}
\tensor{T} = 
\begin{bmatrix}
\pdd{^2\phi}{r^2} & 0 & 0\\
0 & \frac{1}{r}\pdd{\phi}{r} & 0\\
0 & 0 & \frac{1}{r}\pdd{\phi}{r}
\end{bmatrix}
= 
\begin{bmatrix}
- \pdd{g_r}{r} & 0 & 0\\
0 & - \frac{g_r}{r} & 0\\
0 & 0 & - \frac{g_r}{r}
\end{bmatrix}
\end{equation}

\subsection{Collapse condition}\label{sec:estimate_tidal_force}

Our clump will feel a total tidal force that is the sum of three contributions: the central protostar, the envelope and the clump itself. 
We now derive a condition for the clump to collapse.

We first need to compute the gravitational field of the protostar. It is simply that of a point source
\begin{equation}
g_{*}(r) = - \frac{G M_{*}}{r^2}
\end{equation}
resulting in a tidal tensor of
\begin{equation}
\tensor{T}_{*} = \frac{G M_{*}}{r^3}
\begin{bmatrix}
- 2 & 0 & 0\\
0 & 1 & 0\\
0 & 0 & 1
\end{bmatrix}
\end{equation}
The tidal force is strongest in the direction towards the star.
The clump is stretched in the radial direction and contracts in the two tangential directions, turning a spherical object into an elongated ellipsoid.                                                                                                                       

The gravity of the envelope with density profile from Equation~(\ref{eq:rho_profile}) is given by
\begin{equation}
g_{\mathrm{gas}}(r) = -\frac{4 \pi}{3 - \alpha} G \rho_{\mathrm{gas}} (r) r
\end{equation}
for $\alpha < 3$.
Its tidal tensor thus becomes
\begin{equation}
\tensor{T}_{\mathrm{gas}} = \frac{4 \pi}{3 - \alpha} G \rho_{\mathrm{gas}}(r)
\begin{bmatrix}
-(\alpha-1) & 0 & 0\\
0 & 1 & 0\\
0 & 0 &  1
\end{bmatrix}
\end{equation}
If $\alpha=2$, this generates a contraction in both tangential directions which is equally strong to the expansion in the radial direction.

The clump itself is considered to be homogeneous, with
\begin{equation}
g_{\mathrm{clump}}(r) = \frac{4 \pi}{3} G \rho_{\mathrm{clump}} r
\end{equation}
which results in a tidal tensor which is proportional to the clump density and contraction equally strong in all directions
\begin{equation}
\tensor{T}_{\mathrm{clump}} = \frac{4 \pi}{3} G \rho_{\mathrm{clump}}
\begin{bmatrix}
1 & 0 & 0\\
0 & 1 & 0\\
0 & 0 & 1
\end{bmatrix}
\end{equation}
Adding up these three contributions, the non-zero components of the total tidal tensor are
\begin{align}\label{eq:theory_eigenvals}
T_{rr} = - \frac{2 G M_{*}}{r^3} - \frac{4\pi (\alpha - 1)}{(3 - \alpha)} G \rho_{\mathrm{gas}} (r) &+ \frac{4 \pi}{3} G \rho_{\mathrm{clump}} \\
T_{\theta \theta} = T_{\phi \phi} = \frac{G M_{*}}{r^3} + \frac{4 \pi}{(3 - \alpha)} G \rho_{\mathrm{gas}}(r) &+ \frac{4 \pi}{3} G \rho_{\mathrm{clump}}
\end{align}
which are also its eigenvalues.
As noted before, a positive eigenvalue implies collapsing in the associated direction, a negative one means expanding.
To resist the effect of tidal stripping, and thus to be able to form a star, all three eigenvalues must be positive.
The tangential eigenvalues cannot be negative for $0 \leq \alpha < 3 $, so in order to be collapsing a clump must satisfy the condition $T_{rr}>0$ or
\begin{equation}\label{eq:collapse_condition}
\boxed{
\rho_{\mathrm{clump}} > \frac{3 M_{*}}{2 \pi r^3} + \frac{3(\alpha - 1)}{(3-\alpha)} \rho_{\mathrm{gas}}(r)
}
\end{equation}

\subsection{Tidal radius and tidal mass}
\label{sec:tidal_radius}

Once a first Larson core has formed inside its envelope, we know $M_*$ and $\rho_{\mathrm{gas}}(r)$. 
For a given clump density, the collapse condition is a function of the distance to the central protostar.
One can define the tidal disruption radius $r_{\mathrm{tidal}}$ as the radius for which the inequality in equation (\ref{eq:collapse_condition}) becomes a strict equality.
In other words, if the clump with density $\rho_{\mathrm{clump}}$ is located within $r_{\mathrm{tidal}}$, it will be tidally stripped. 
If it is outside this radius, it will resist the tidal forces.
As a consequence, all the gas mass within the tidal radius will eventually end up in the central object.
We call this mass the tidal mass, which gives us an estimate for the end mass of the protostar
\begin{align}
M_{\mathrm{tidal}} = M_{\mathrm{gas}}(<r_{\mathrm{tidal}}) &= \int_0^{r_{\mathrm{tidal}}} \rho_{\mathrm{gas}}(r) \: 4 \pi r^2 dr\\
                    &= \frac{4 \pi}{(3 - \alpha)} \rho_{\mathrm{gas}}(r_{\mathrm{tidal}}) \: r_{\mathrm{tidal}}^3
\end{align}

Contrary to the other theories discussed so far, it is difficult to estimate this tidal mass analytically.
The typical clump density can be estimated using gravitational fragmentation theory, considering the smallest collapsing clumps,
namely clumps with radius equal to the sonic length. Under typical Milky Way conditions, this gives $\ell_{\rm sonic} \simeq 0.04$~pc,
and $\alpha_{\rm vir} < 1$ translates into $\rho_{\rm clump} > 3.8\times 10^{-18}~{\rm g~cm}^{-3}$.
As discussed earlier, the first Larson core has a well defined mass with $M_{\rm 1LC}=0.01 M_\odot$. 
The gaseous envelope properties, on the other hand, emerge from the supersonic turbulence in the cloud, and are not uniquely defined.
We assume here that the envelope is a strict singular isothermal sphere, connected to the first Larson core with the Jeans density at the Jeans radius,
both defined in Section~\ref{sec:jeans_mass}, with $\alpha=2$, $\rho_0=\rho_{\rm J}=5 \times 10^{-14}{\rm g~cm}^{-3}$ and $r_0=R_{\rm J} \simeq 12.5$~AU.

If we first ignore the tidal forces from the envelope, we can apply the collapse condition from Equation~\ref{eq:collapse_condition},
and find $r_{\rm tidal}=656$~AU for $\rho_{\rm clump}=10^{-17}{\rm g~cm}^{-3}$
and $r_{\rm tidal}=305$~AU for $\rho_{\rm clump}=10^{-16}{\rm g~cm}^{-3}$. 
We see that denser clumps can survive tidal stripping closer to the first Larson core.
If we consider now only the gaseous envelope, we find similar values, with 
$r_{\rm tidal}=1531$~AU for $\rho_{\rm clump}=10^{-17}{\rm g~cm}^{-3}$
and $r_{\rm tidal}=484$~AU for $\rho_{\rm clump}=10^{-16}{\rm g~cm}^{-3}$. Overall, this gives a rather large range of values between 300 and 1500~AU.
The corresponding tidal masses range from $0.05 M_\odot < M_{\rm tidal} < 0.25 M_\odot$, in good agreement with the characteristic mass of the IMF.
A complete view of the parameter dependence of the tidal radius and tidal mass is shown in Fig. \ref{fig:dependence}.
For envelope profiles with a slope of $\alpha = 2$, we recover the previous estimates, with tidal masses around 0.1 \Msun.
For shallower profiles with the same normalisation, it is much more difficult for clumps to collapse close to the centre.
The tidal mass is thus larger.
The opposite is true for steeper profiles, where the clump density quickly becomes much larger than that of the background envelope.
In this case the tidal mass typically becomes smaller than 0.1 \Msun. 

Following \cite{Lee_Hennebelle2018b}, we can also consider that the clump density is proportional to the gaseous envelope,
with
\begin{equation}
\rho_{\rm clump} = A \rho_{\rm gas}(r) 
\end{equation}
Injecting this into Equation~\ref{eq:collapse_condition}, we see that a condition for the clump to collapse is $A>\frac{3(\alpha - 1)}{(3-\alpha)}$.
Only clumps significantly denser than the background gas envelope can collapse, less dense clumps are always tidally stripped.
If we consider again the singular isothermal sphere and choose $A=4$ as a minimum clump overdensity, the collapse condition from Equation~\ref{eq:collapse_condition}
becomes identical to our simple derivation of Section~\ref{sec:tidal_theory_quick} and the tidal mass becomes $M_{\rm tidal}=6 M_{\rm 1LC}$.

The arguments discussed here give us some analytical estimates for the tidal mass. Our next step is to use simulations to try and estimate the tidal mass directly from the actual gas distribution.

\section{Simulation setup}\label{sec:setup}
\begin{figure*}
\center
\includegraphics[scale=0.75]{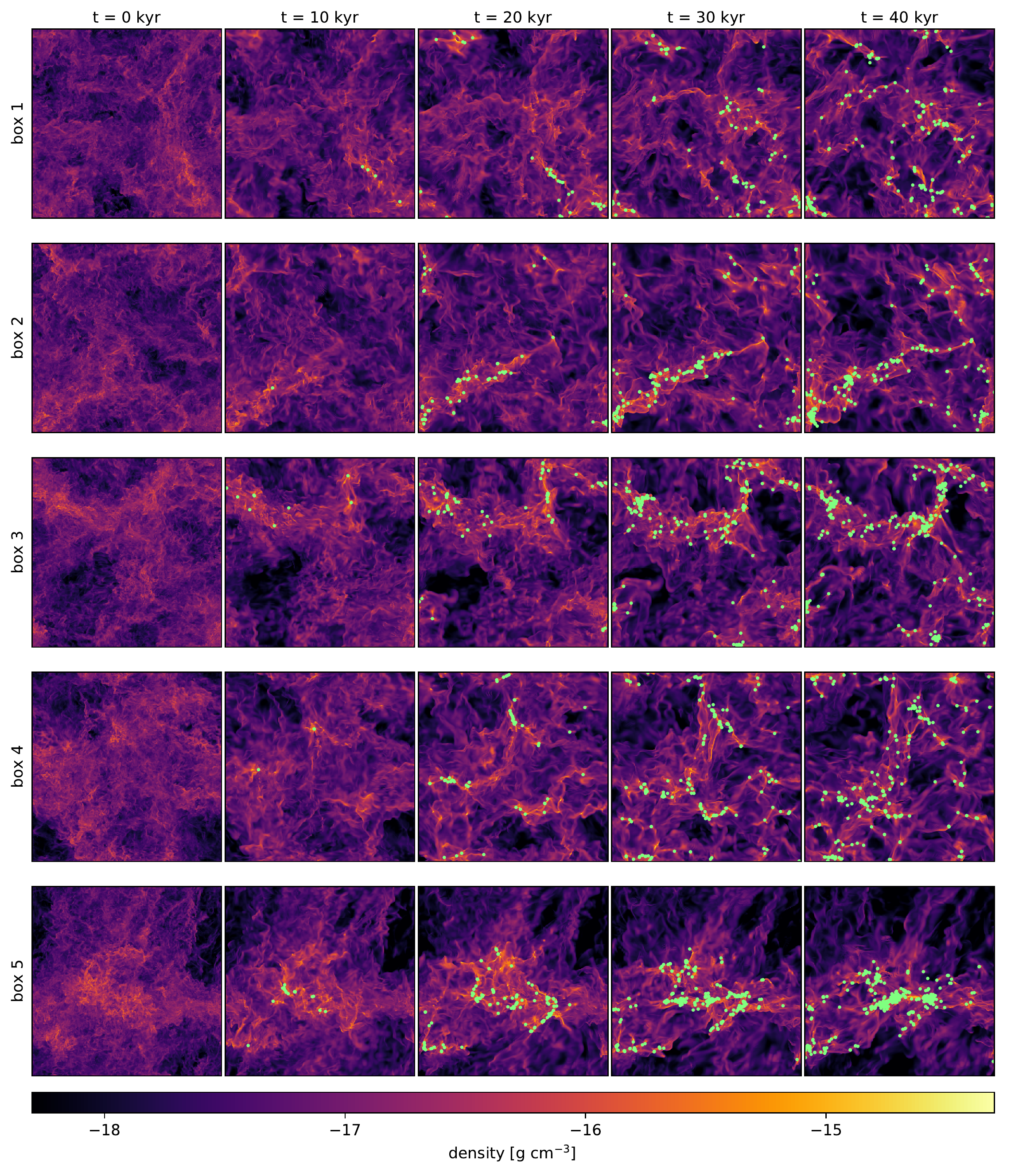}
\caption{Time evolution of the five different clouds. Each panel is a density map averaged along the line of sight. The green dots mark the sink positions.}
\label{fig:maps_all_boxes_evo}
\end{figure*}

To explore the idea that tidal forces play an important role in setting the peak of the IMF further, we run a suite of turbulent box simulations.
The goal is to study the tidal stripping in a realistic turbulent setting.
The simulations are designed to form many stars while maintaining a resolution for which individual star can be distinguished.
This allows us to investigate specific locations of star formation in detail, as well as providing good statistics on the cloud scale.
The simulations are performed with the Adaptive Mesh Refinement (AMR) code \texttt{RAMSES} \citep{Teyssier2002} taking into account 
hydrodynamics, self-gravity and a polytropic equation of state to mimic the effect of radiation.
Sink particles are used to represent stars.
We choose a base refinement level of 8 covering a periodic box of size 0.25~pc and a maximum refinement level 12, which gives us a maximal spacial resolution of 12.6 AU.
This maximum resolution corresponds exactly to $R_{\rm J}$, the Jeans length at the opacity limit.
This is also roughly the size of the first Larson core.  
In Appendix \ref{app:resolution_effects}, we show that this resolution is good enough for the purpose of this study and that our results are converged.
We refine the grid based on a Jeans-length refinement criterion, namely we always resolve the local Jeans length with at least 4 cells until we reach the maximum level of refinement.

\subsection{Initial conditions}\label{sec:ICs}

Our simulation setup consists of a periodic box of size 0.25 pc containing 260 $\mathrm{M}_{\odot}$ with fully developed turbulence with a Mach number $\mathcal{M}_{1\mathrm{D}} 
= \mathcal{M}_{3\mathrm{D}} / \sqrt{3} \simeq 7.3$.
In this simulation suite, we use five initial conditions, which only differ by the random seeds used to generate the Gaussian random field for the turbulent velocity.
All the other average properties are strictly identical.
We will refer to these as box 1 to box 5.
Using different randomised turbulence gives us different large scale structures in each box.
This allows us to examine many configurations with different tidal fields and gives us an idea of how the direct environment of the protostar can influence its growth.
For details about how the initial conditions are generated, we refer to Appendix \ref{app:ICs}.

Note that we have no external turbulence driving force. During the run, the turbulence only evolves through the combined effects of hydrodynamics and self-gravity.
Note also that our box is quite far from typical Milky-Way conditions and Larson's relation.
This small 0.25~pc box should be considered as a small, compact, high density region within a much bigger molecular cloud for which the 1D velocity dispersion approaches 1.5~km/s.
Using a periodic box without driving means this piece of cloud is considered to be fully decoupled from the parent cloud.
This simplified set-up is valid for the purpose of our experiment, which is to focus on the small mass end of the IMF and investigate the effect of local tidal forces around first Larson cores.

\subsection{Equation of state}\label{sec:EoS}

Initially the temperature is set to $T_0 = 10$~K everywhere in the box.
Since we do not include radiative transfer in our simulations, we adopt instead a simple recipe based on a polytropic equation of state (EOS)
$$
T = \left\{
    \begin{array}{ll}
        T_0 & \mathrm{if} \, \rho < \rho_{\mathrm{knee}} \\
        T_0 \left( \frac{\rho}{\rho_{\mathrm{knee}}} \right)^{\Gamma - 1} & \mathrm{if} \, \rho > \rho_{\mathrm{knee}}
    \end{array}
\right.
$$
For densities below a certain threshold $\rho_{\mathrm{knee}}$, the gas is strictly isothermal.
Above this density, $T(\rho)$ follows a power law with slope $\Gamma - 1$.
This mimics the fact that, when the optical depth becomes higher than 1 and the density exceeds the critical density at the opacity limit, radiation gets trapped and the evolution 
becomes adiabatic. 
We use a value of $\rho_{\mathrm{knee}}=5 \times 10^{-14} \, \mathrm{g}/\mathrm{cm}^3$ or $n_{\mathrm{knee}}=2.5 \times 10^{10} \, \mathrm{H}/\mathrm{cm}^3$ 
and a polytropic index of $\Gamma=5/3$ consistent with results found in RHD simulations \citep{Vaytet_et_al2013}.
It has been pointed out that the EOS is important in shaping the IMF \citep{Kitsionas2007}.
Some form of polytropic EOS is certainly needed, since it was shown that pure isothermal simulations produce an IMF peak which directly depends on the resolution \citep{Guszejnov_et_al2018}.
Without a transition to the adiabatic regime, there is no mechanism that allows for the formation of first Larson cores with a well-defined physical mass.

\subsection{Sink particle formation}
\label{sec:sinks}

When a density peak reaches conditions for which it is expected to form a first Larson core that will trigger the second collapse towards the protostar, 
we introduce a sink particle \citep{Bate_at_al1995, Bleuler2014}.
Because our simulations are highly resolved, we believe sink particles can represent individual stars (see Appendix \ref{app:resolution_effects}).
Resolving the second collapse is however quite demanding in term of spatial resolution. 
It has been shown by \cite{Lee_Hennebelle2018b} that the internal structure of the first Larson core
allows one to adopt a relatively low density threshold, slightly larger than the critical density corresponding to the opacity limit,
and still get the right statistics for the second collapse sites. The main requirement for us is thus to resolve the spatial extent of the first Larson cores,
detect when the peak density of the first Larson core exceed a density threshold $\rho_{\mathrm{threshold}} \ge \rho_{\rm crit}$ and introduce a sink particle.
These are the key elements which allow us to investigate the true IMF (instead of binary or cluster IMF).

The details of our sink particle formation criteria are as follows.
Sinks are formed within density peaks, called here clumps, found by the native \texttt{RAMSES} clumpfinder \texttt{PHEW} \citep{Bleuler2015}.
Individual cells are therefore not all eligible for sink formation, only cells located at density maxima.
A clump must fulfill the following criteria to be eligible for sink formation:
\begin{itemize}
\item it has to be dense enough: $\rho_{\mathrm{peak}} > \rho_{\mathrm{threshold}}$
\item it has to be relevant enough: $\rho_{\mathrm{peak}} > 2 \rho_{\mathrm{saddle}}$ \citep[see][for the exact meaning of relevance]{Bleuler2015}
\item the sphere of radius $R_{\rm acc}$ and centred on $r_{\mathrm{peak}}$ has to contain a mass larger than $M_{\mathrm{seed}}$
\item it is not kinetically supported: $E_{\mathrm{kin}} < - E_{\mathrm{grav}}$
\item it is not thermally supported: $E_{\mathrm{therm}} < - E_{\mathrm{grav}}$.
\item there is no existing sink nearby: $|r_{\mathrm{peak}} - r_{\mathrm{sink}}| > 2 R_{\mathrm{acc}}$
\end{itemize}
The density threshold and seed mass are chosen to match the conditions in the first Larson cores that trigger the second collapse. We adopt here 
and at our fiducial resolution $\rho_{\mathrm{threshold}} = 5 \times 10^{-14}\, \mathrm{g/cm}^3$ and $M_{\mathrm{seed}} = 5 \times 10^{-3}$ \Msun.
Once a sink is formed it will accrete gas at the Bondi rate \citep{Bleuler2015}, with all Bondi flow parameters estimated within the accretion radius $R_{\mathrm{acc}}=50$~AU,
or equivalently 4 cells at the maximum level of refinement.
In practice, after the sink is created with an initial mass equal to $M_{\mathrm{seed}}$, the rest of the first Larson core mass is quickly accreted and the pressure support disappears, 
mimicking quite realistically the second collapse. 

Another important parameter in our sink particle algorithm is the sink merging time scale. 
New sinks are allowed to merge with another older sink only during a time scale $t_{\rm merge}$ after their birth  \citep{Bleuler2015}.
This time scale is associated with the first Larson core lifetime, and is usually chosen around 1000~yr, which corresponds to the dynamical time of the first Larson core \citep{Bhandare_et_al2018}.
During this time scale, we expect the first Larson core to still be gaseous and a collision with another core will lead to one protostar rather than a binary.
If the merging timescale is chosen too small, one can get too many small mass stars in multiple systems.
If the chosen value is too high, not a single binary will form.
In our simulation, this parameter is set to $t_{\rm merge}=1500$~yr.

The resolution of our fiducial set of simulations is fixed to 12.6~AU, corresponding to AMR level 12. In Appendix~\ref{app:resolution_effects}, we perform a resolution study 
to quantify the robustness of our results. In this study, we adopt the same polytropic equation of state, so that the properties of our first Larson cores are preserved. 
We do however adjust the sink particle formation parameters, so that the density threshold for sink formation is larger and closer to the onset of the second collapse, and
the seed mass at sink creation and the sink accretion region are smaller with increasing resolution. The conclusion of this resolution study is that our fiducial case ($\ell_{\rm max}=12$) is properly converged.

\section{Results}\label{sec:cloud_results}


\begin{figure}
\center
\includegraphics[scale=0.65]{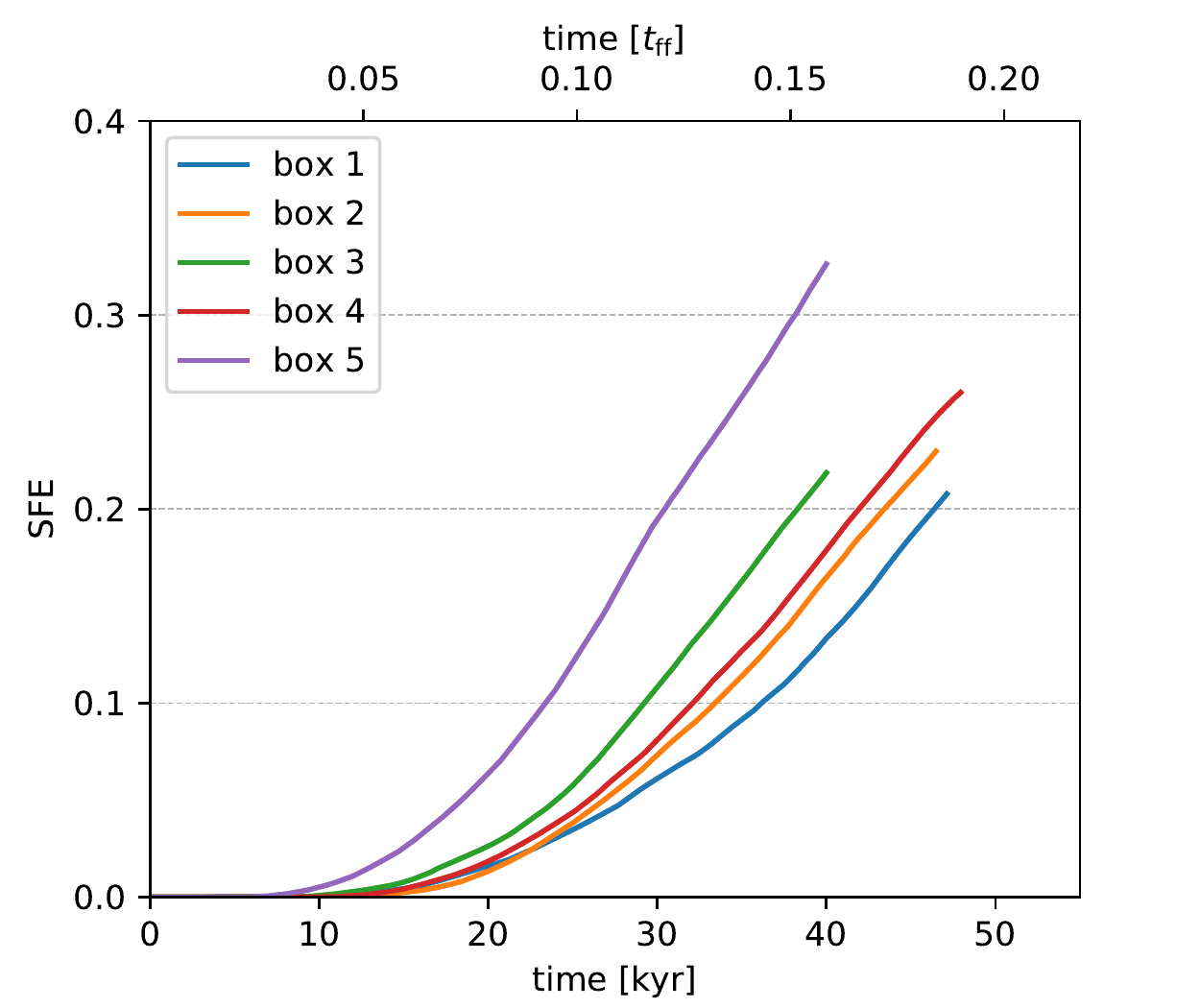}\\
\caption{Comparison between the five ICs of the time evolution of the star formation efficiency.}
\label{fig:SFE_compare}
\end{figure}

\begin{figure*}
\center
\includegraphics[scale=0.70]{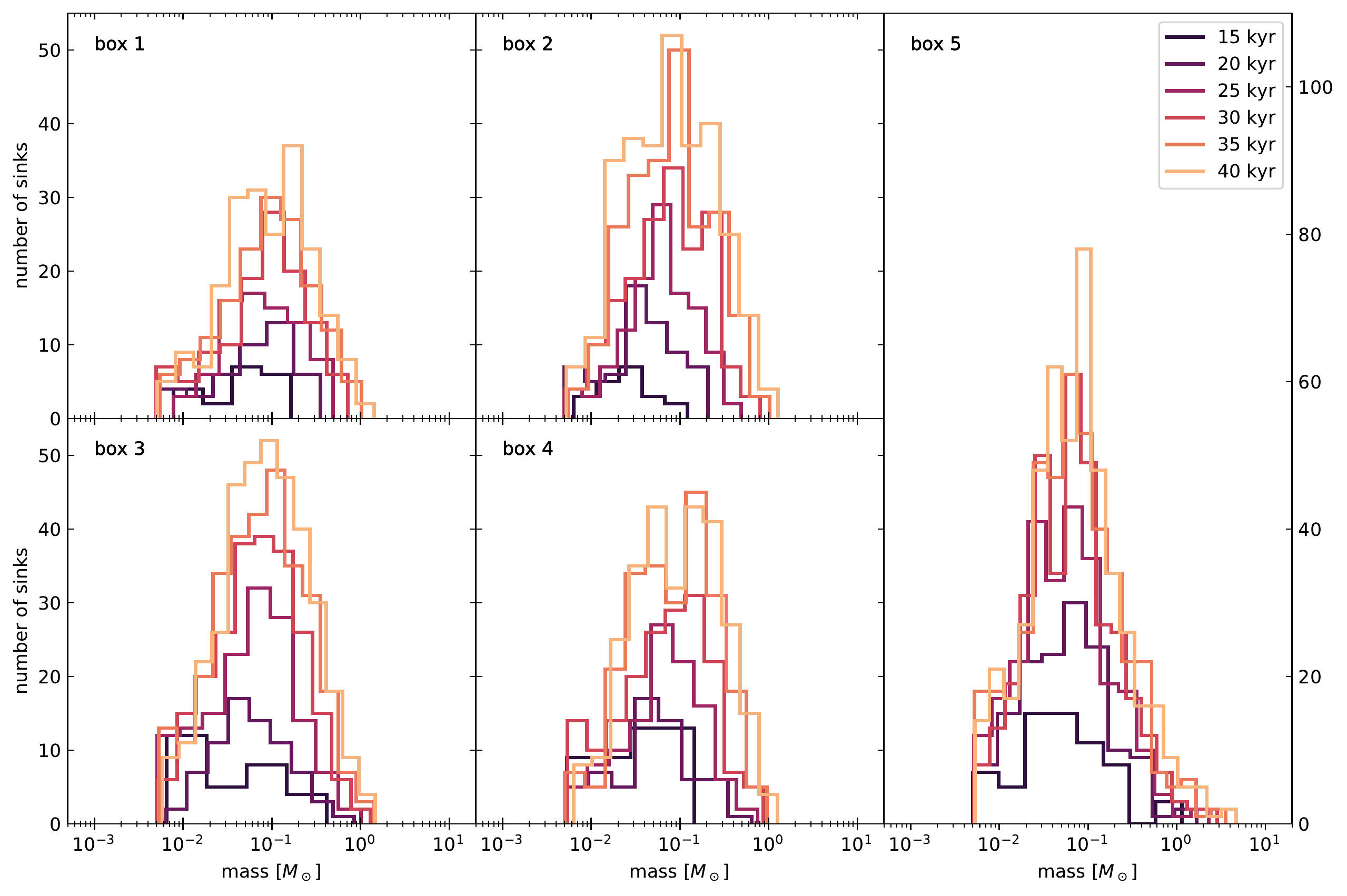}
\caption{Time evolution of the IMF for the five different boxes.}
\label{fig:IMF_all_boxes_evo}
\end{figure*}


\subsection{Evolution of the clouds}

The evolution of the five clouds is visualised in Fig.~\ref{fig:maps_all_boxes_evo}.
In an initial stage, large-scale over-dense regions start to collapse under gravity, forming filaments which are the locations of the formation of the first sinks.
More and more stars form as the simulation continues and the mass in the box is slowly depleted.
Fig.~\ref{fig:SFE_compare} shows the evolution of the total mass in stars compared to the initial cloud mass, also called the star forming efficiency (SFE)
\begin{equation}
\mathrm{SFE} = \frac{M_*}{M_{\mathrm{gas,init}}}
\end{equation}
Once sink formation starts, the SFE initially evolves roughly parabolic and later transitions into a linear regime.
What is striking is that the different SFEs vary by a factor of up to 3 for our different random seeds, even though the boxes have the same exact global mass, size and Mach number.
Box 5 is most notable, with a SFE much higher than all the other boxes at all times.

As one can see in the density maps (Fig.~\ref{fig:maps_all_boxes_evo}), the primary locations of sink formation are the filaments formed from large scale over-densities.
We see that box 5 has one massive filament in the centre of the box, while in box 1 for example, there are many small filaments which are spread out in a more uniform way over the entire volume.
The presence of large scale filaments thus has a strong effect on the overall star formation rate.

The origin of this difference in geometry might be that the turbulent energy is in different modes for the various initial conditions.
It is indeed well known that star formation is more efficient in presence of strong compressive modes, than for pure solenoidal turbulence \citep{Federrath_et_al2010, Federrath_Klessen2012, Orkisz2017}
but this is not the topic of this paper so we don't elaborate further. 
We only use this diversity of compressive and solenoidal modes to deliver a variety of possible environments in our star forming regions.

\subsection{The simulated IMF}
\begin{figure*}
\center
\includegraphics[scale=0.70,trim={0cm 0cm 0cm 0cm},clip]{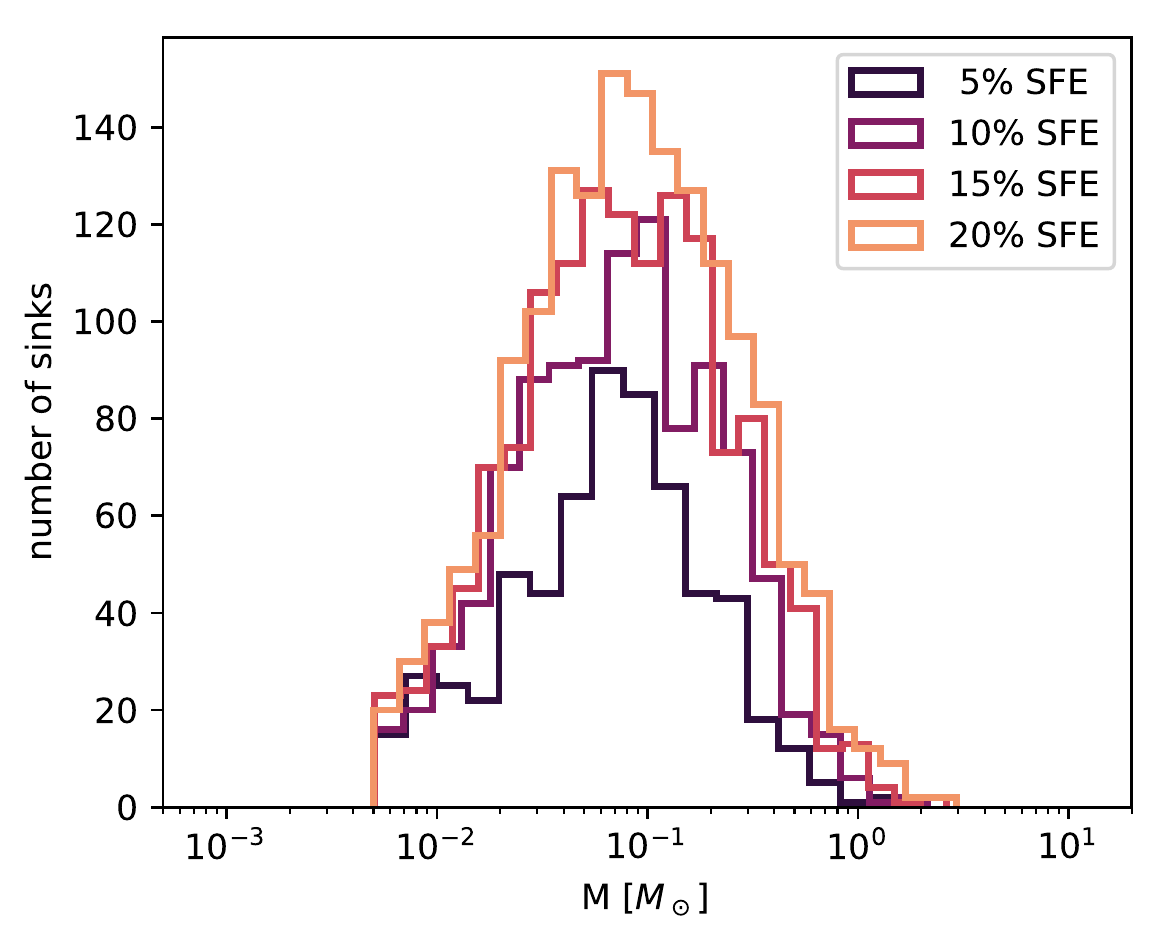}
\includegraphics[scale=0.70,trim={0cm 0cm 0cm 0cm},clip]{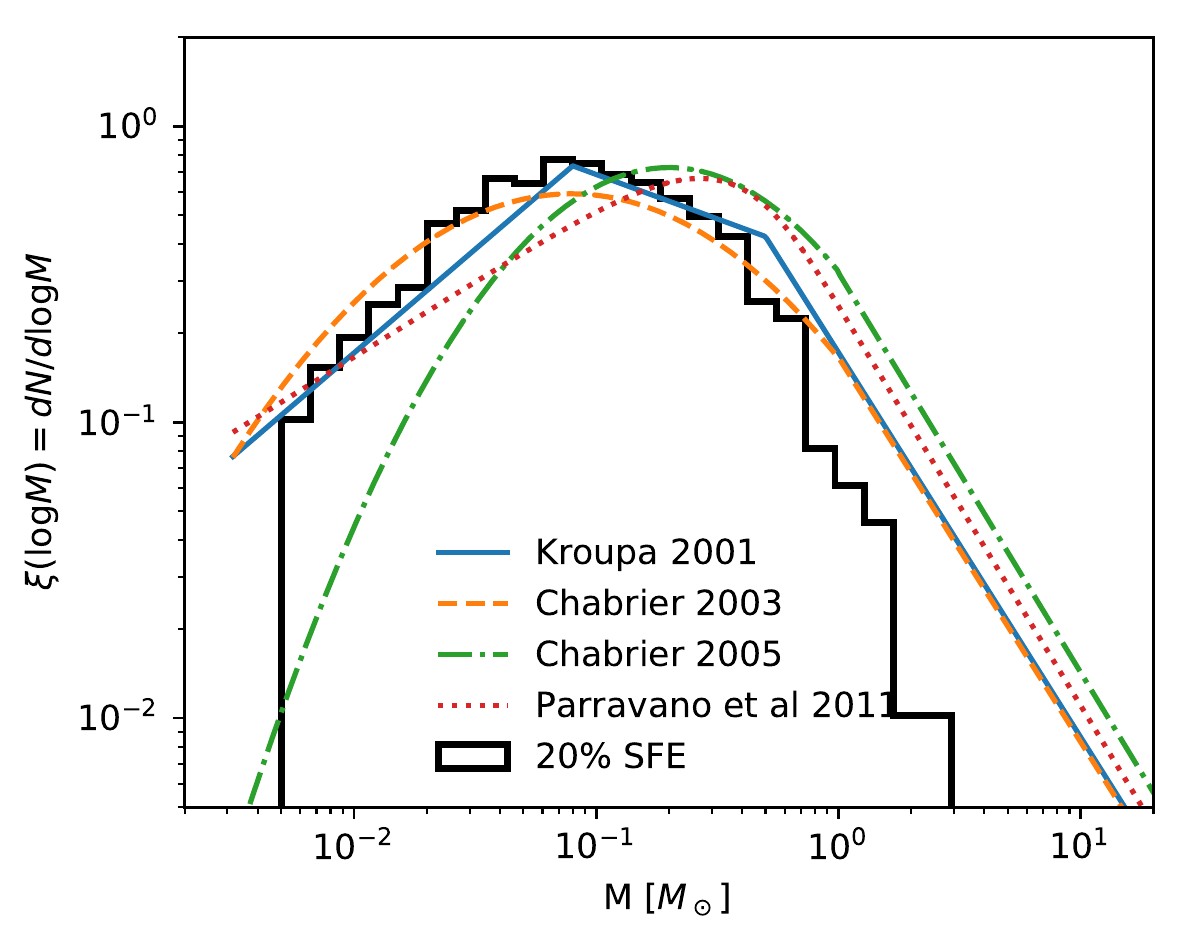}
\caption{Top: Stacked IMF at different SFE. Bottom: Stacked IMF at 20\% SFE compared to the analytical forms.}
\label{fig:stacked_IMF_lvl12}
\end{figure*}

Fig.~\ref{fig:IMF_all_boxes_evo} shows the number of sinks as a function of mass for the five different boxes.
Since in our simulations, a sink particle represents an individual star, this is equivalent to the IMF.
Even though they have very different SFEs, for each box there is a clear peak around 0.1~\Msun.
To improve the statistics, we can stack the data of the five boxes together and plot the resulting mass function, which is shown in Fig.~\ref{fig:stacked_IMF_lvl12}.
The stacking is done at different times, when the simulations all have the same SFE, to not let one simulation dominate the result.
Also here, after stacking, we see a clear peak around 0.1~Msun.
Remark that the position of the peak remains roughly constant during the simulation. The exact moment we evaluate the IMF is thus not important.

Comparing our sink mass function with some of the commonly adopted analytic forms of the IMF (Fig.~\ref{fig:stacked_IMF_lvl12}, bottom panel), 
we see that the simulated peak agrees quite well with the observed one, especially for the Kroupa and Chabrier IMFs.
This indicates that even though our simulation only contains gravity and hydrodynamics, with a simplified EOS to account for radiation, we manage to capture a physical process that sets a clear peak in the IMF. Note that the agreement between our simulated results and the observed IMFs is not so good at the high mass end.
Only in box 5 do we find stars with a mass significantly higher than 1~\Msun.
This might be due to our limited box size and total mass, giving us poor statistics at the high mass end.
It has also been shown that radiation and magnetic field are key ingredients for the formation of massive stars \citep{Hennebelle_Commercon2012, Tan2014}.
Since we don't include these in our simulations, it is quite possible that the most massive cores fragment too much and do not allow for the formation of massive enough massive stars.
This is not important here, as the goal of this paper is to focus on the characteristic mass of the IMF, not the high mass end.
On the low mass end, however, we have a clear cutoff at a mass slightly lower than the first Larson core mass.
This could be associated with our adopted numerical seed mass, which is, for our fiducial runs, only half the first Larson core mass.
We show however in Appendix~\ref{app:resolution_effects}, that this cut-off persists even with a much lower seed mass 
(see Fig.~\ref{fig:IMF_res_stacked}).

Can any of the theories discussed in Section~\ref{sec:theories_peak} explain the characteristic masses we observe in these simulations?
First, we see very few sinks with masses lower than the first Larson core mass, around 0.01 $M_{\odot}$, 
so this characteristic mass seems to corresponds to a lower limit for the mass of stars.
The high-resolution run shows a few sinks with mass below the first Larson core mass, because our seed mass is significantly smaller (by a factor of 4) in this case. 
But these very low mass sinks are rapidly accreting their surrounding first Larson cores and quickly reach the low mass end cut-off.
Given our initial condition parameters, the Bonnor-Ebert mass is $M_{\rm BE} = 0.26~M_\odot$, where we use the average box density $M_{\rm box}/L^3$ as the external density, 
namely $\rho_{\rm ext}=1.1 \times 10^{-18}{\rm g~cm}^{-3}$. 
This is almost a factor of 3 too large to explain our peak value which lies around 0.1~$M_{\odot}$. 
As discussed in Section~\ref{sec:theories_peak}, a better approach in such turbulent environments is to use 
the compressed post-shock density as external density. This boils down to dividing the previous mass by the 1D Mach number of the turbulence ${\cal M} \simeq 7.3$. 
We now obtain $M_{BE}^{\mathrm{turb}} = 0.03~M_{\odot}$ which is a factor of 3 too low. The reason for this disagreement comes from our adopted cloud properties
that do not match the typical cloud conditions in the Milky Way. Our set-up is more compact, or for a fixed size, more turbulent. 
But this analysis also shows that the origin of the peak in the IMF in our simulations cannot be the turbulent Bonnor-Ebert mass.
Most of the other theories discussed in Section~\ref{sec:theories_peak} give similar characteristic masses, and therefore fail to explain our simulations.
The only theory that could still work here is the tidal screening theory.

\subsection{Studying the local tidal field}
\label{sec:tidal_exact}

The tidal screening theory states that because of the presence of large tidal forces around young protostars, no other stars can form within a certain volume,
called the tidally screened region.
To study this in our simulations, we look at the tidal tensor and its eigenvalues in the envelope around newly formed sink particles.

In each AMR cell, we compute the three eigenvalues and the corresponding eigenvectors of the tidal tensor, using the computed gravitational force stored in the \texttt{RAMSES} outputs.
If the eigenvalue is negative, we know that the gas at this location is currently being stretched in the direction of the corresponding eigenvector.
If the eigenvalue is positive, the gas is collapsing along the direction of the corresponding eigenvector.
To be able to collapse and form a first Larson core, a clump of gas needs to be collapsing in all three directions, otherwise it will simply turn into a long thin spaghetti,
fall towards the centre of the envelope and be accreted by the older first Larson core sitting there.

If we order the eigenvalues, it is thus enough to look at the smallest (or most negative) eigenvalue.
If this one is positive, we know that the others will also be positive and the gas parcel will collapse.
We can express the eigenvalue in units of density by dividing by $4\pi G/3$.
This is handy, because if we were to place a small homogeneous clump of certain density in the tidal field, 
we can immediately see whether it will be tidally stripped or not, since this simply adds a constant positive term to the most negative eigenvalue (see Eq.~\ref{eq:theory_eigenvals}).


Using the definition of the tidal tensor (Eq.~\ref{eq:def_tidal}), the most direct approach is to calculate the spatial derivatives of the gravitational acceleration.
Taylor expanding the acceleration around a fiducial point $\vec{x}_0$, we can write
\begin{align*}
a_i(x_1, x_2, x_3) &- a_i(x_{1,0}, x_{2,0}, x_{3,0}) \approx\\
&+ \frac{\partial a_i}{\partial x_1} (x_1-x_{1,0})
+ \frac{\partial a_i}{\partial x_2} (x_2-x_{2,0})
+ \frac{\partial a_i}{\partial x_3} (x_3-x_{3,0})
\end{align*}
or
\begin{align}\label{eq:tidal_comps}
\Delta a_i =
- T_{1 i} \Delta x_1
- T_{2 i} \Delta x_2
- T_{3 i} \Delta x_3
\end{align}
For each grid cell in the gas envelope around the first Larson core, we estimate the local tidal tensor's components $T_{ij}$ using a least square approach
on the closest neighbouring cells, using the relative positions and accelerations and fitting the above equation in each spatial direction.
We then symmetrise the obtained tensor and calculate its eigenvalues using \texttt{numpy}'s \texttt{linalg.eigh}.

An example of this tidal tensor eigenvalue analysis for an isolated sink is shown in Fig.~\ref{fig:tidal_bubble_sink1_A}.
The maps shows the smallest (most negative) of the 3 eigenvalues in a 2D slice around the sink.
Red colours are for a negative eigenvalue, corresponding to tidal stripping in the direction of the eigenvector, 
while blue colours are for a positive eigenvalue, corresponding to fully collapsing fluid elements.

Looking at the map, we can see the presence of a strong tidal field in a region of a several hundreds of AU (up to 1000~AU in diameter) around the sink particle, which is located in the centre of the image.
The three gray contours mark constant value of $-10^{-16}$, $-10^{-17}$ and $-10^{-18} \mathrm{g}/\mathrm{cm}^3$ respectively. 
If we were to place a new, potentially collapsing clump within one of these contours, it would need to have a density higher than this value to be able to collapse and resist tidal stripping.
Closer to the sink, the density has to be so high to overcome the tidal forces that it is virtually impossible for a potentially collapsing clump to form a star. 
These measurements confirm our analytical calculations in Section~\ref{sec:tidal_radius}, where we have estimated the tidal radius and the corresponding tidally screened mass
for typical potentially collapsing clump densities.

\subsection{Computing the tidally screened mass}

In the previous section, we have computed the tidal field around an isolated first Larson core, and showed that {\it potentially} collapsing clumps
would be tidally stripped and won't collapse. In this section, we want to go one step further and compute the actual mass that is tidally screened around the same first Larson core.
We will then apply the same method around {\it all} first Larson cores, right after they form during the simulation. Since this requires analysing all the simulation snapshots,
we need a fast method to measure the tidal field around sink particles.

\subsubsection{Spherical approximation of the tidal field}
\label{sec:tidal_spherical}

Instead of measuring the tidal force directly from the simulation like described in the previous section (which is computationally expensive), 
we want to develop and test a simple model inherited from our spherical analysis of Section~\ref{sec:tidal_theory}.
The tidal field in our star forming regions consists of the contribution of the sink and its surrounding turbulent envelope.
We approximate the contribution of the gaseous envelope using a spherical model with a power law density profile of the form $\rho(r)= A r^{-\alpha}$.

\begin{figure}
\center
\includegraphics[scale=0.60]{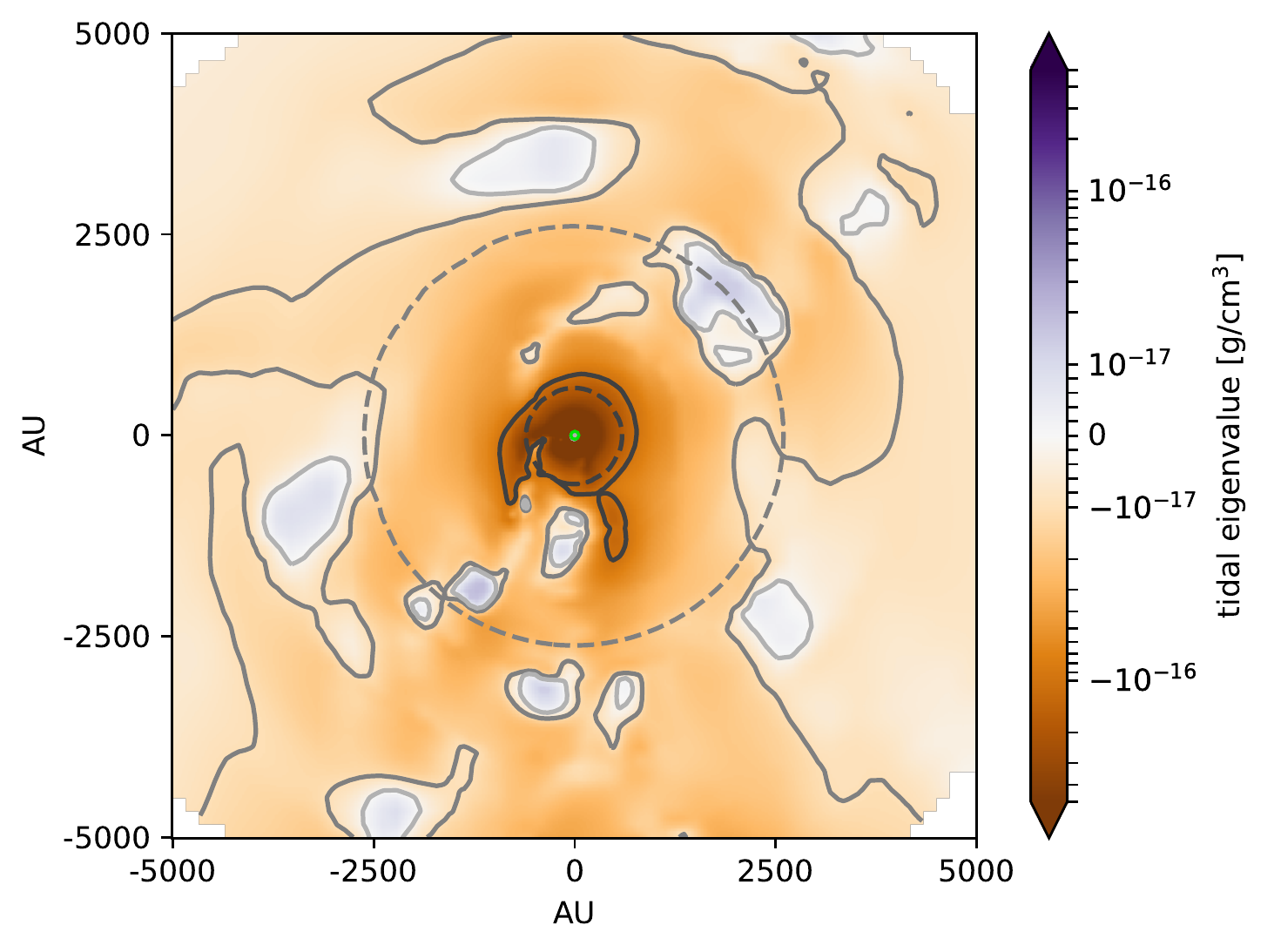}
\caption{Minimum tidal tensor eigenvalue as directly measured from the simulation output using the method described in the text (section \ref{sec:tidal_exact}). The map shows a 2D slice around an isolated sink, in the plane of the sink. The green circle in the centre represents the sink accretion radius. The solid contours show the eigenvalue thresholds of $-10^{-16}$, $-10^{-17}$ and $-10^{-18}$ g/$\mathrm{cm}^3$, while the dashed contour show the same thresholds for the spherical model (section \ref{sec:tidal_spherical}).}
\label{fig:tidal_bubble_sink1_A}
\end{figure}

To estimate the power law slope and the normalisation of the profile, we measure the enclosed mass at two different radii ($R_1=$ 300 AU and $R_2=$ 3000 AU).
From this we then find
\begin{align}
\alpha &= 3 - \frac{\log (M_1/M_2)}{\log (R_1/R_2)}\\
A&= \frac{3 - \alpha}{4 \pi} \frac{M_1}{R_1^{3- \alpha}}
\end{align}
where $M_1$ and $M_2$ are the masses enclosed in radius $R_1$ and $R_2$ respectively.
Note that this spherical fit can be a poor approximation of the actual density distribution, if the sink is not in the centre of the envelope or the region is dominated by a filament.
The corresponding gravitational acceleration is usually a much better fit, even in these unfavourable cases.
The most negative eigenvalue of the tidal tensor is then
\begin{align}
\lambda_\mathrm{model}[\mathrm{g}/\mathrm{cm}^3] &= \lambda_{r,*} + \lambda_{r,\mathrm{env}}\\
&= - \frac{3 M_{*}}{2 \pi r^3} - \frac{3(\alpha - 1)}{(3 - \alpha)} \rho (r)
\end{align}
The dashed circles in Fig.~\ref{fig:tidal_bubble_sink1_A} compares this simplified model to the exact tidal field.
We can see it compares quite well with the exact measurement. More importantly, the volume enclosed by the isocontours
matches quite closely to the actual tidally screened region, for a given clump density.

\subsubsection{Applying the collapse condition on the actual density distribution}

In our current simplified picture, we now want to estimate a local collapse criterion, using the gas density in each cell and comparing
it to the eigenvalue of our spherically symmetric tidal tensor.
We don't consider the density of a potentially collapsing clump virtually located at a given position like before, 
but we use instead the actual gas density at that position to compute the collapse condition described in Section~\ref{sec:estimate_tidal_force}. 
Usually, the dense filaments generated by the turbulence provide the over-densities within the envelope that will survive the tidal forces and collapse.
The remainder of the gas in the envelope will be tidally stripped (negative collapse criterion) and will define the tidally screened region.
Fig.~\ref{fig:tidal_bubble_sink1_B} visualises this collapse condition in a 2D slice around our isolated example.
In this map, the gas that is expected to be tidally stripped is shown in light to dark orange, while the gas that can collapse is shown in light to dark green.
When comparing with the density map in Fig.~\ref{fig:tidal_bubble_sink1_C}, we can see that collapsing regions closely follows the filaments surrounding the first Larson core.
A large positive value for the collapse criterion (the units are again in density units) means the gas will collapse in a short time scale, while a large negative value
corresponds to strong tidal stripping.

\subsubsection{The tidal bubble and the corresponding tidal mass}

\begin{figure}
\center
\includegraphics[scale=0.60]{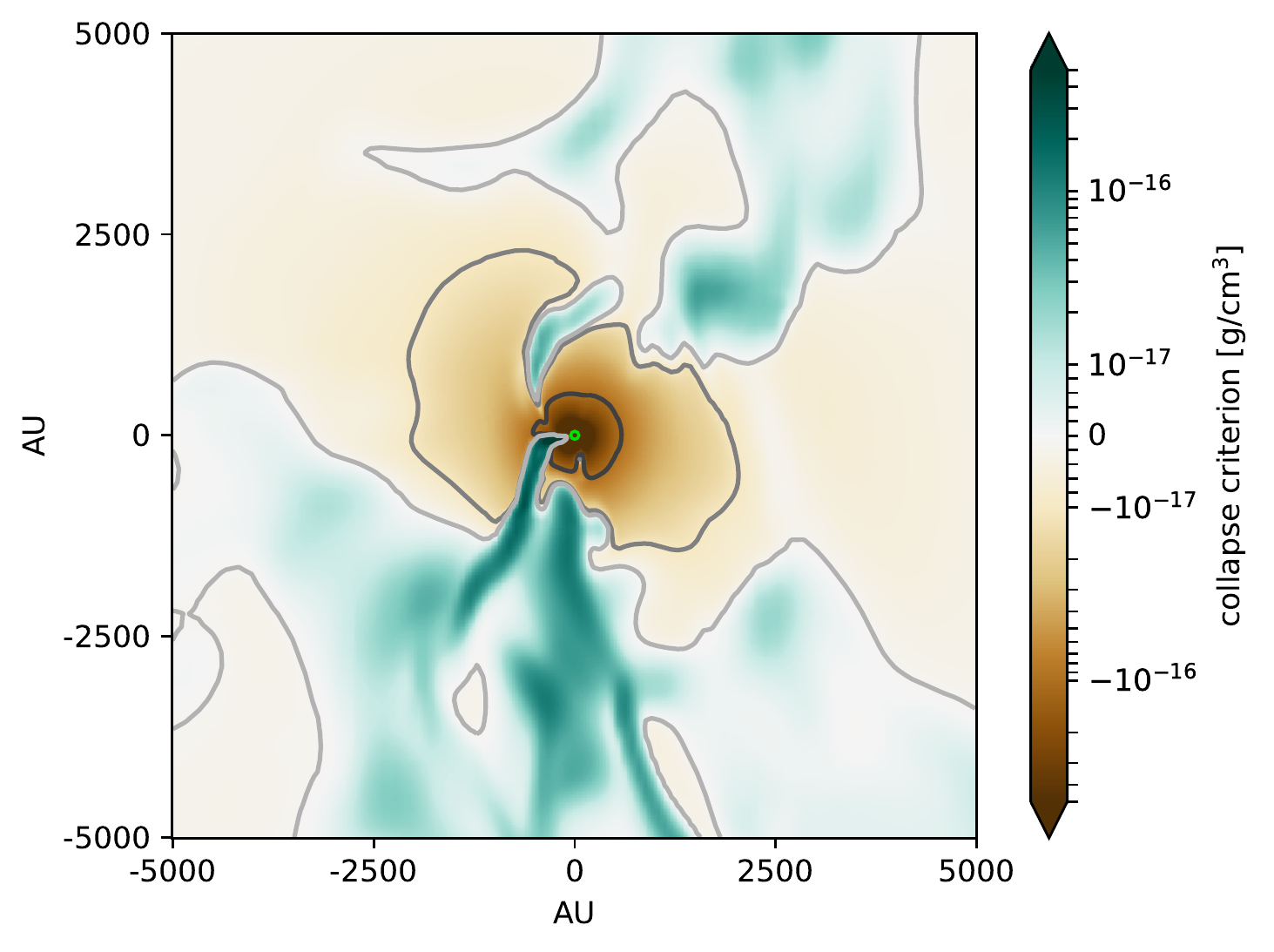}
\caption{Map of the collapse criterion of a 2D slice around an isolated sink, in the plane of the sink. The green circle in the centre represents the sink accretion radius. The contours mark a value of $-10^{-16}$, $-10^{-17}$ and $-10^{-18} \mathrm{g}/\mathrm{cm}^3$. Everything which is green is dense enough to collapse.}
\label{fig:tidal_bubble_sink1_B}
\end{figure}

\begin{figure}
\center
\includegraphics[scale=0.60]{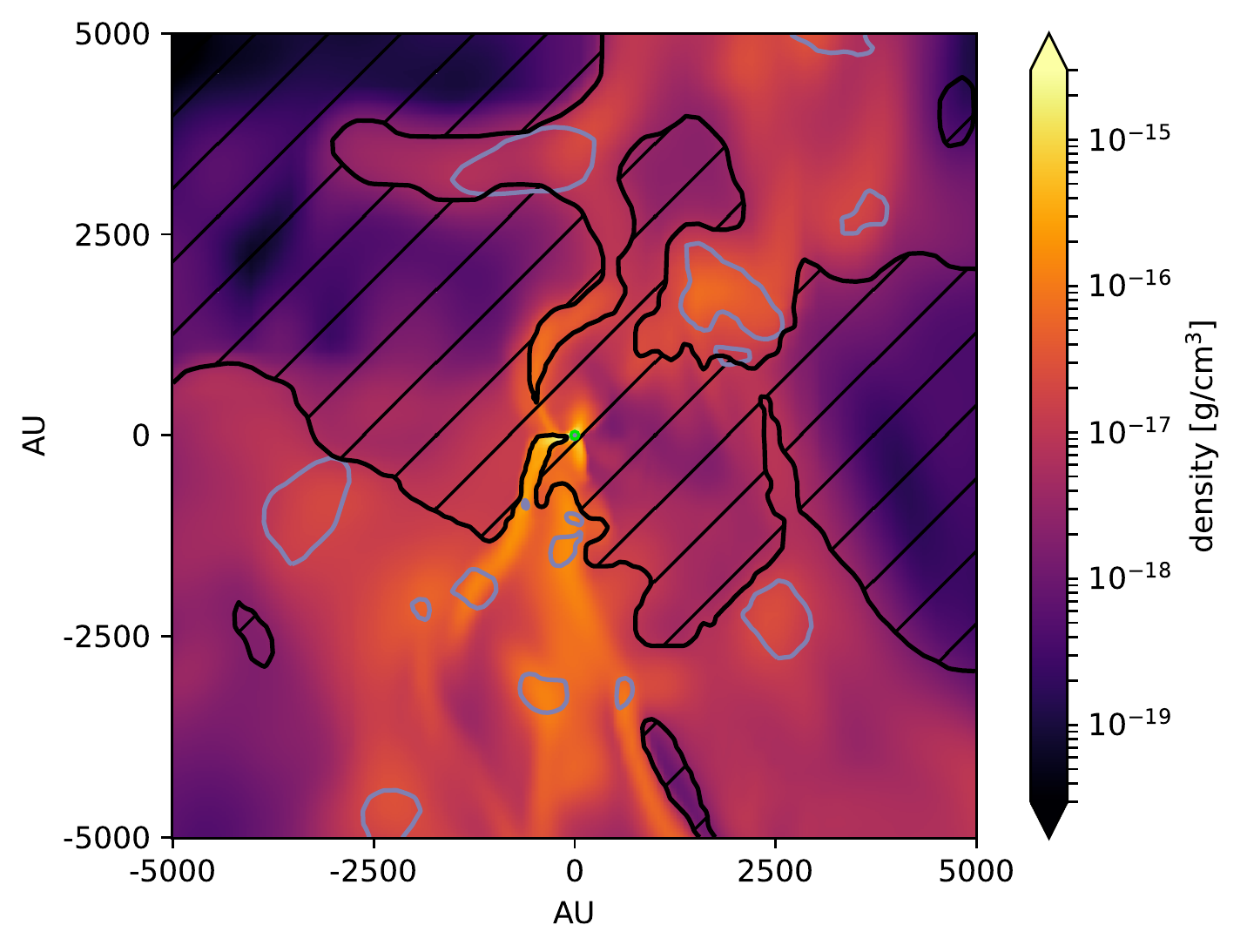}
\caption{2D slice of the gas density around an isolated sink, in the plane of the sink. The green circle in the centre represents the sink accretion radius. The tidal bubble is marked in black. The grey contours show the places where the exactly measured tidal eigenvalue tends toward collapsing ($-10^{-18}$ g/$\mathrm{cm}^3$ in Fig.~\ref{fig:tidal_bubble_sink1_A}).}
\label{fig:tidal_bubble_sink1_C}
\end{figure}

Note that the previous analysis is performed right after the formation of the first Larson core.
Our objective is therefore to predict at this very specific moment in time the mass that will be efficiently tidally stripped 
and end up in the central first Larson core, while the rest of the mass can eventually collapse and form other first Larson cores later on.
We will call the region which is dominated by tidal stripping the \textit{tidal bubble}.
While a natural definition for the tidal bubble would be the region for which the collapse condition is negative, not all of this gas will make it to the centre.
If the tidal field is only weakly stripping, a parcel of gas will not have the time to react to the effect before it is swept away by the turbulence or collapses onto another clump.
In practice, we thus have a stricter criterion and a threshold value for the collapse criterion which is smaller than zero.
In Fig. \ref{fig:tidal_bubble_sink1_B}, we overplot several possible threshold values.
We see that only for more negative thresholds, the tidal bubble is closed.

The exact threshold value chosen, is based on a timescale argument, as we will discuss in section \ref{sec:bubble_edge}.
In Fig.~\ref{fig:tidal_bubble_sink1_C}, we show a map of the gas density, with superimposed the region corresponding to the tidal bubble, shown as a black hatched area.
The total gas mass within the tidal bubble is the tidal mass (including the initial mass of the first Larson core) and, in the tidal screening theory, corresponds to the final mass of the star. 
We have also indicated as a grey contour the collapsing regions identified using the exact tidal tensor analysis performed in the previous section.
We see that they correspond nicely to all regions outside of the tidal bubble. 
As a caveat of our current approach, note that there is no guarantee that the gas within the tidal bubble at the formation time of the first Larson core will indeed end up in the 
final star, since the subsequent accretion will proceed in a highly dynamical environment. 

\subsubsection{Multiple systems and competitive accretion}
\label{sec:neighbors}

\begin{figure}
\center
\includegraphics[scale=0.60]{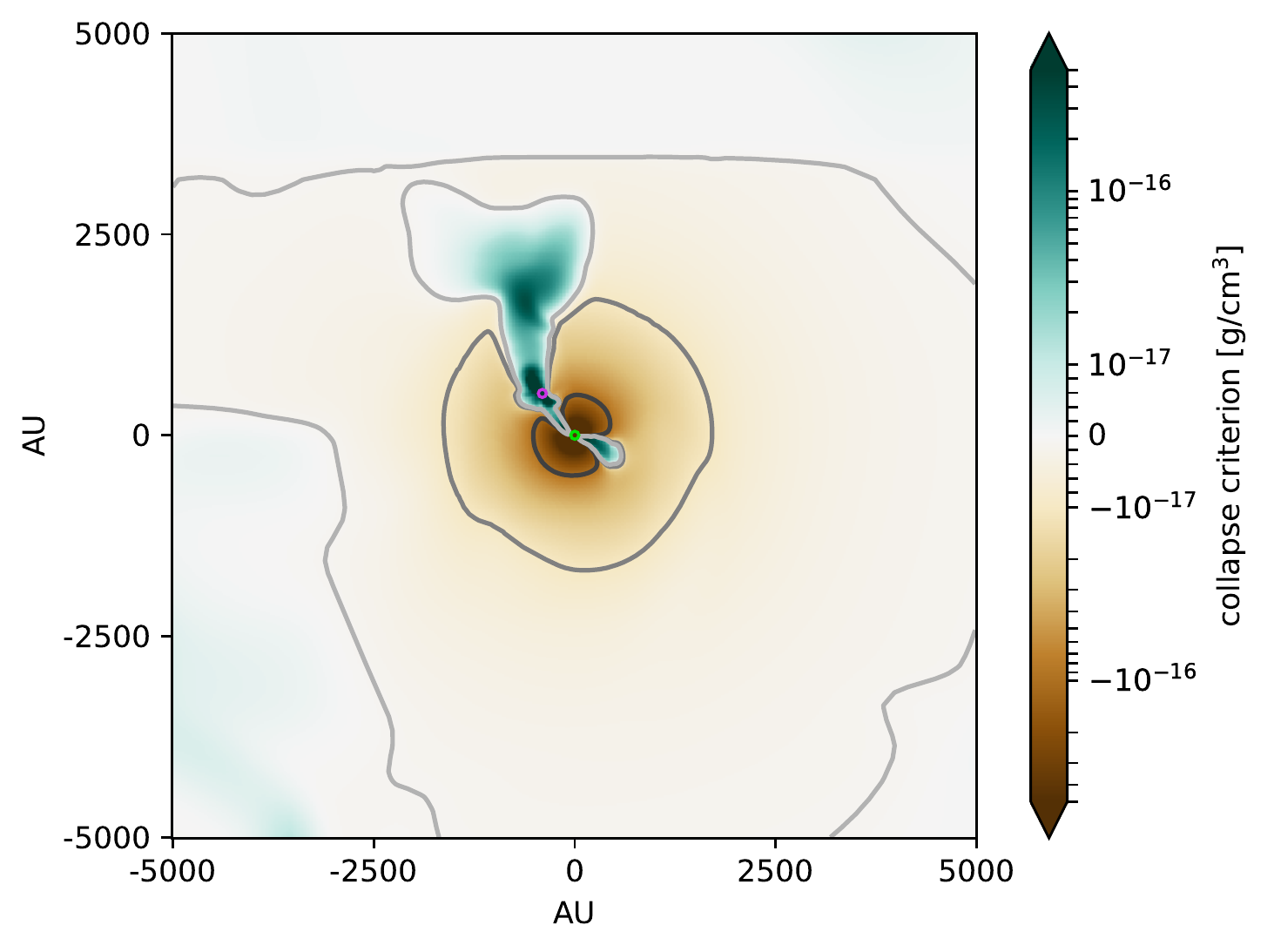}
\includegraphics[scale=0.60]{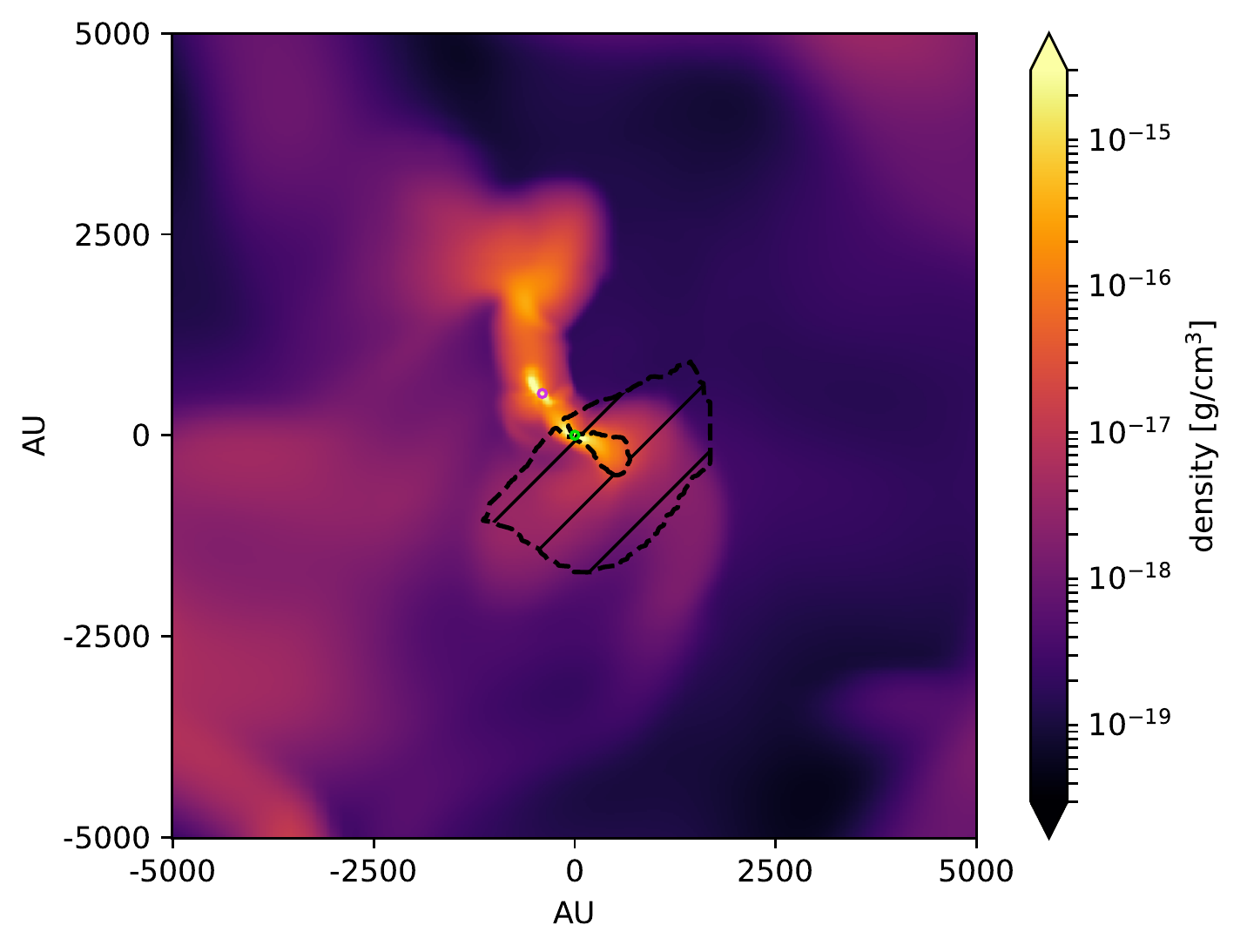}
\caption{Same as Fig.~\ref{fig:tidal_bubble_sink1_B} and \ref{fig:tidal_bubble_sink1_C} but now for a binary in the plane of both sinks. The calculation is centred on the green sink (the companion is shown in magenta). The hatched area corresponds to the tidal bubble of this green sink.}
\label{fig:tidal_bubble_sink7}
\end{figure}

In the previous sections we have analysed the simple case of an isolated, newly born first Larson core.
Many first Larson cores are however born in a much more complex environment, with multiple sinks forming a star cluster and sharing the same gas reservoir.

The first difficulty is to estimate the tidal field from the common gaseous envelope.
A young core appearing in a crowded envelope will see its gaseous envelope severely depleted of its mass by the older sink nearby.
The gas reservoir will therefore be significantly less massive compared to the first sink that formed in the same envelope. Its shape might also have changed.
In order to estimate the envelope parameters, we apply the same technique as before, using as centre of the density profile the position of the newly born core.

The second difficulty is to compute the tidal field resulting from all the sink particles in the vicinity of the newly born first Larson core.
Note that our goal is to assign to each new born core a tidally screened region, the tidal bubble.
In case of a multiple system, we do so by assigning a gas cell to a tidal bubble depending on the relative strength of the different sink plus envelope tidal fields.
Large sink particles with a strong tidal field will be assigned a larger tidal bubble.

In Fig.~\ref{fig:tidal_bubble_sink7}, we show an example of a binary system, with a younger core forming close to an older one.
The young sink is shown as a green circle while the old one is shown as a red circle.
The collapse criterion is shown on the top panel, and again, like for the isolated case, dense clumps and filaments are collapsing, while the rest of the envelope
appears as tidally stripped. In the bottom panel, we show the map of the gas density in the same region, showing the tidal bubble as the black hatched area.
We see that our analysis only assigns to the young sink the region which is dominated by the tidal field of the young sink.
The rest is assigned to the old sink and will end up accreted by the old sink according to the theory.

The same caveats we discussed in the previous section apply here, probably even more.
These often crowded and highly non-spherical environments are difficult to describe, so that complex numerical simulations are fully justified.
The theory of competitive accretion, as discussed in section \ref{sec:competitive_accretion}, is probably very useful in such cases. Although we don't explore it in the present paper,
combining tidal screening theory and competitive accretion appears as an interesting idea.
Overall, our current approach allows us to estimate roughly the final mass of the stars, by matching it with the tidal bubble at formation time.
We will see in the next section whether this can indeed explain the numerical IMF we have obtained.

\subsection{Predicting the IMF}
\label{sec:prediction}

\begin{figure}
\center
\includegraphics[scale=0.70]{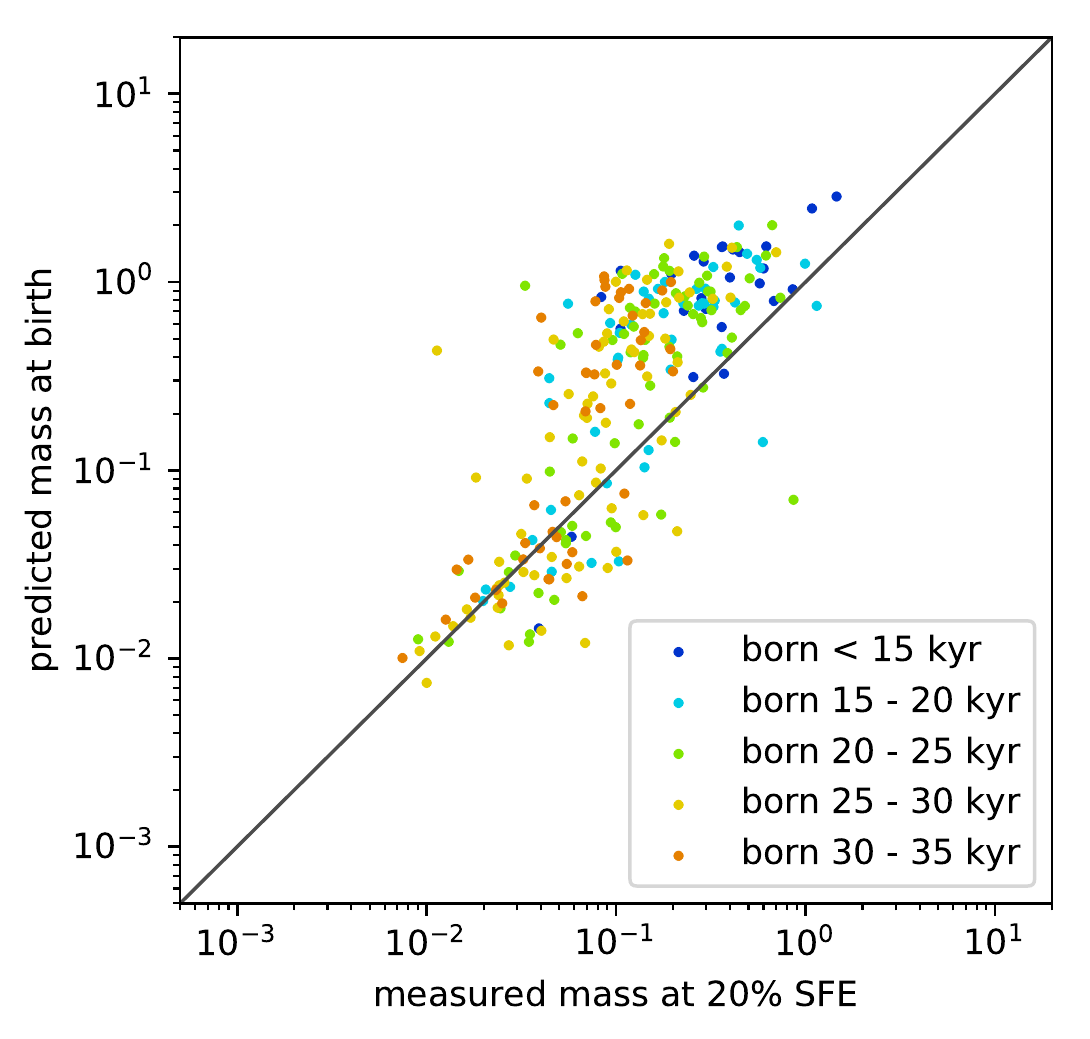}
\caption{Measured mass versus the mass contained in the tidal bubble defined as all the gas for which the collapse condition is negative, for sinks born at least 5 kyr before the end of the simulation (box 3).}
\label{fig:default_bubble_correlation}
\end{figure}

\begin{figure}
\center
\includegraphics[scale=0.63]{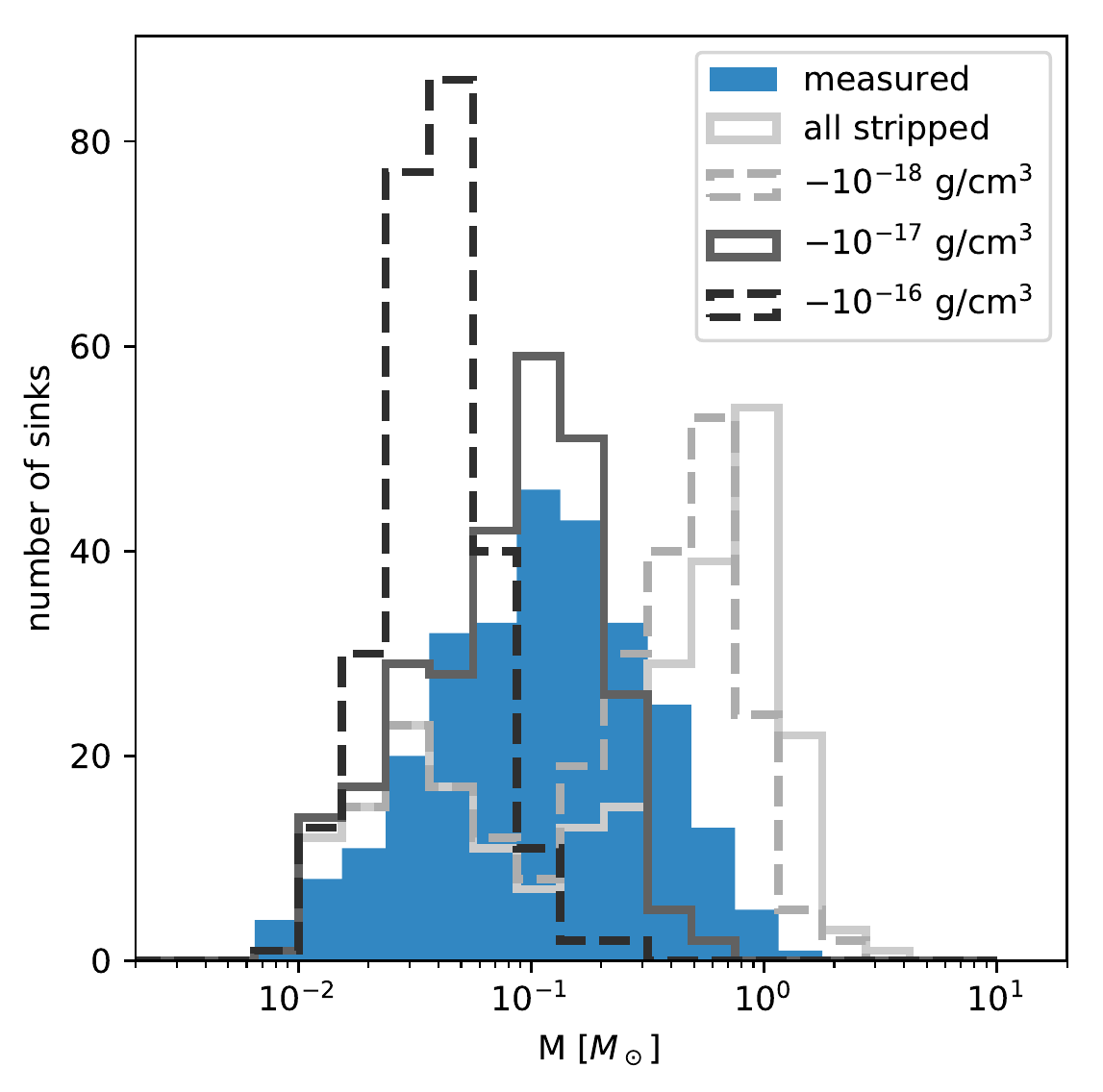}
\caption{Measured and predicted IMF of the sinks in Fig.~\ref{fig:default_bubble_correlation} using different definitions for the edge of the tidal bubble. There is a bimodality for a collapse condition threshold close or equal to zero.}
\label{fig:IMF_thresholds}
\end{figure}

\begin{figure*}
\center
\includegraphics[scale=0.60]{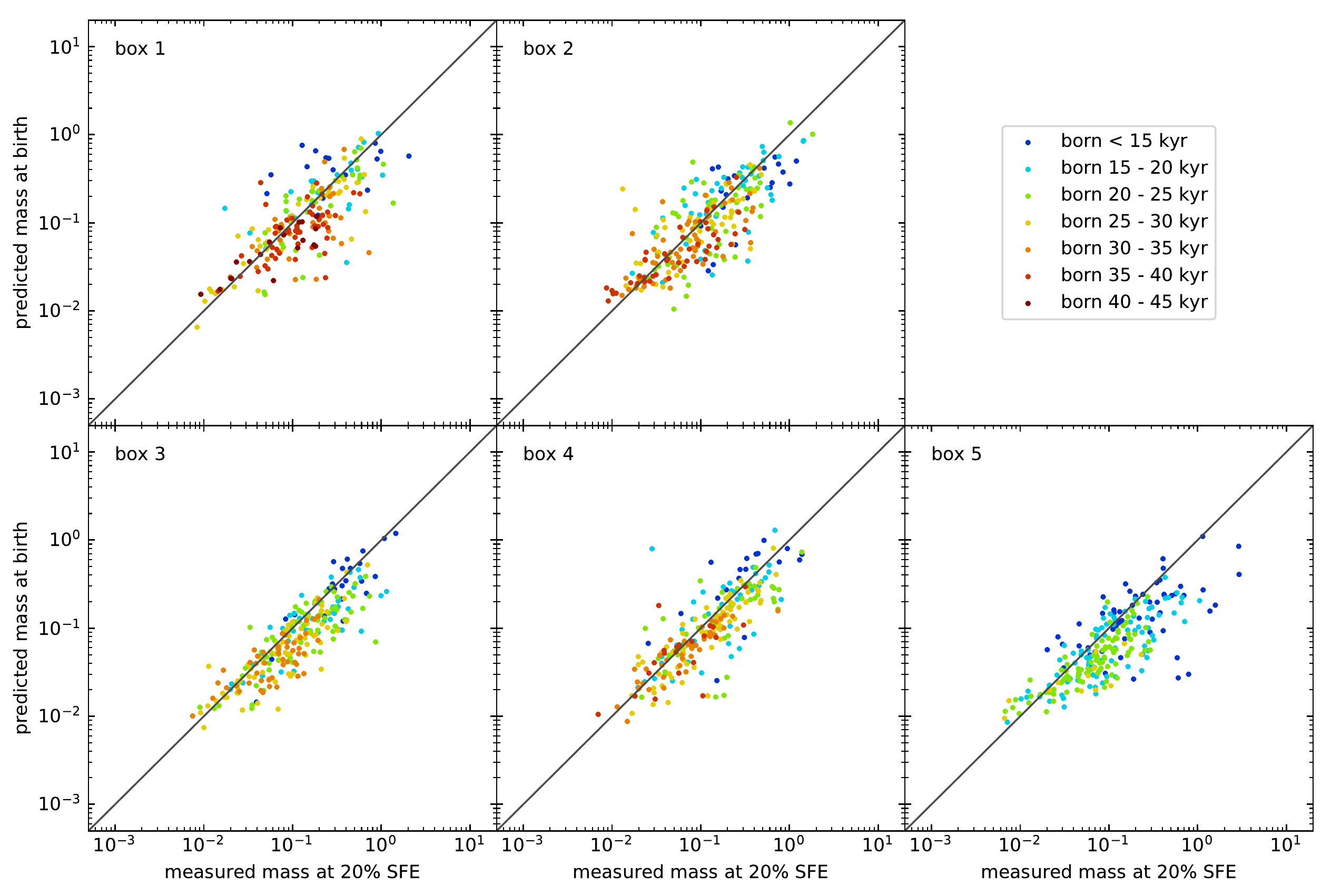}
\caption{Measured versus predicted sink mass for all sinks born at least 5 kyr before the end of the simulation.}
\label{fig:correlation}
\end{figure*}

\begin{figure*}
\center
\includegraphics[scale=0.60]{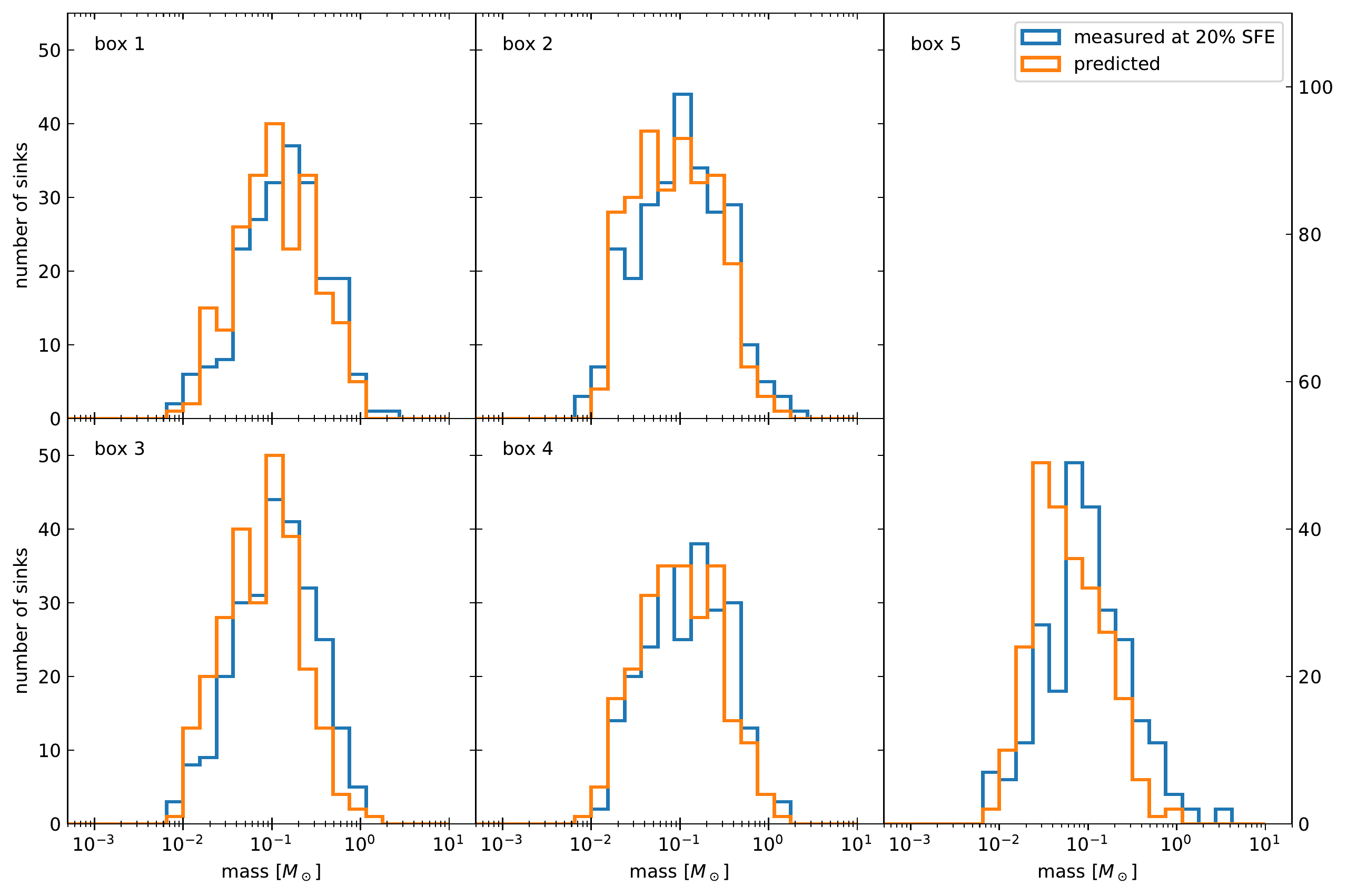}
\caption{Predicted and measured IMF of the sinks in Fig.~\ref{fig:correlation}.}
\label{fig:prediction_all_boxes}
\end{figure*}

\begin{figure}
\center
\includegraphics[scale=0.70,trim={0cm 0cm 0cm 0cm},clip]{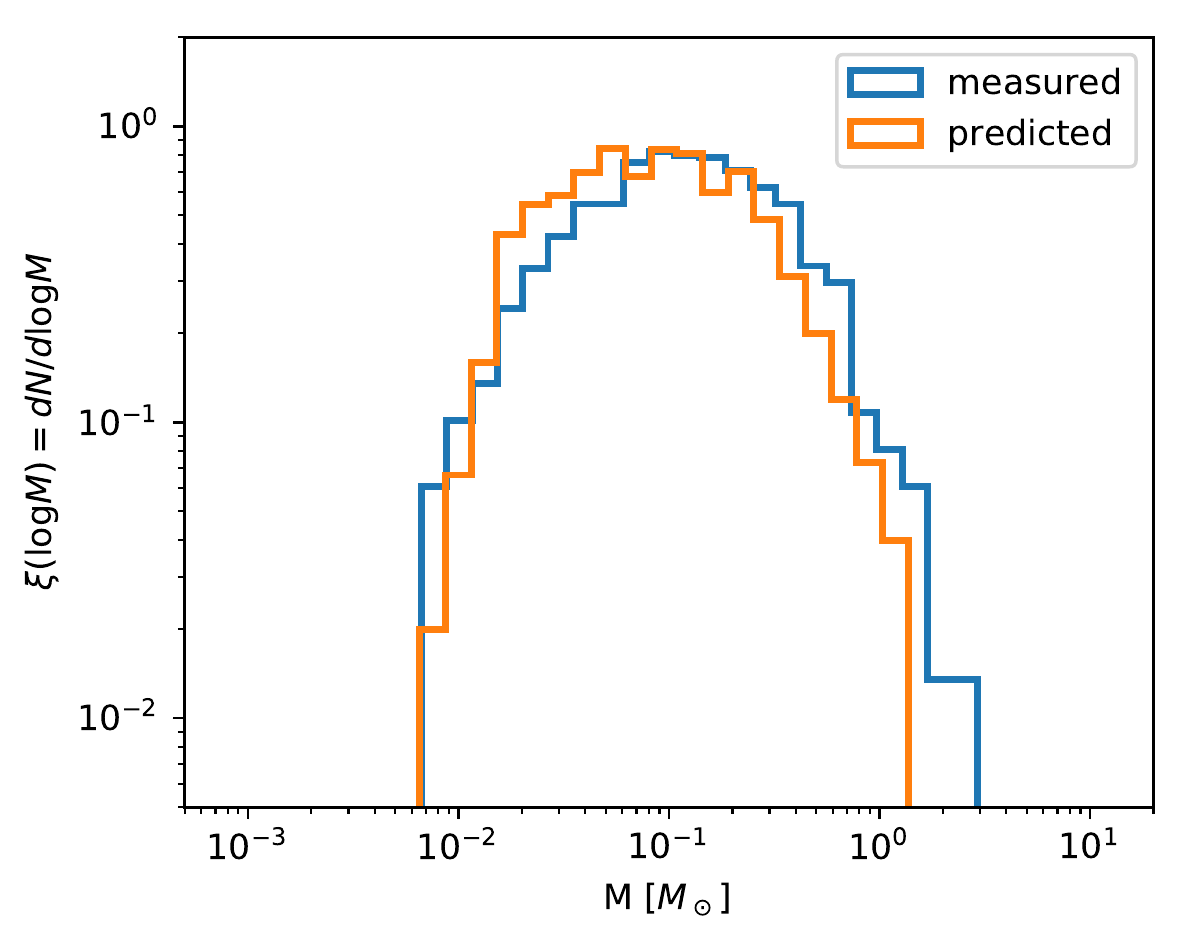}
\caption{Combining all the data from Fig.~\ref{fig:prediction_all_boxes} for a comparison of the stacked measured IMF at 20\% SFE and predicted IMF.}
\label{fig:prediction_stacked}
\end{figure}

Now that we are equipped with the previously described techniques to compute the tidally stripped mass in the region surrounding newly born first Larson cores
(the region we called the tidal bubble), we can repeat the procedure for every new sink in the simulation.

\subsubsection{The edge of the tidal bubble}
\label{sec:bubble_edge}
First, defining the edge of the tidal bubble as a collapse condition of exactly zero, corresponding to all the tidally stripped gas around the sink, 
we plot in Fig.~\ref{fig:default_bubble_correlation} the relation that compares the tidal mass with the actual final mass for each sink for one of the simulations.
We immediately see that there are two groups of sinks: the low mass sinks, for which the prediction is very good, and the higher mass sinks, for which the mass is overestimated.
Instead of taking a threshold of zero, we can use a stricter definition for the edge of the tidal bubble and choose a threshold for the collapse condition equal to some negative value.
In Fig.~\ref{fig:IMF_thresholds}, we show the corresponding mass functions and compare our predicted IMF when using different values for the tidal bubble threshold, corresponding to the contours in Fig.~\ref{fig:tidal_bubble_sink1_B} and the top panel of Fig.~\ref{fig:tidal_bubble_sink7}.
We see that there is a clear bimodality when using all the tidally stripped gas or a threshold value which is close to zero to define the edge of the tidal bubble (light grey histograms).
Interestingly, the minimum between the two peaks is located exactly at the position of the measured peak.
When taking a stricter threshold for the tidal bubble edge, the high mass peak shifts to the left, until the bimodality disappears (dark grey histograms).

The presence of this bimodality indicates there is a fundamental difference between the two groups of sinks.
When we inspect several of these sinks more closely, we find that practically all sinks in the low mass group form in crowded regions deviating strongly from spherical symmetry.
This situation is very dynamical, making it difficult for newly born sinks to accrete mass.
The other members of the star cluster have also already partially depleted the gas in the shared envelope.
In this case, the tidal mass is dominated by the effect of the neighbours.
The fact that in this regime our prediction is quite good, tells us that the procedure for dividing the mass between neigbouring sinks as discussed in section \ref{sec:neighbors} is reasonable.
The sinks in the high mass peak, tend to form in relative isolation and can accrete undisturbed for a significant amount of time.
The few sinks in between the peaks have configurations which are neither of the two extremes; isolated but with another star forming region close by or forming in a binary or small star cluster.

A threshold for the collapse criterion close to zero clearly overestimates how much mass an isolated sink can accrete and thus how big the tidal bubble is.
A threshold of $-10^{17}~\mathrm{g}/\mathrm{cm}^3$ results in a peak in the same location as the measure one, but also underestimates the number of high mass sinks.
As we can see in Fig. \ref{fig:default_bubble_correlation}, high mass sinks generally form early on in the simulation.
They have more time to accrete mass and this needs to be taken into account.
In order to estimate the mass that will be accreted by isolated first Larson cores, we need to apply this additional selection criterion
within the tidally stripped region.
We require the tidal stripping time scale, evaluated at the birth epoch, to be shorter than the time during the 
initial and the final age of the sink particle. Converted in density units, this translates into
\begin{equation}
\mathrm{threshold} = C \times \frac{3 \pi}{32 G (t_{\mathrm{end}} - t_{\mathrm{form}})^2}
\end{equation}
This threshold defines a region in which gas is expected to be tidally stripped at a fast enough rate so that it can be accreted onto the sink before the end of the simulation.
Gas experiencing weaker tidal forces is kept outside of the tidal bubble, as it will not react to the tidal field fast enough.
Turbulent motions might have moved the gas parcel away from the sink or gravity might have made it fall onto another clump.
We have adjusted the fudge factor to $C=0.5$, which seems to give us the best match between the tidal mass and the final mass of the sink.

\subsubsection{Final prediction for the IMF}
We now finally analyse all the sinks in our five different simulations.
In Fig.~\ref{fig:correlation}, we plot the comparison of the tidal mass as defined with the timescale dependent threshold, with the actual final mass for each sink.
We see that both mass are well correlated, supporting the idea that the tidal mass, estimated at the birth epoch of the first Larson core,
is a good proxy for the final mass of the star.
The values of the Pearson correlation coefficient are listed in Table~\ref{table:pearson} 
and show that, although there is certainly some scatter, the correlation is very good.
We also see in this figure, as pointed out before, the emerging of different families of sink particles forming at different epochs. 
There is a general trend that sinks which form early in the simulation end up having larger final masses,
with values at or above the peak mass of the IMF.
They are seen as blue circles in the top right part of each diagram.
These stars are the formed in isolation but typically end up being the most massive member of a relatively large star cluster.
Sinks that form late in the simulation are shown as orange to dark red circles.
They tend to have masses lower or equal to the peak mass of the IMF.
These stars generally formed in relatively crowded environments or in already significantly depleted envelopes.
Intermediate cases form the bulk of the stellar population in our simulations.
They form in relative isolation, with typical zero or one neighbour.

In Fig.~\ref{fig:prediction_all_boxes}, we plot the mass functions of our estimated tidal masses.
They match quite nicely the measured final IMFs.
Not only are the predicted peak positions quite close to the measured ones (except in box 5 where there is a larger offset), 
also the high and low mass end are properly recovered.
Since the individual box IMFs are quite noisy, in order to improve the statistics, we have again stacked them at the same SFE in Fig.~\ref{fig:prediction_stacked}. 
Here we see that the agreement is quite good, with a small shift towards lower masses. This confirms our visual impression in Fig.~\ref{fig:correlation} that our predicted masses
using the tidal bubble are slightly (by 0.1 dex) underestimating the true final mass.

\begin{table}
\begin{center}
\begin{tabular}{l | c }
IC code & Pearson coefficient \\
\hline
box\_1 & 0.7494 \\
box\_2 & 0.7795 \\
box\_3 & 0.8647 \\
box\_4 & 0.8104 \\
box\_5 & 0.7888 \\
\end{tabular}
\end{center}
\caption{Pearson correlation coefficients for the predicted mass at birth using tidal screening theory versus the mass measured at the end of the simulation (corresponding to Fig.~\ref{fig:correlation}).}
\label{table:pearson}
\end{table}
\section{Discussion}
\label{sec:discussion}

We now discuss our results in light of previous works on the same topics. We will also discuss caveats in our modelling approach,
as well as remaining issues of the tidal screening theory.

\subsection{Problems with established IMF theories}

The fundamental issue in classical IMF theories is that the predicted peak mass depends on global cloud properties, e.g. the cloud average density and  the turbulence Mach number.
Using these theories, one would thus expect different values for the characteristic star mass if these global properties were to vary between different star forming clouds.
Observations in the Milky way show, however, a universal IMF for all clouds, even though they have a large variety of properties.
As discussed in the the beginning of this paper, the Larson relation comes to the rescue and imposes a magic cancellation of these possible variations.
Whether the IMF is also universal in other galaxies is still unsure, since one expects the normalisation of the Larson relation to differ widely in other galaxies.
It remains unfortunately quite difficult to accurately measure the IMF, as well as the Larson relation, in distant galaxies \citep{Pan2016, Conroy2017, Gennaro2018}.
A recent theoretical study using cosmological galaxy formation simulations predicts extreme variations in these cloud properties \citep{Guszjnov2019}.
If the IMF truly depends on the global cloud properties as predicted by classical theories, the observed variations in the IMF would be much larger than what current observational
evidence of IMF variations suggest \citep{vanDokkum&Conroy2012}.
Note however that small-scale predictions of current galaxy formation models must be taken with a grain of salt.

Unlike the classical theories, tidal screening theory does not depend on global cloud properties but only depends on the small scale physics close to the opacity limit.
The characteristic star mass is thus determined based on {\it local} gas conditions and not on {\it global} cloud conditions,
in particular the properties of the first Larson cores and of their gaseous envelopes are what matters.
In this case, a universal IMF is a much more naturally outcome.

Many groups have also studied the IMF directly in numerical simulations.
Early results from \cite{Bonnell&Bate2006} and \cite{Nordlund_Padoan2003} confirmed both competitive accretion theory and turbulent fragmentation theory
(the turbulent Bonnor-Ebert hydrostatic view). Note that the initial conditions in their simulated clouds were chosen to satisfy Larson relation, 
and under these strict Milky-Way-like  conditions, the emerging IMFs were found to reproduce both the predicted IMF and the observed IMF.
These early simulations were not as resolved as the more recent ones. Our resolution study in Appendix \ref{app:resolution_effects}, for example, strongly suggest that our IMF is fully converged.

More recently, \cite{Lee_Hennebelle2018a} performed simulations of clouds with very different properties, but with the same polytrope and first Larson core mass.
They found that the resulting IMF characteristic mass was not evolving with the parent cloud properties, in contradiction with classical theories but in agreement with the tidal screening view.

Using a complementary approach, suites of simulations have been performed using different polytropic equation of states, 
corresponding to different first Larson core masses, but with identical initial turbulent cloud properties \citep{Motta_et_al2016, Lee_Hennebelle2018b}.
According to classic theories, this should have led to an identical IMF characteristic mass,
since the global cloud properties were unchanged.
Surprisingly, they obtained instead a characteristic mass growing proportionally to the first Larson core mass.
This behaviour is exactly what one would expect for the tidal screening theory.

These series of recent numerical results, including the one presented in this paper, show us that the characteristic mass of the IMF is still an open question.
The tidal screening theory appears to alleviate a number of problems in the classic theories.
It is however a relatively young theory and probably needs further numerical and analytical studies to fully understand what determines the peak mass of the IMF.

\subsection{Open questions in tidal screening theory}

Tidal screening theory appears as a nice alternative to turbulent fragmentation and competitive accretion.
We believe that tidal stripping from the first Larson core and its surrounding envelope could be combined with these older theories
to eventually deliver a fully consistent theory for the IMF.

In this paper, we have analysed in great details the tidal field around a newly formed first Larson core embedded in its gaseous, quasi-isothermal envelope.
The magnitude of the tidal eigenvalue was used to estimate the region within which clumps of a given density are not able to collapse anymore.
Using the lognormal distribution of the density fluctuations inside our turbulent box, we could estimate the probability of finding a collapsing clump of a certain density 
and at a certain distance from the central core. This is precisely what \cite{Lee_Hennebelle2018b} and \cite{Lee_et_al2019} did in their earlier, albeit recent study using an analytical approach,
providing also an analytical estimate of the tidal mass.
In this paper, instead of adopting a probabilistic approach, we used the actual gas density field in the simulation to compute the tidal mass.
We are however still missing a very important step: namely a complete statistical description of the IMF that accounts for tidal screening effects,
We speculate that combining the Press and Schechter approach used in the theory of turbulent fragmentation of \cite{Hennebelle_Chabrier2008} and \cite{Hopkins2012}
with tidal screening theory could be a promising path.  The resulting collapse probability would depend on the distance of a proto-clump 
to a nearby existing core through the two-point correlation function of the turbulent density fluctuations. We postpone the exploration of these ideas for future works.

A second issue we have also addressed here still remains an open question: this is the real impact of the presence of two or more neighbouring first Larson cores
on their respective tidal masses. It is indeed unclear how much mass will be assigned to each core in the final state. In Section~\ref{sec:neighbors}, 
we have addressed this issue by comparing the relative strength of the tidal fields and defining unambiguously one tidal bubble per Larson core.
This approach turns out to be a rather good predictor of the final mass for small mass stars. We still believe however that this recipe is too simplistic. 
A better approach to deal with such cases is the theory of competitive accretion \citep{Bonnell_et_al2001b, Bonnell&Bate2006}. 
This theory predicts that the accretion rate for stars in the centre of a star cluster is highest, and thus they become most massive.
This is not unlike what we see in our tidal bubble scenario.
Stars that form in already existing star clusters have small tidal masses because the tidal field of the older sinks and their envelope is quite strong.
The precise connection between tidal screening and competitive accretion remains to be explored in future studies.
One option could be to explicitly follow the gas using tracer particles in our simulations.

\subsection{Caveats in our numerical methodology}

Modern simulations of turbulent, self-gravitating molecular clouds have come a long way since the early work of 
\cite{Bate1998}.
Although resolution has increased significantly, thanks to better numerical techniques and more powerful computers, these simulations will never be a faithful representation of the reality.
For example, as mentioned in Section~\ref{sec:setup}, our computational boxes are periodic, without external driving to preserve the kinetic energy of turbulence on large scales.
Since we aim at studying the effect of the local tidal field around first Larson cores, we believe our conclusions are not affected by this simplification.

A major caveat of our present study is the absence of magnetic fields or radiation fields. While the effect of the latter is captured roughly by our polytropic equation of state,
the absence of magnetic field could be problematic in the vicinity of the first Larson core and could affect the dynamics of the surrounding envelope. 
Magnetic fields in the ideal MHD limit are known to prevent the formation of protostellar disks altogether 
\citep{Hennebelle_Teyssier2008},
which may have an effect on the stellar masses.
Fortunately, since what really matters is the geometry of the field lines during the collapse, rather than the
strength of the field close to the first Larson core, \cite{Lee_Hennebelle2018c} have shown that the effect of magnetism on the IMF is minor.
Moreover, 
\cite{Masson2016}
 have shown that non-ideal effects play a very important role in the vicinity of the first Larson cores, and restore some of the properties (but not all) of the zero magnetic field limit.
We therefore believe that it is probably more realistic to ignore magnetic fields in the present study, than include them in the ideal  MHD limit, and take the risk of grossly overestimating their effect on our results. A fully consistent, non-ideal MHD approach would ideally be needed but clearly lies beyond the scope of this study.

Feedback effects from the protostars are also completely ignored in our simulations. Both radiation from the massive end of the stellar population
\citep{Dale_2011, Gavagnin2017}
and jets launched from magnetically supported proto-stellar disks 
\citep{review_jets2014, outflow_review2016}
could have a strong effect.
In this paper, we use the simple approach of terminating the simulation early,
when only 20\% of the mass has been converted into star, as a proxy for the regulating effect of these feedback processes. We also believe that the characteristic mass
of the IMF is set up early and by dynamical processes only. This would ultimately need to be tested by implementing these feedback processes in similar simulations.
Note however that, during all our simulation, the peak of the IMF remains roughly constant.
This suggests that the characteristic mass of the IMF does not depend on the exact termination time of star formation and therefore our conclusions would not be affected by 
adding feedback processes.

While the missing physical effects described here might nevertheless alter the predicted IMF and bring it even closer the the observed one, 
we want to stress that tidal screening theory was used in this paper to explain the IMF that emerged from our simulations.
The fact that it seems to match the observed IMF peak came as a bonus,
Any additional physics will thus not change our conclusions that tidal screening seems to be the main factor in determining the peak of the IMF in our current simulations.
Note  however that different physical effects play a role at different scales or evolutionary stages in the star formation process.
Missing physics might be, for example, the reason why the high mass tail in our IMF does not match the observed IMF.
It has been pointed out before that high mass star formation does not necessarily follow the same principles than low mass star formation
\citep{review_massive_stars2012, Tan_et_al2014}
.

Finally, we discuss here possible caveats in our analysis of the simulation outputs and how it seems to support tidal screening theory.
First, as presented in the previous sections, the tidal mass can predict the mass of individual stars quite well.
The largest caveat here is that there is no guarantee that the mass inside the tidal bubble will end up fully in the sink particle.
Star formation in a turbulent cloud is a complex dynamical process and the local environment of the first Larson core 
might change drastically from the moment of birth to the moment the star reaches its final mass.
Individual cores within their gaseous envelopes are also not isolated and might interact with one another, which can strongly affect the local tidal field.
Another caveat to discuss here concerns the tight correlation we have observed between the tidal mass and the final star mass, but correlation is not causation.
While the good values for the Pearson correlation coefficient are surely encouraging to support tidal screening theory, 
there might be other confounding variables in the problem. More studies are therefore required to check for sure whether the tidal field in responsible for the characteristic mass
of stars.

\section{Conclusions}
\label{sec:conclusion}

In conclusion, we summarise our findings as follows.
\begin{itemize}
\item Observations show a universal IMF, something which is difficult to explain using classical IMF theories (turbulent Bonnor-Ebert spheres, turbulent fragmentation and competitive accretion) which predict an IMF peak which scales with global cloud properties.
Recent, high resolution numerical experiments show that, indeed, the peak of the IMF does not scale with cloud properties, but does scale with the polytropic EOS critical density, and therefore with the first Larson core mass.\\

\item The recently proposed tidal screening theory, states that strong tidal fields around first Larson cores prevent clumps from collapsing within a certain radius called the tidal radius.
As a result, the mass within this radius will be accreted by the central first Larson core.
This theory predicts an IMF peak around $10 \times M_{\mathrm{1LC}}$ and independent of the global cloud properties .\\

\item We explored this theory further and derived a collapse criterion for clumps that form close to an existing star (or first Larson core).
Whether a clump can collapse or not depends on its density, the distance to the star, the mass of the star and the profile of the envelope around the star.\\

\item We performed a series of numerical simulation of a small star forming turbulent box representing a collapsing region inside a larger molecular cloud.
We measure the IMF and find it has a peak at about 0.1 \Msun. Classical theories do not explain this result.\\

\item We studied the tidal fields around first Larson cores (sink particles) and find they are strong enough to prevent potential clumps from collapsing unless they have high enough densities.
Since these clumps are rare due to the form of the density PDF generated by turbulence, tidal forces effectively screen a large region around the first Larson core.\\

\item Using the actual density field of the simulation and a local collapse criterion, we construct a tidal bubble around every star at the moment of its birth,
in which tidal forces are strong enough to strip anything within a timescale shorter than the remaining simulation time.
We then use the gas mass in this tidal bubble as a predictor for the final mass of the star.\\

\item We found a very good correlation between our predicted mass and the true final mass of the star.
Stars formed in isolation generally have masses at or above the characteristic mass.
Low mass stars are formed in regions that already contain other stars.\\

\item Our predicted IMF, based on the tidal mass, matches the simulated one, based on the sink particle final mass, quite well, with a peak slightly shifted to lower masses.
The predicted IMF also reproduce the simulated IMF at both the high and the low mass end. Note that both IMF, predicted and simulated, also match quite well the observed IMF.\\

\end{itemize}
We conclude that tidal forces are an important ingredient in determining the final mass of a star.
We speculate that a complete theory of the IMF would combine tidal screening theory with competitive accretion and turbulent fragmentation,
possibly leading to more accurate predictions.
It remains to confirm that the tidal screening formalism presented in this paper can still predict the IMF when additional physics is included.
This will be the topics of a follow up paper.



\section*{Acknowledgements}
The simulations performed for this work were executed on Piz Daint at the Swiss Supercomputing Center CSCS in Lugano and analysed on the hydra cluster of the University of Zurich.
We thank Patrick Hennebelle, Yueh-Ning Lee, {\AA}ke Nordlund and Troels Haugb{\o}lle for interesting discussions related to the theory and simulations presented in this work.

\bibliographystyle{mnras}
\def\apj{ApJ}
\def\apjs{ApJS}
\def\apjl{ApJL}
\def\aj{AJ}
\def\mnras{MNRAS}
\def\aap{A\&A}
\def\nat{Nature}
\def\pasj{PASJ}
\def\prd{PRD}
\def\physrep{Physics Reports}
\def\jcap{JCAP}
\bibliography{paper1}

\appendix
\newpage
\section{Details of the Initial conditions}\label{app:ICs}

\begin{figure}
\center
\includegraphics[scale=0.6]{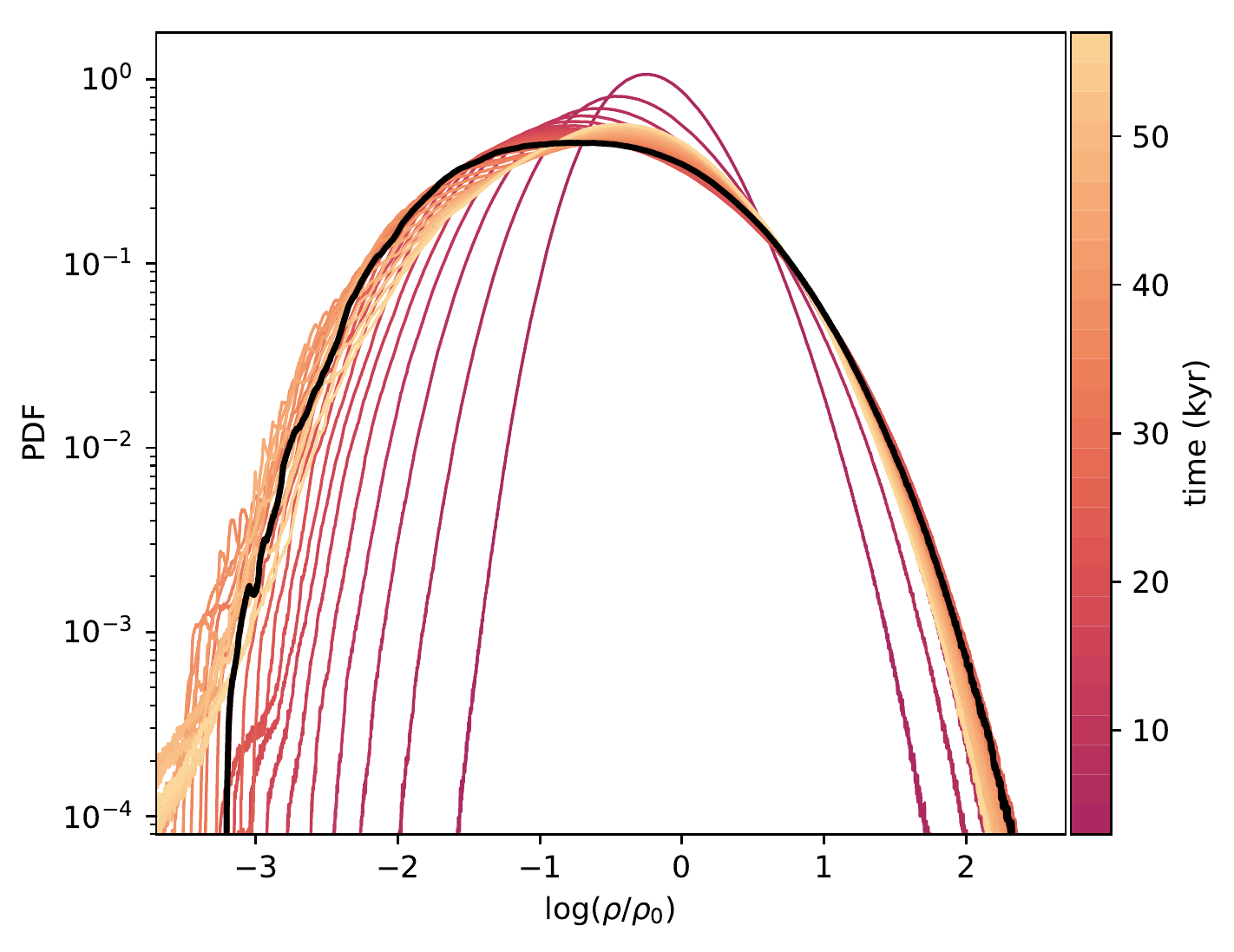}
\caption{Evolution of the volume-weighted density PDF in the initialisation phase of box 1. The black curve is the PDF at the moment where the kinetic energy is half.}
\label{fig:PDF}
\end{figure}

\begin{figure}
\center
\includegraphics[scale=0.60]{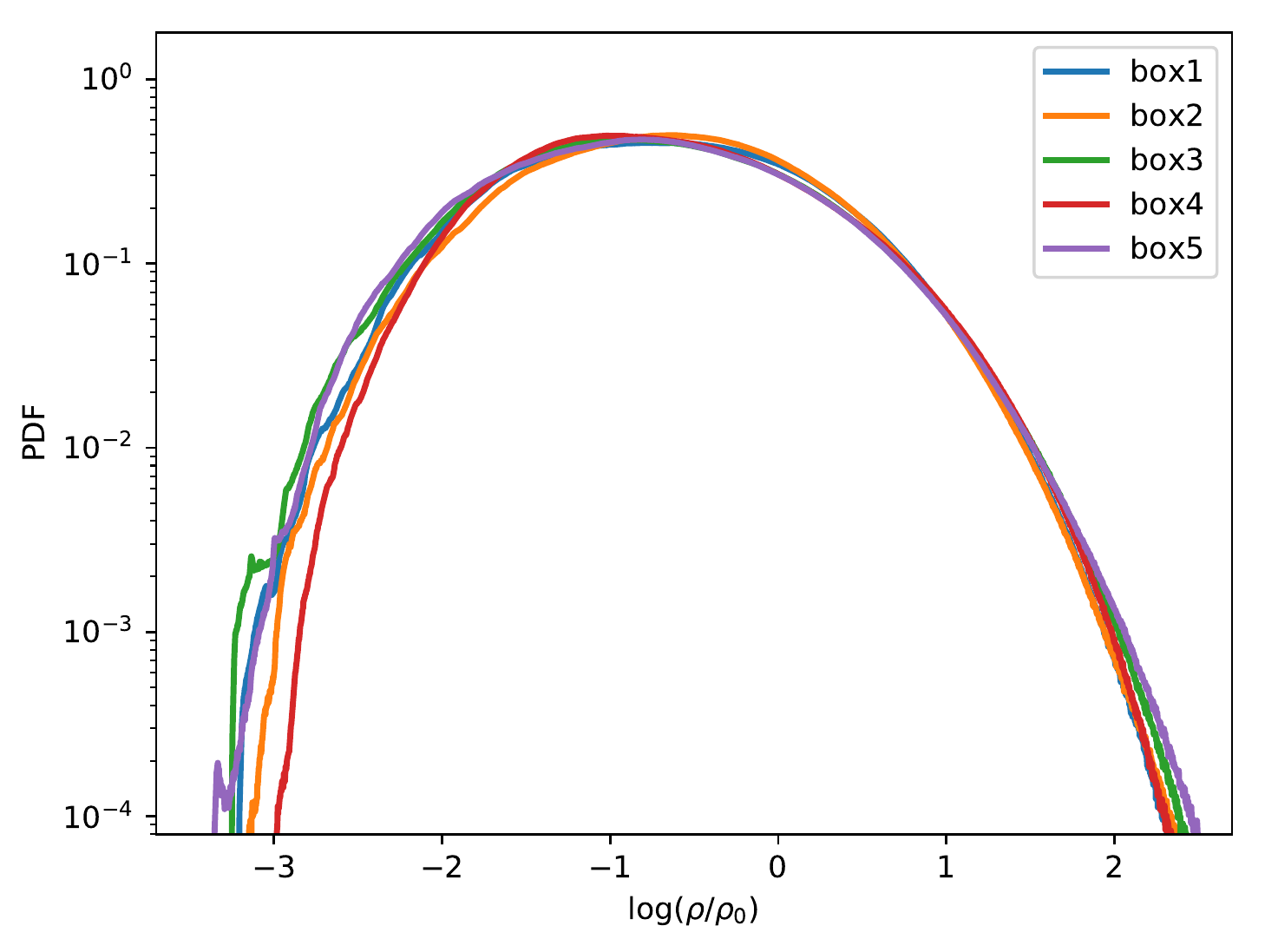}
\caption{The volume-weighted density PDF at the end of the initialisation for each box.}
\label{fig:PDF_compare}
\end{figure}

\begin{table}
\begin{center}
\begin{tabular}{l | c c c}
\hline
IC code & seed $v_x$ & seed $v_y$ & seed $v_z$\\
\hline
box\_1 & 10668 & 20036 & 72523 \\
box\_2 & 28794 & 82562 & 12770 \\
box\_3 & 35588 & 03573 & 81985 \\
box\_4 & 49835 & 11371 & 07657 \\
box\_5 & 55773 & 62846 & 84953 \\
\hline
\end{tabular}
\end{center}
\caption{Random number generator seeds used for the generation of fluctuations by \textsc{music}.}
\label{table:seeds_ic}
\end{table}

\begin{table}
\begin{center}
\begin{tabular}{l | c c c c c}
\hline
IC name & $\mathcal{M}_x$ & $\mathcal{M}_y$ & $\mathcal{M}_z$ & $\mathcal{M}_{3 \mathrm{D}}$ & $t_{1/2 E_{\mathrm{kin}}}$ [kyr]\\
\hline
box\_1 & 7.81 & 7.55 & 6.43 & 12.62 & 27.3 \\
box\_2 & 7.14 & 7.79 & 7.04 & 12.70 & 25.0 \\
box\_3 & 6.91 & 7.39 & 7.84 & 12.80 & 22.0 \\
box\_4 & 7.22 & 7.41 & 7.28 & 12.66 & 23.0 \\
box\_5 & 7.51 & 6.95 & 7.64 & 12.77 & 22.0 \\
\hline
\end{tabular}
\end{center}
\caption{Properties of each box at the and of the initialisation phase.}
\label{table:ic_properties}
\end{table}

To generate our turbulence initial conditions, we start with a uniform density and imprint random Gaussian fluctuations on the velocity field following the velocity power spectrum of Burger's turbulence $P(k) \propto k^{-4}$.
These fluctuations are generated using \textsc{music} \citep{Hahn2011}, a multiscale initial condition generator originally developed for cosmological simulations.
We rescale the velocity field so that the box has initially a 1D Mach number of $18/\sqrt{3}$ in each component. The gas is then evolved using a uniform $1024^3$ Cartesian grid 
and develops a lognormal density PDF (see Fig.~\ref{fig:PDF}), as expected for isothermal supersonic turbulence \citep[see e.g.][]{Federrath_et_al2008}.
This initialisation phase is stopped when the kinetic energy has decayed to half its initial value.
At this point the turbulence is fully developed and the density PDF is the widest (see Fig.~\ref{fig:PDF}).
The scripts used to generate the initial conditions are available at \url{https://bitbucket.org/TineColman/boxicgen}.

The exact structure of the turbulence generated for the initial conditions depends on three random number generator seeds used by \textsc{music}, one for each velocity component.
The seeds used for the simulations presented in this work are listed in Table~\ref{table:seeds_ic}.
Since the five boxes have different amplitude for the velocity on large scales, they evolve quite differently.
Table~\ref{table:ic_properties} lists some properties of the initial conditions at the end of our initialisation phase.
Note that the 1D Mach number is not necessarily equal in all directions.
Fig.~\ref{fig:PDF_compare} compares the density PDF of these snapshots and one can see they are fairly similar.
\newpage
\section{Resolution effects}\label{app:resolution_effects}

\begin{table}
\begin{center}
\begin{tabular}{l l l l}
\hline
parameter & levelmax 10 & levelmax 12 & levelmax 14\\
\hline
$\rho_{\mathrm{sink}}$ [g $\mathrm{cm}^{-3}$] &  $2.74 \times 10^{-15}$ & $4.39 \times 10^{-14}$ & $7.02 \times 10^{-13}$\\
$m_{\mathrm{seed}}$ [\Msun] & $1.98 \times 10^{-2}$ & $4.94 \times 10^{-3}$ & $1.23 \times 10^{-3}$ \\
\hline
\end{tabular}
\end{center}
\caption{Simulation parameters that vary for different resolutions.}
\label{table:res_params}
\end{table}

\begin{table}
\begin{center}
\begin{tabular}{r r}
\hline
parameter & value\\
\hline
levelmin & 8\\
$T2$ & 4.2194\\
g\_star & 1.4\\
n\_star & $2.5 \times 10^{10}$ H/cc\\
sink merging timescale & 1500 yr\\
sink accretion scheme & bondi\\
\hline
\end{tabular}
\end{center}
\caption{Simulation parameters that are kept fixed for different resolutions.}
\label{table:sim_params}
\end{table}

\begin{figure*}
\center
\includegraphics[scale=0.5]{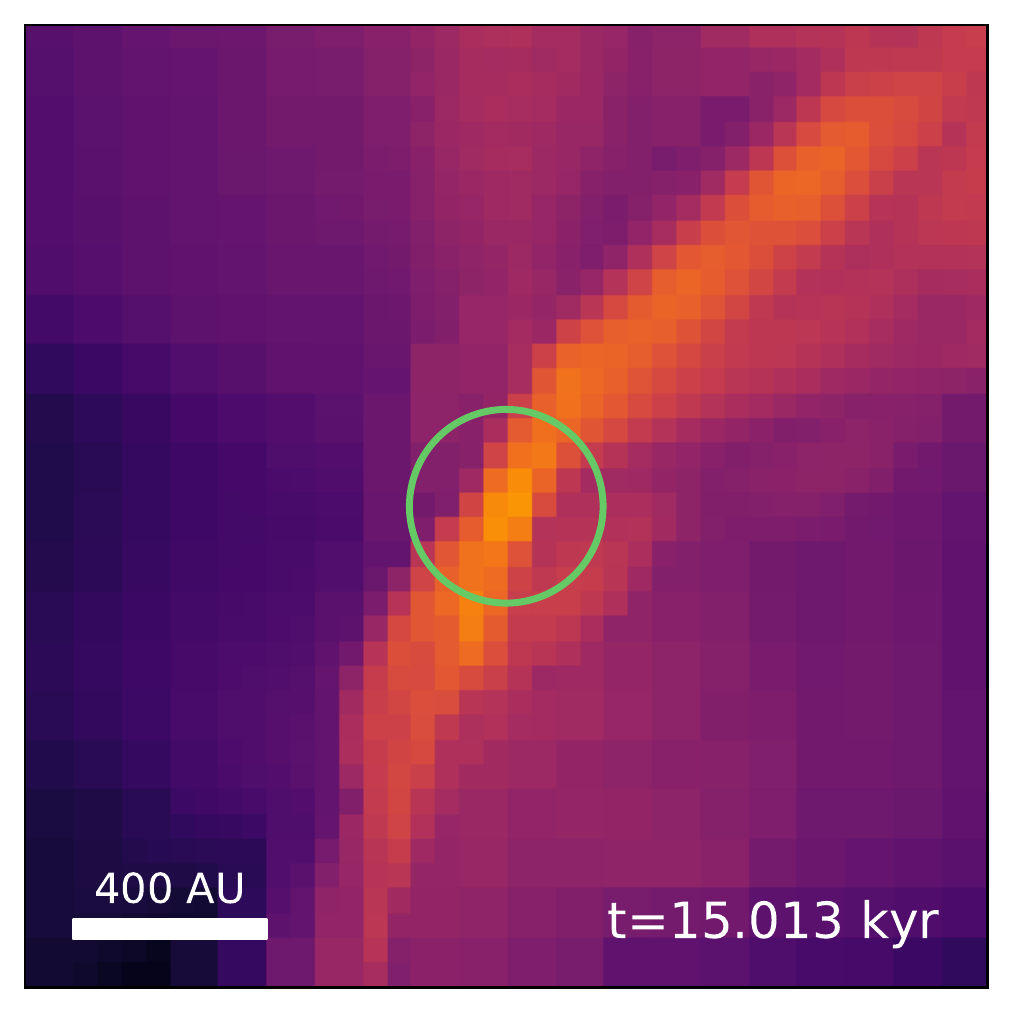}
\includegraphics[scale=0.5]{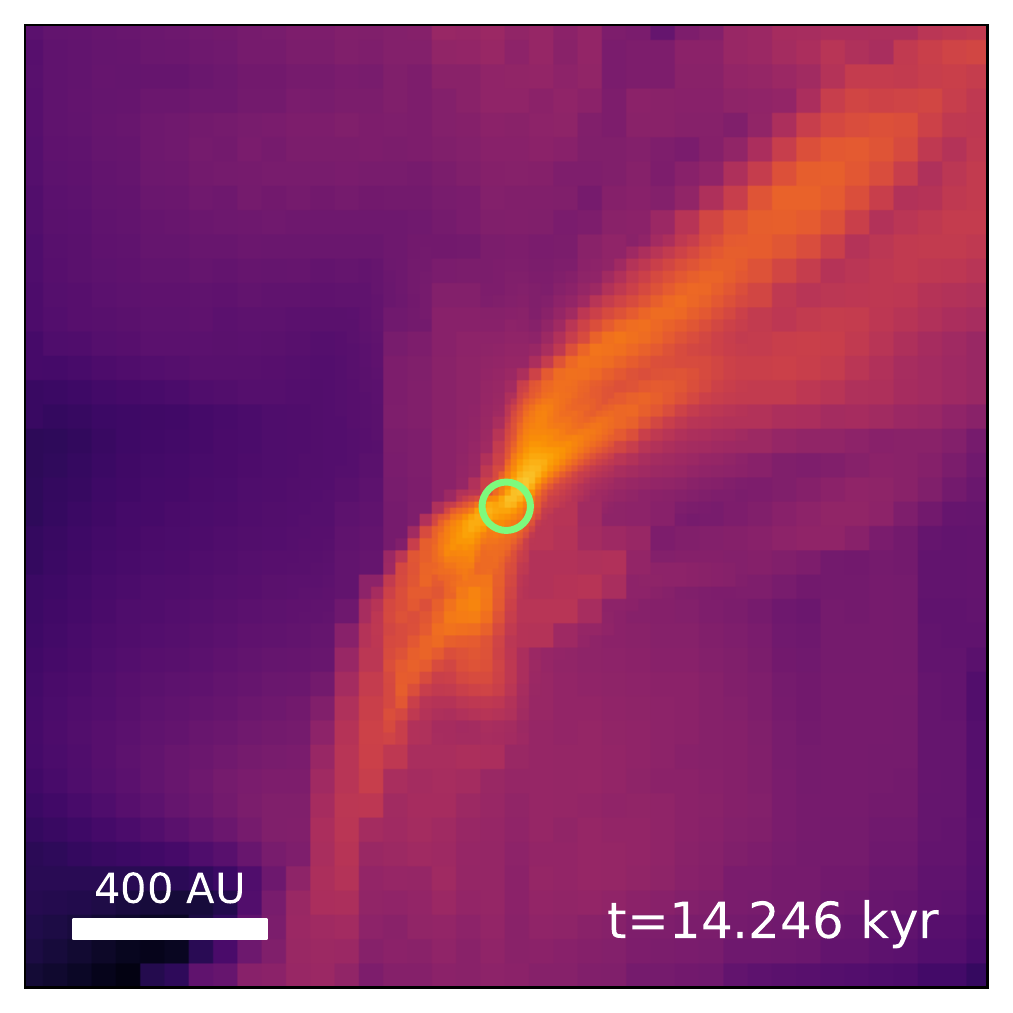}
\includegraphics[scale=0.5]{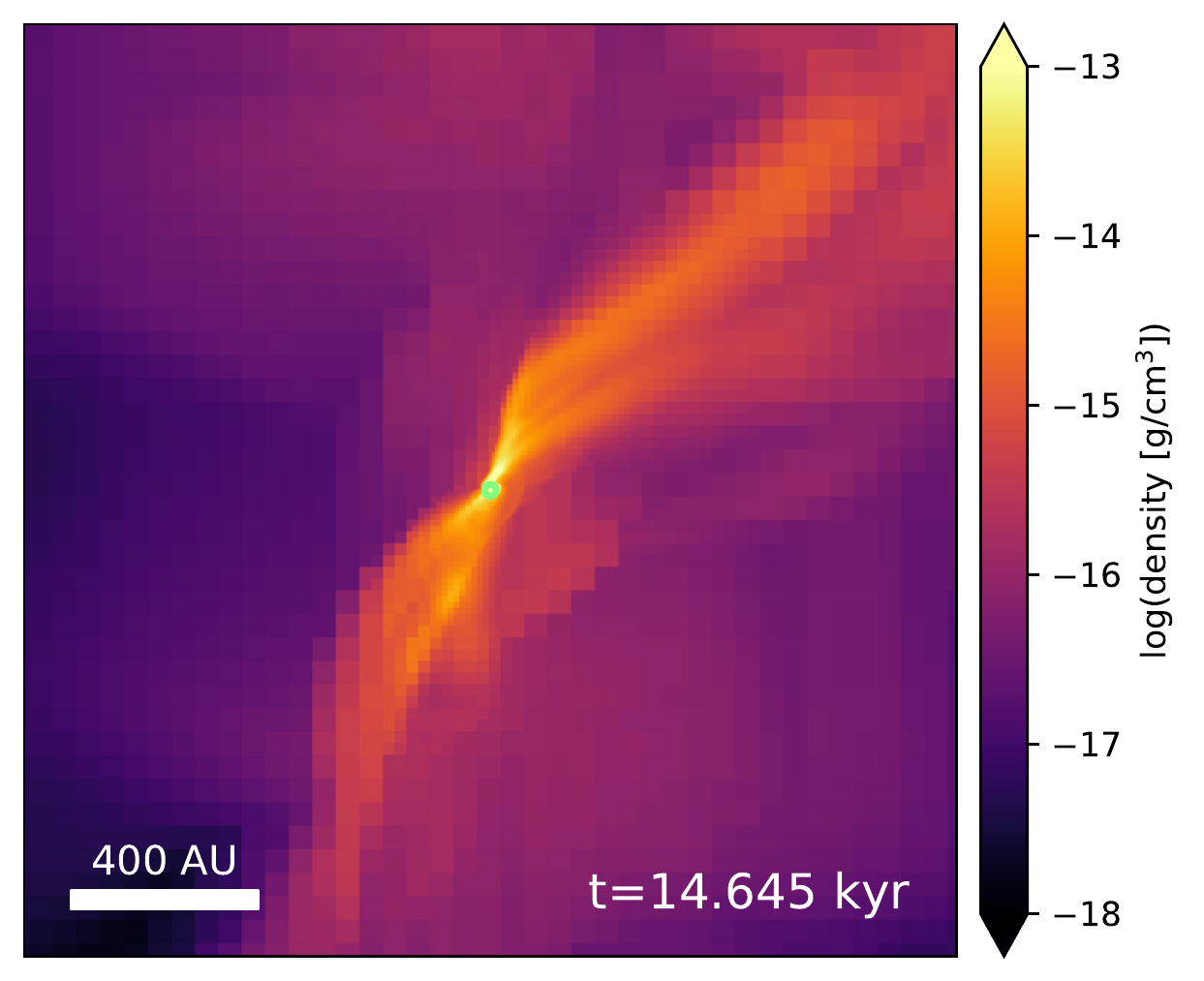}
\\
\includegraphics[scale=0.5]{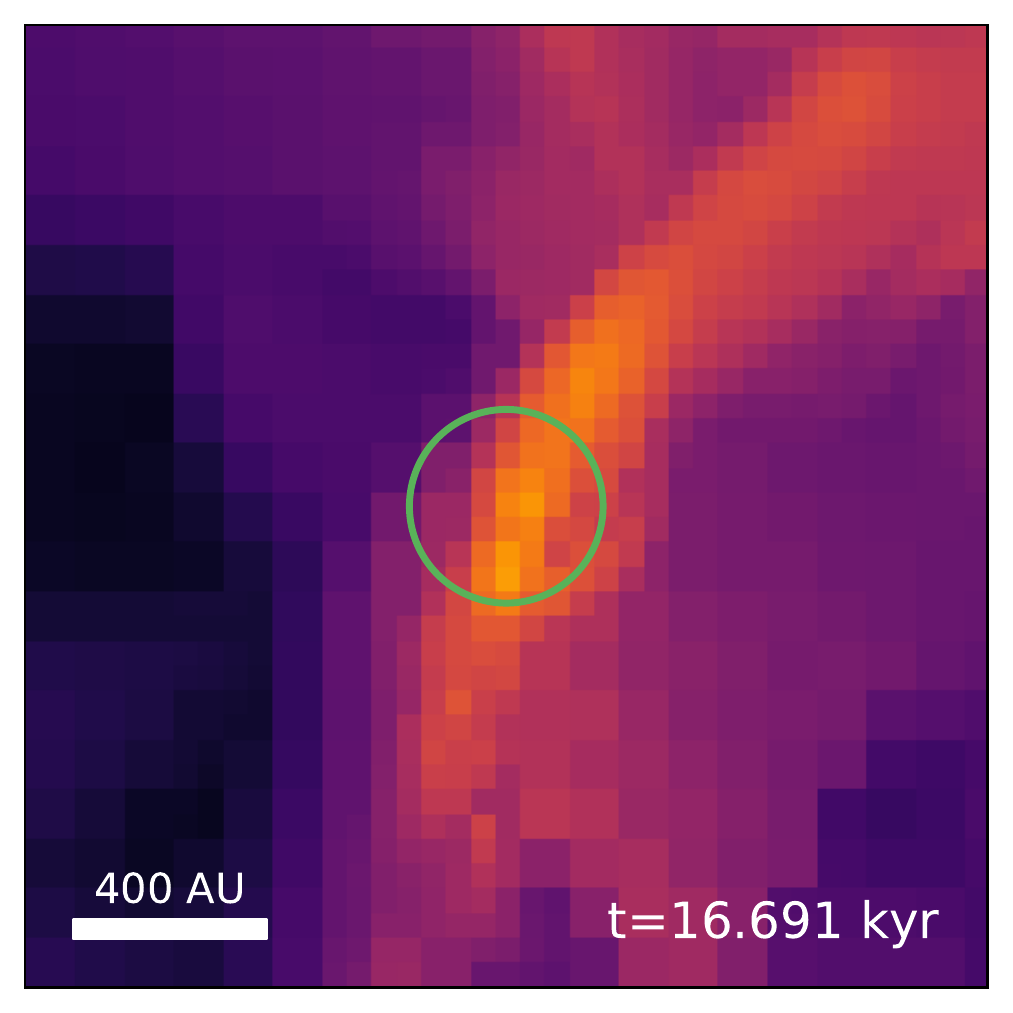}
\includegraphics[scale=0.5]{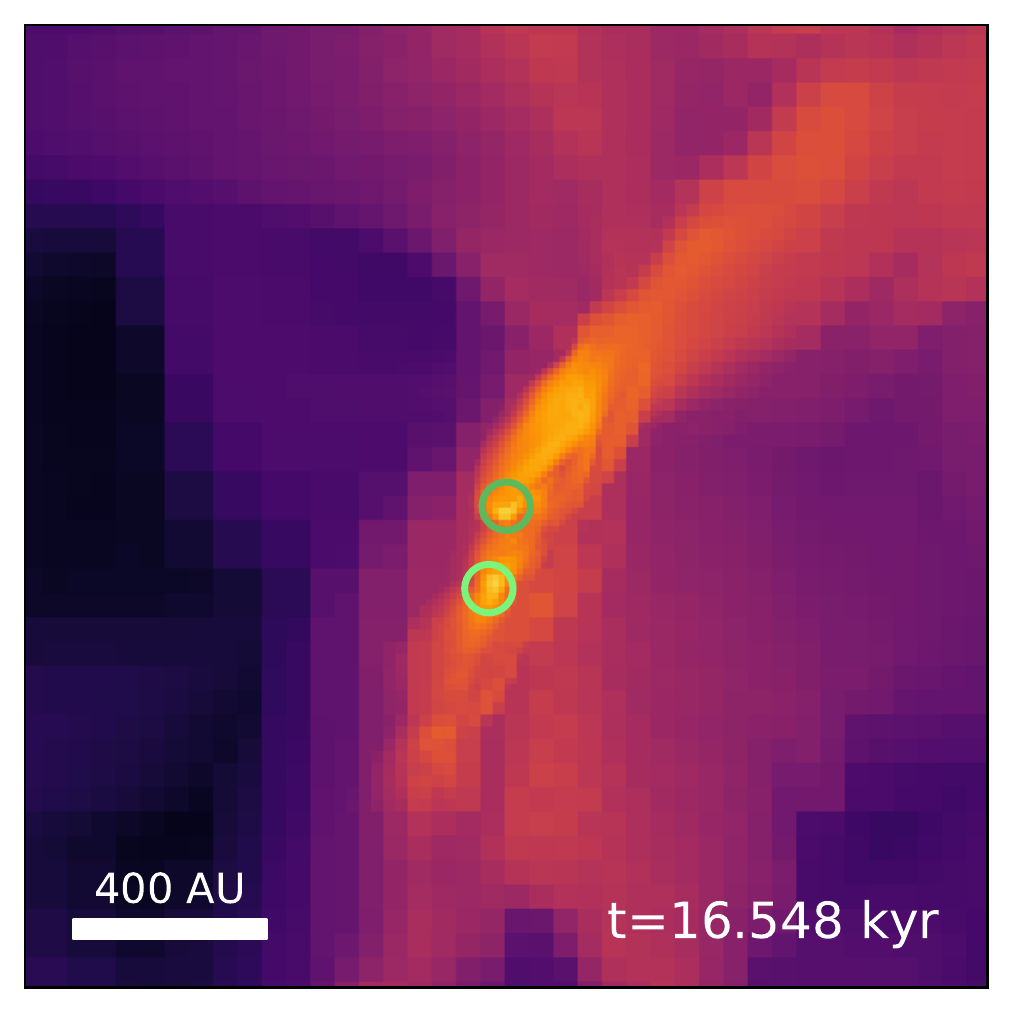}
\includegraphics[scale=0.5]{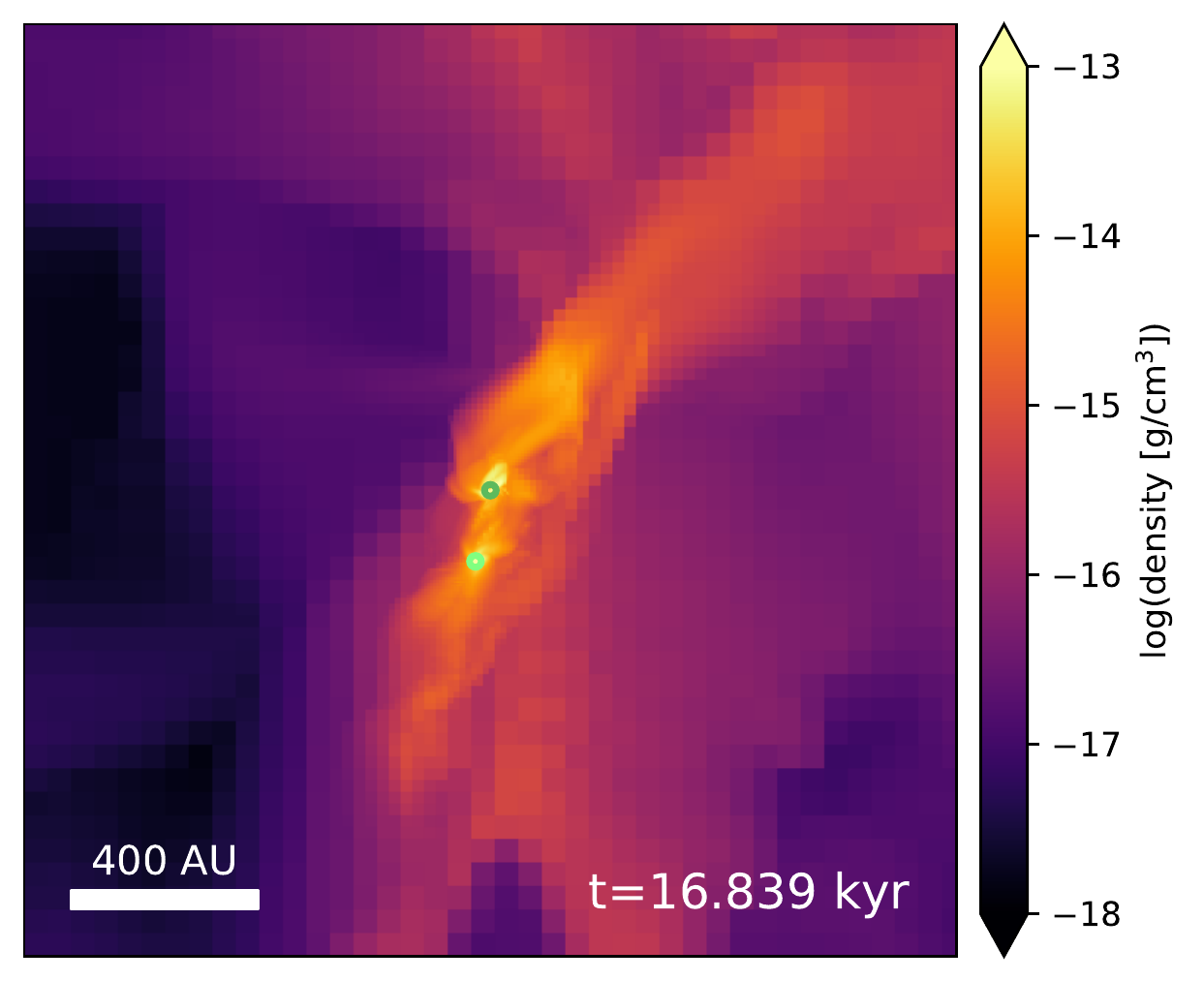}
\\
\includegraphics[scale=0.5]{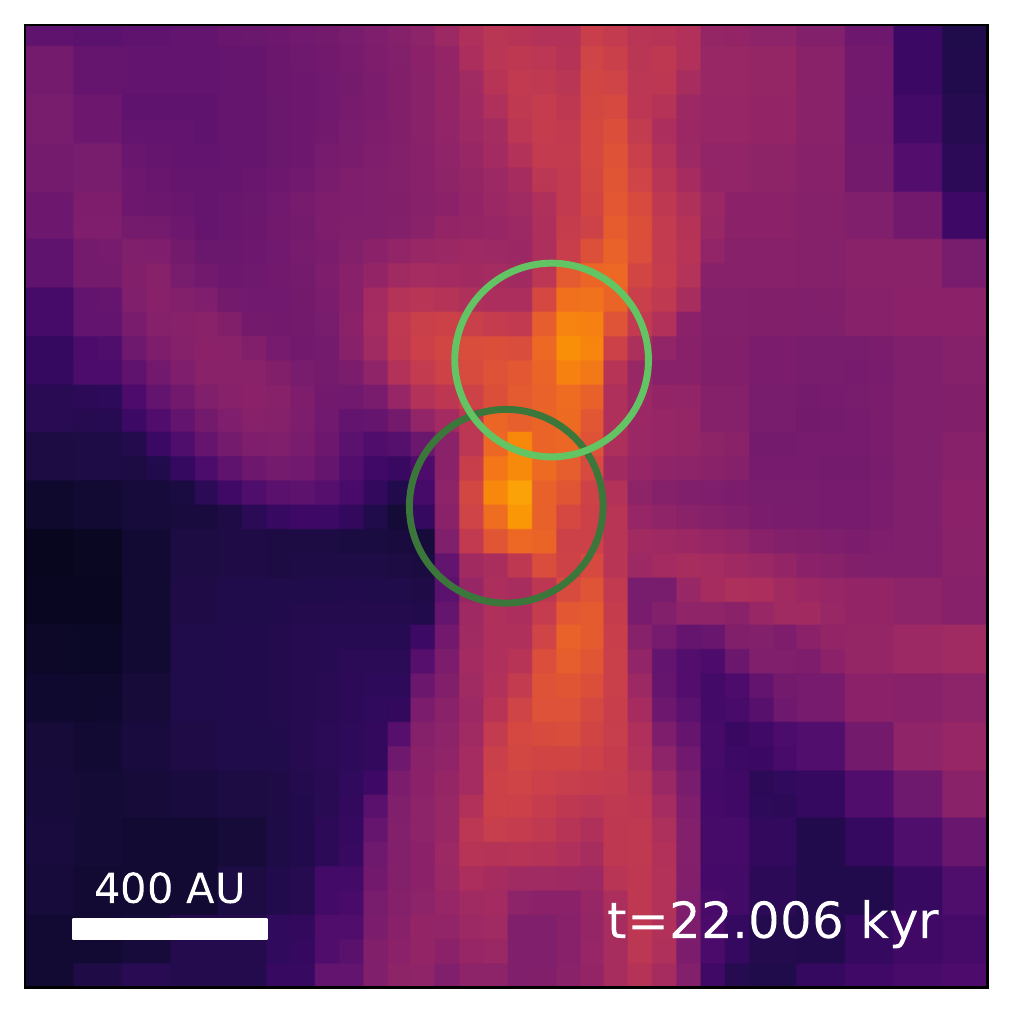}
\includegraphics[scale=0.5]{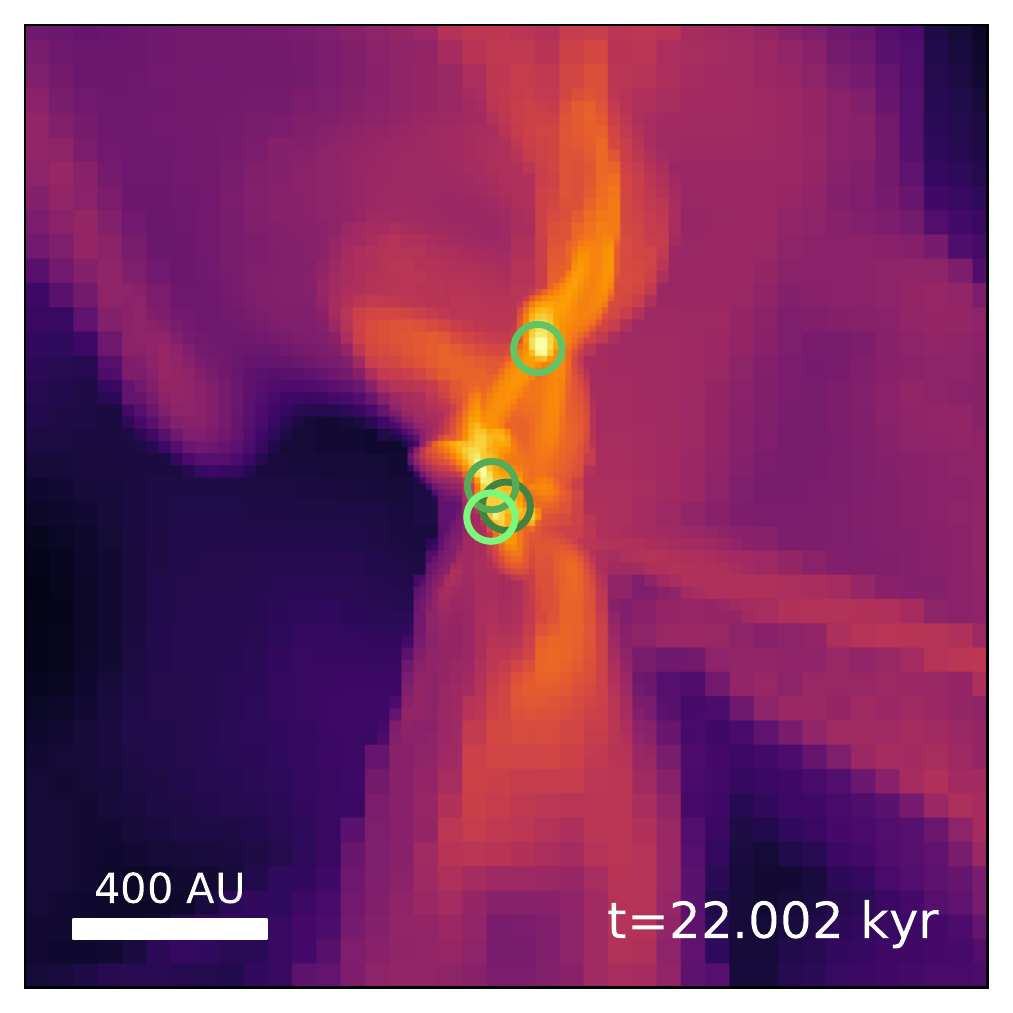}
\includegraphics[scale=0.5]{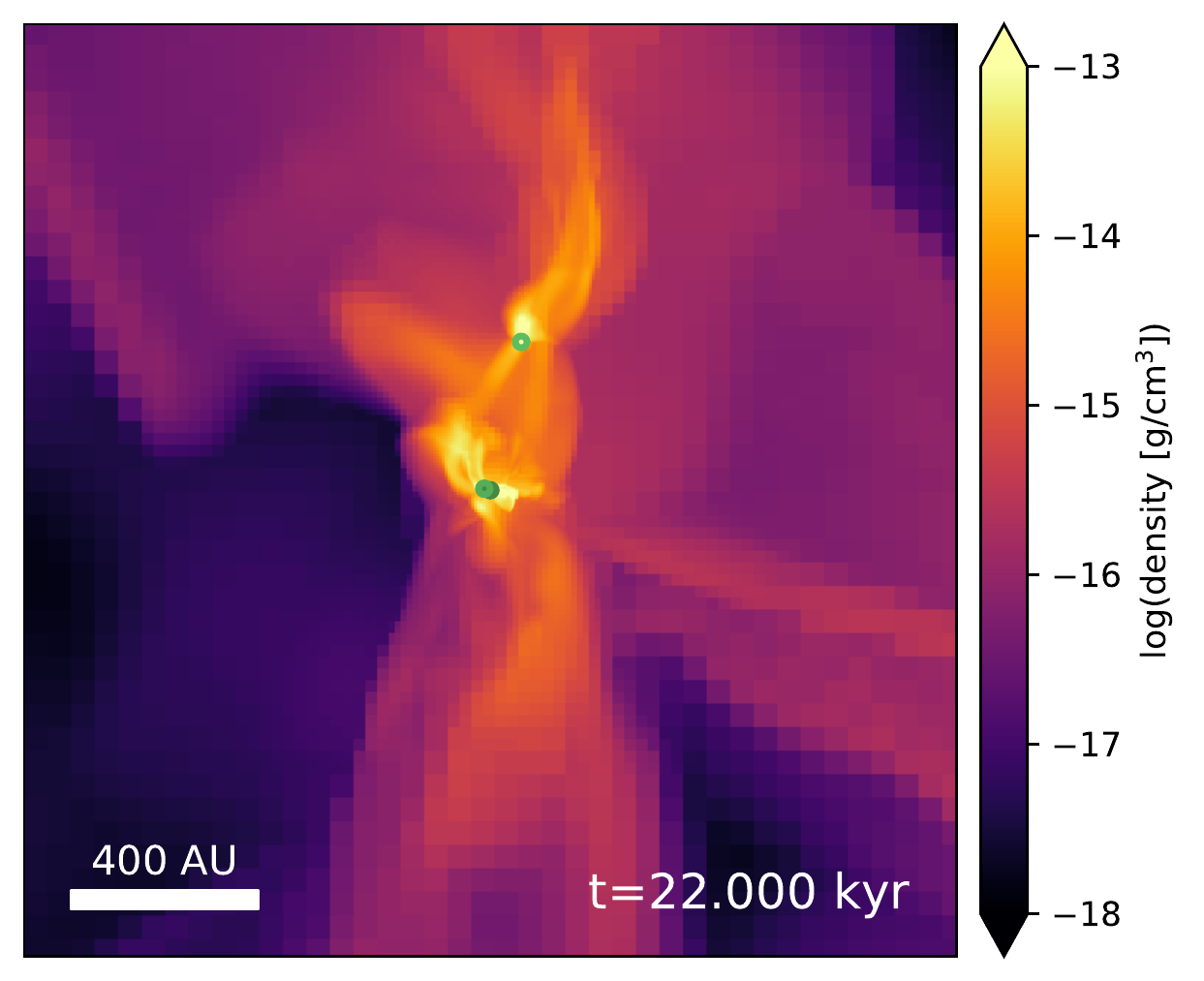}
\\
\includegraphics[scale=0.5]{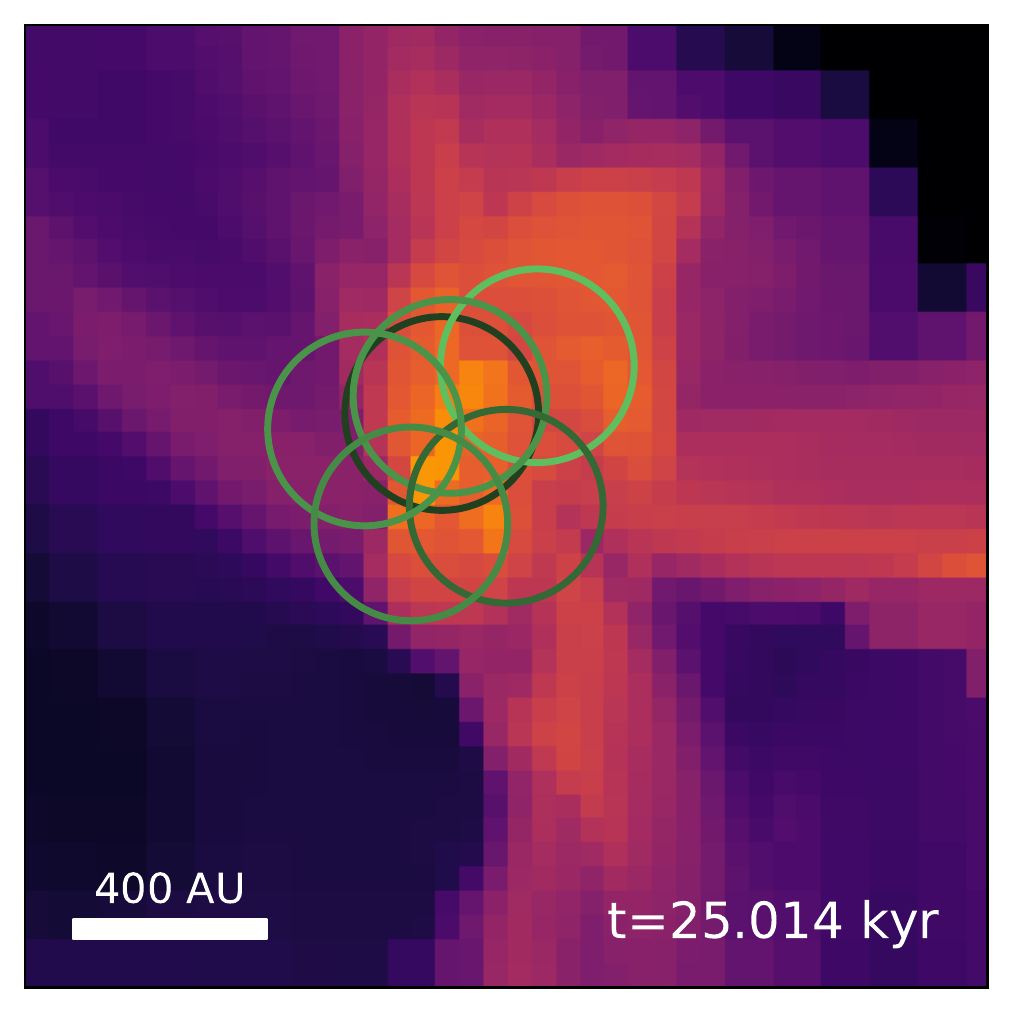}
\includegraphics[scale=0.5]{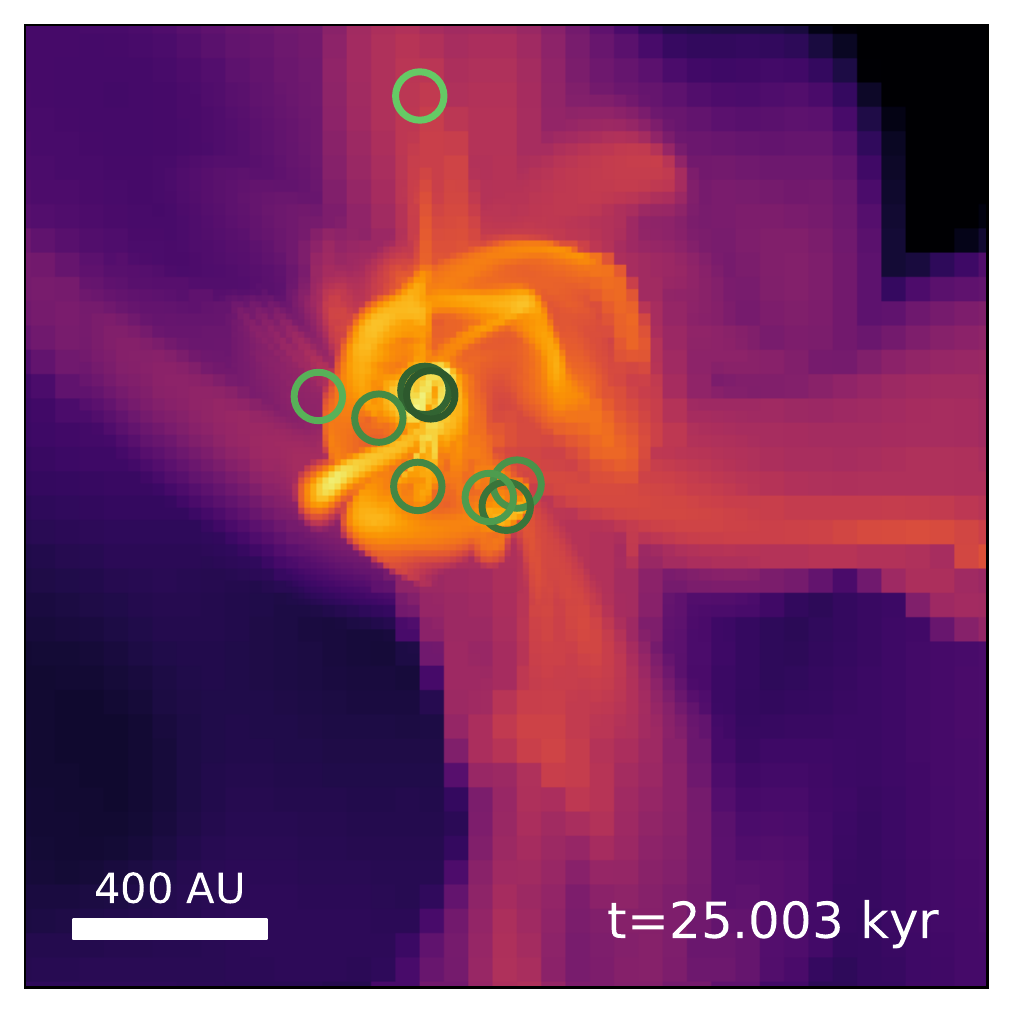}
\includegraphics[scale=0.5]{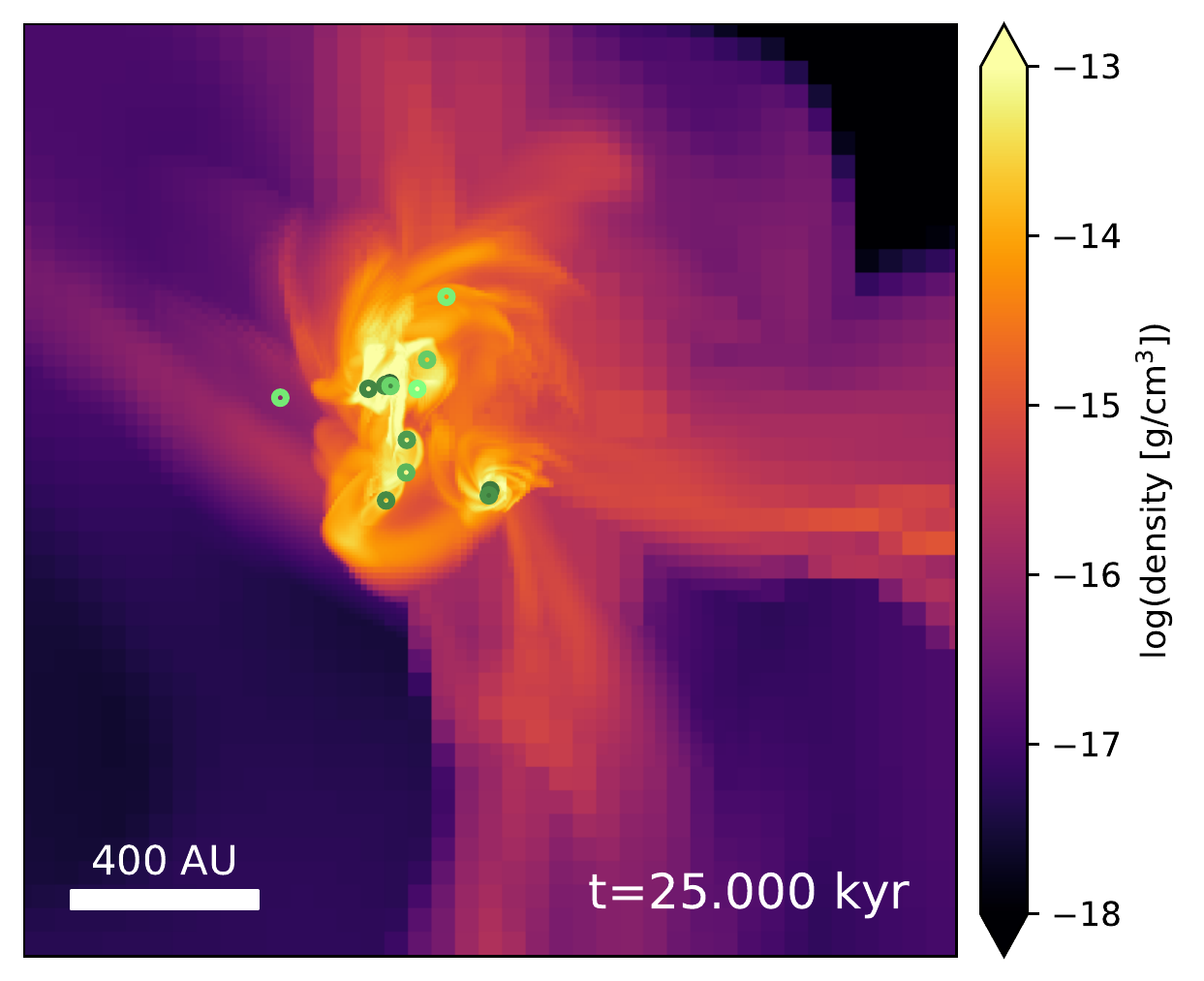}
\caption{Example of an evolution sequence of a star forming region.
The maps shows the maximum density along the line of sight.
Green circles show the sink accretion radius with the color indicating the sink's mass: the darker the circle, the more massive the sink.
From left to right the maximum refinement level is increased: levelmax 10, 12 and 14.
This particular case has an unresolved binary in the low resolution case and merges with another region around 24 kyr. Top row: the primary sink is born. Second row: the companion sink's birth, which is not resolved in the low resolution case. Third row: the formation of another companion, this time present in all resolutions. Bottom row: the region merged with another star forming region.}
\label{fig:unresolved}
\end{figure*}

\begin{figure}
\center
\includegraphics[scale=0.60]{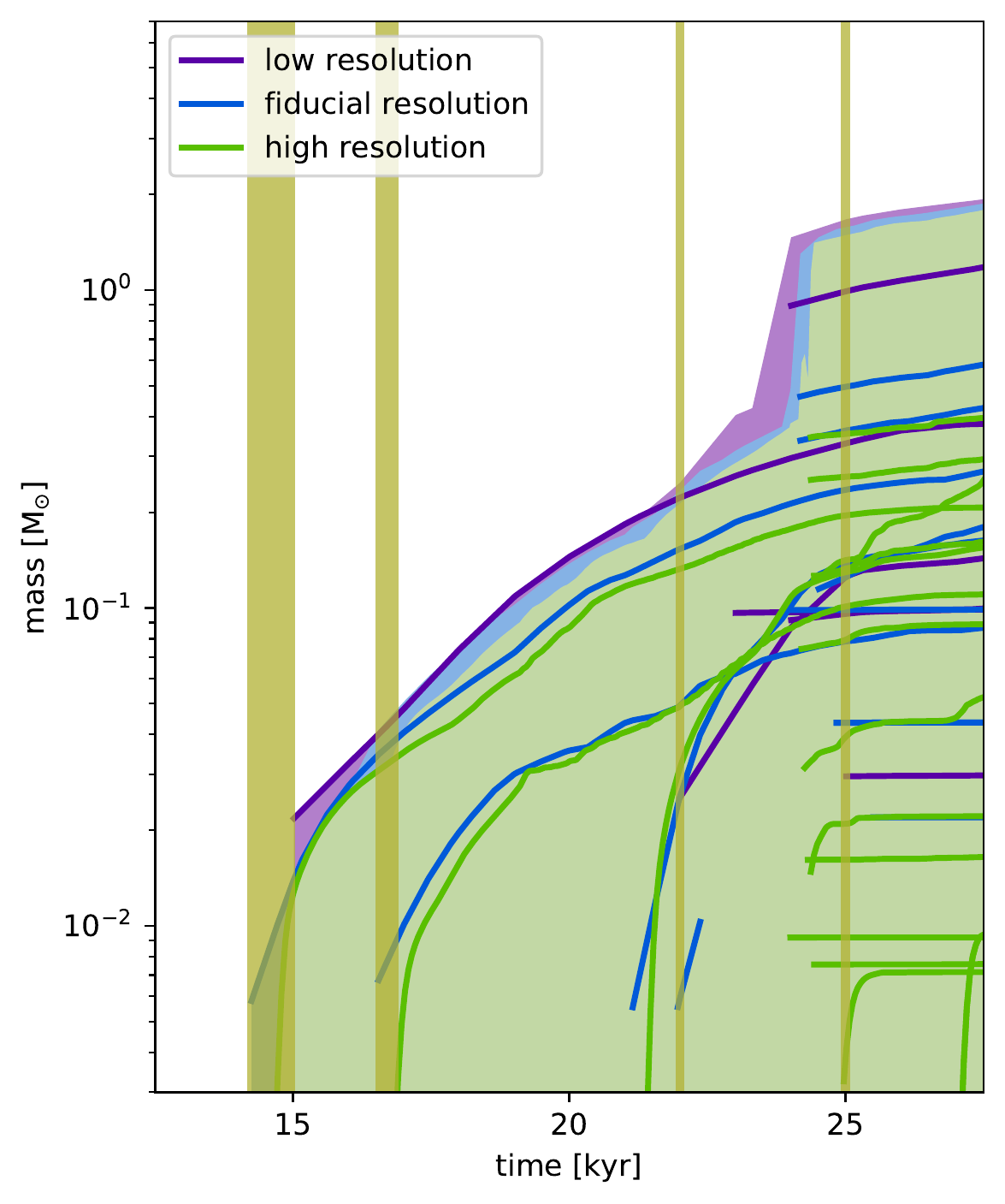}
\caption{The mass evolution of each sink in the star formation region displayed in Fig.~\ref{fig:unresolved}. Each track represents a sink particle, while the colored background shows the total mass in the region. The vertical lines mark the times indicated in Fig.~\ref{fig:unresolved}).}
11\label{fig:unresolved_track}
\end{figure}

\begin{figure*}
\center
\includegraphics[scale=0.5]{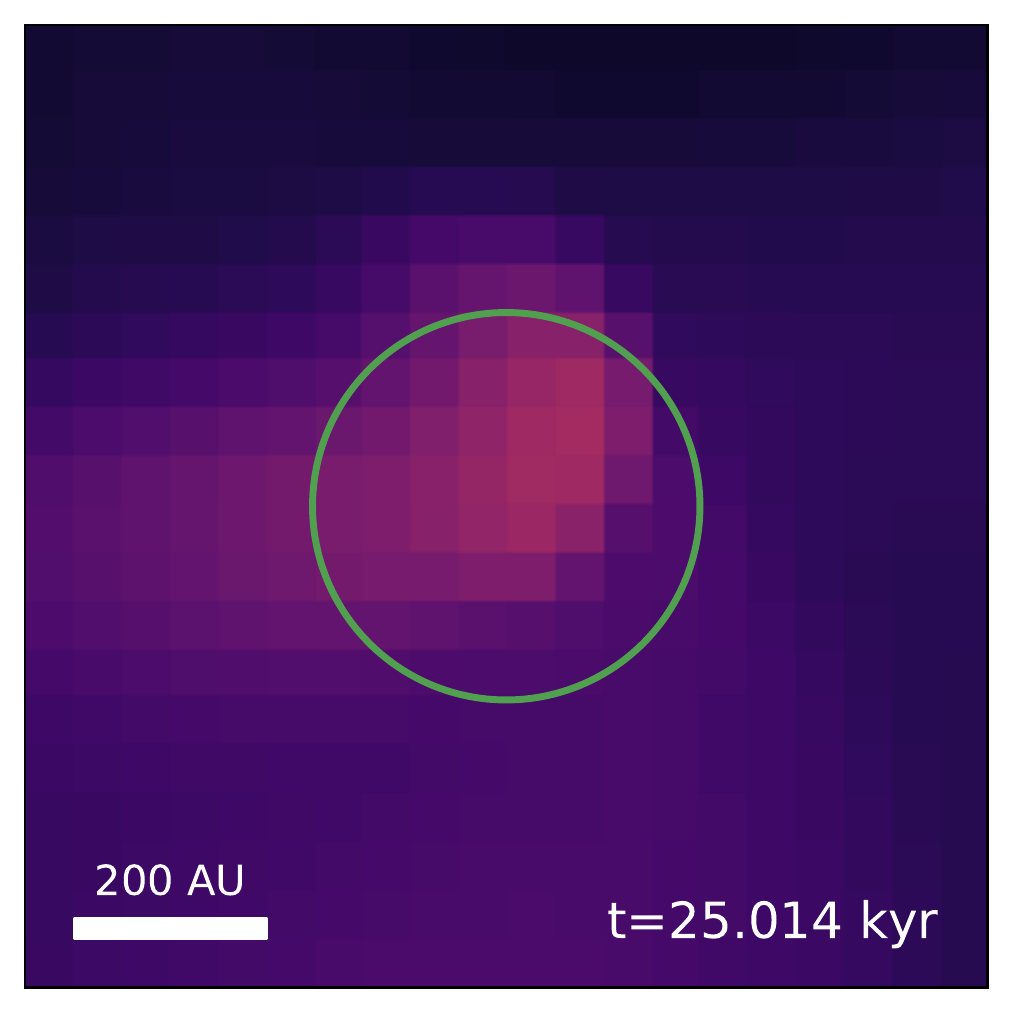}
\includegraphics[scale=0.5]{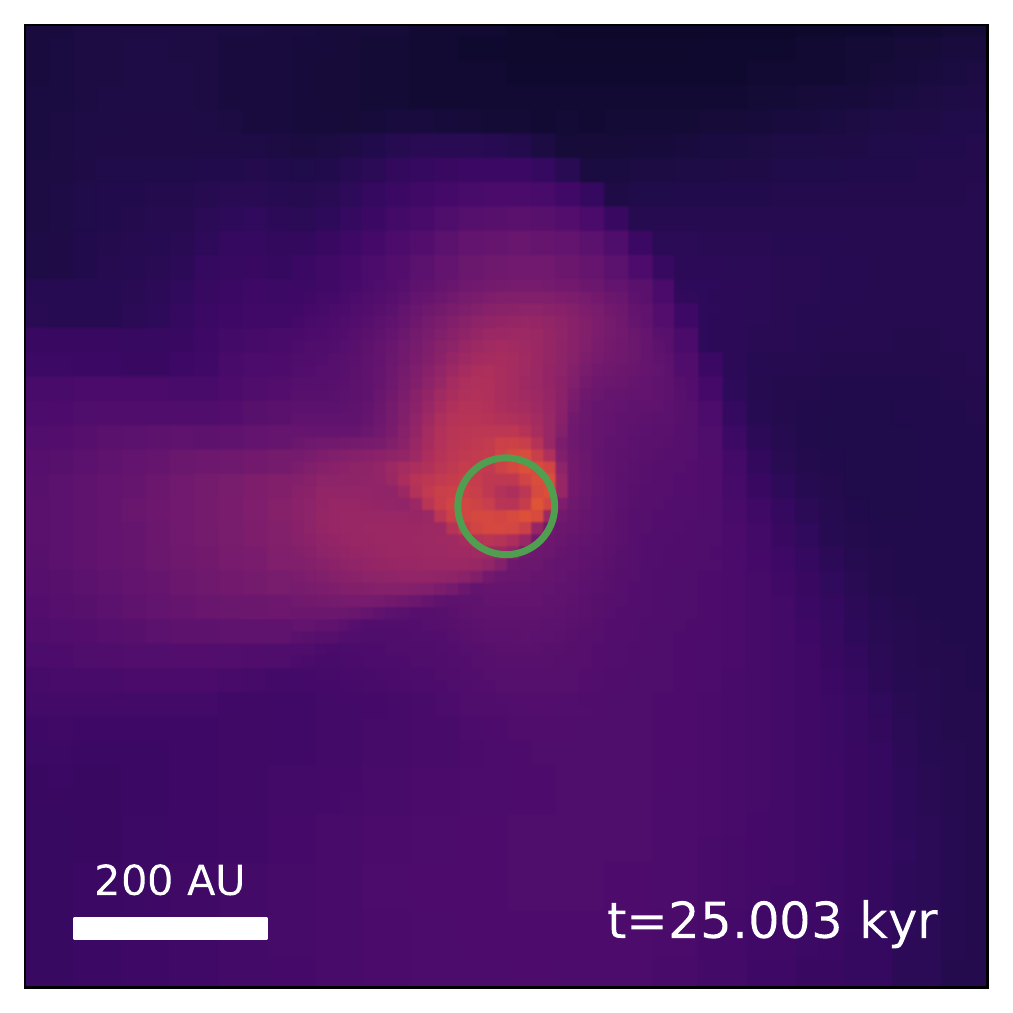}
\includegraphics[scale=0.5]{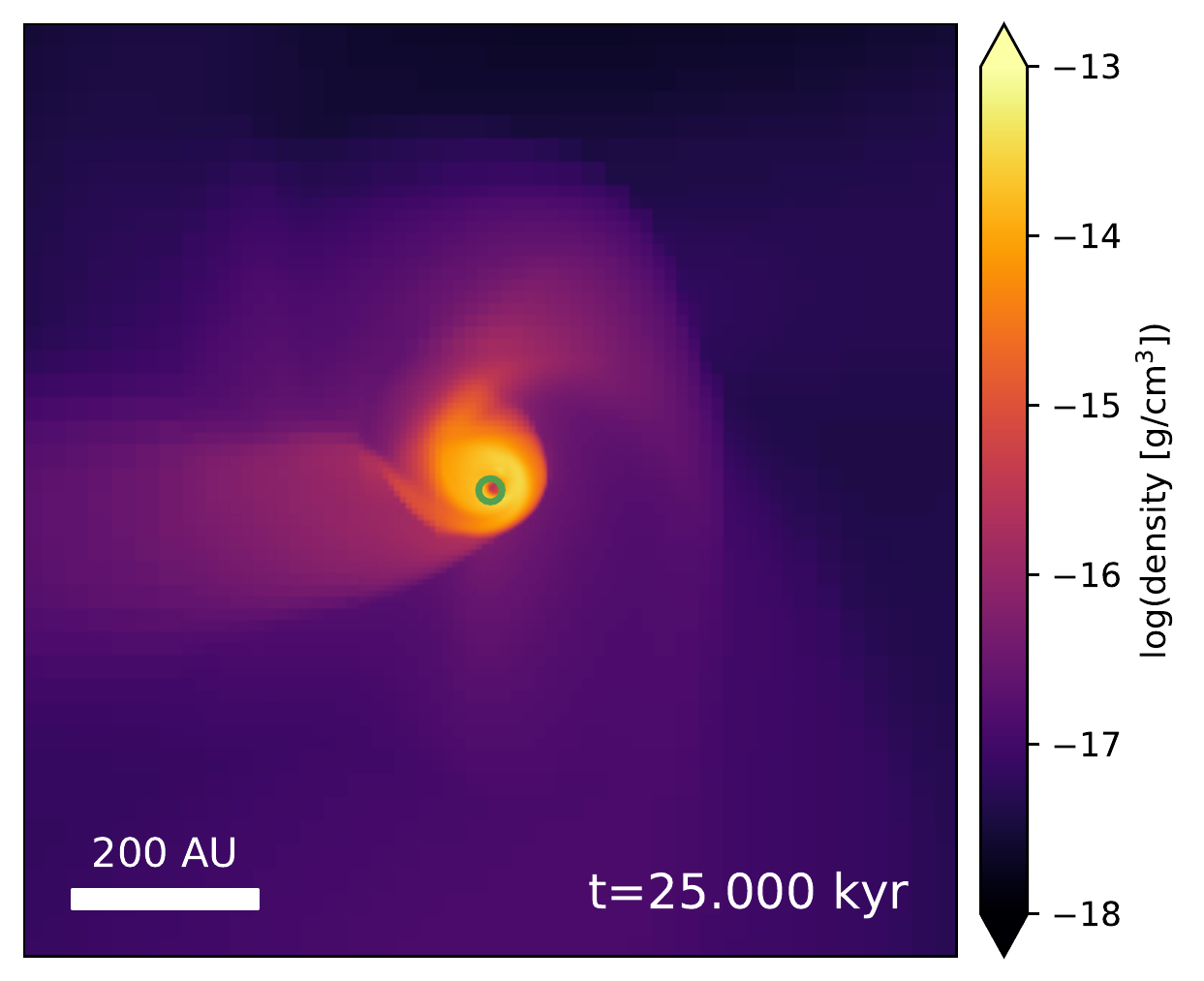}
\\
\includegraphics[scale=0.5]{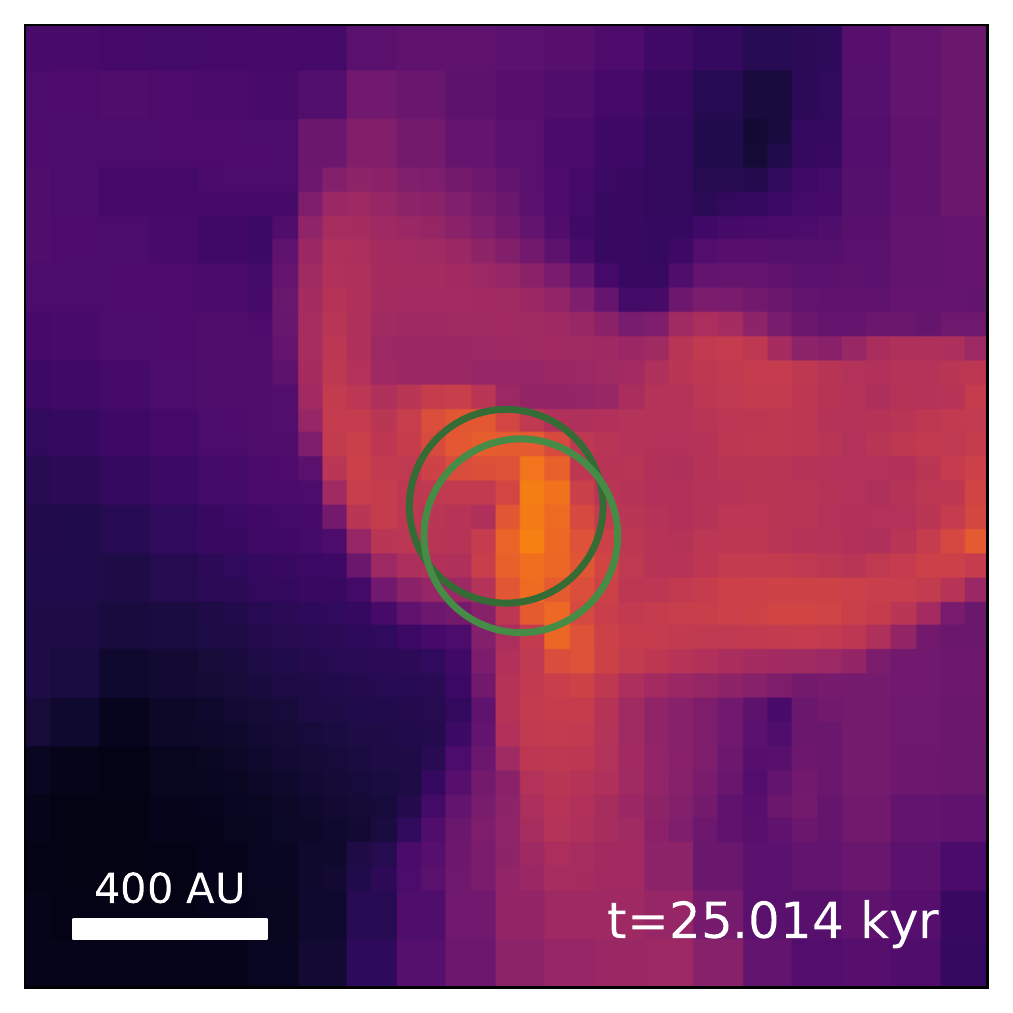}
\includegraphics[scale=0.5]{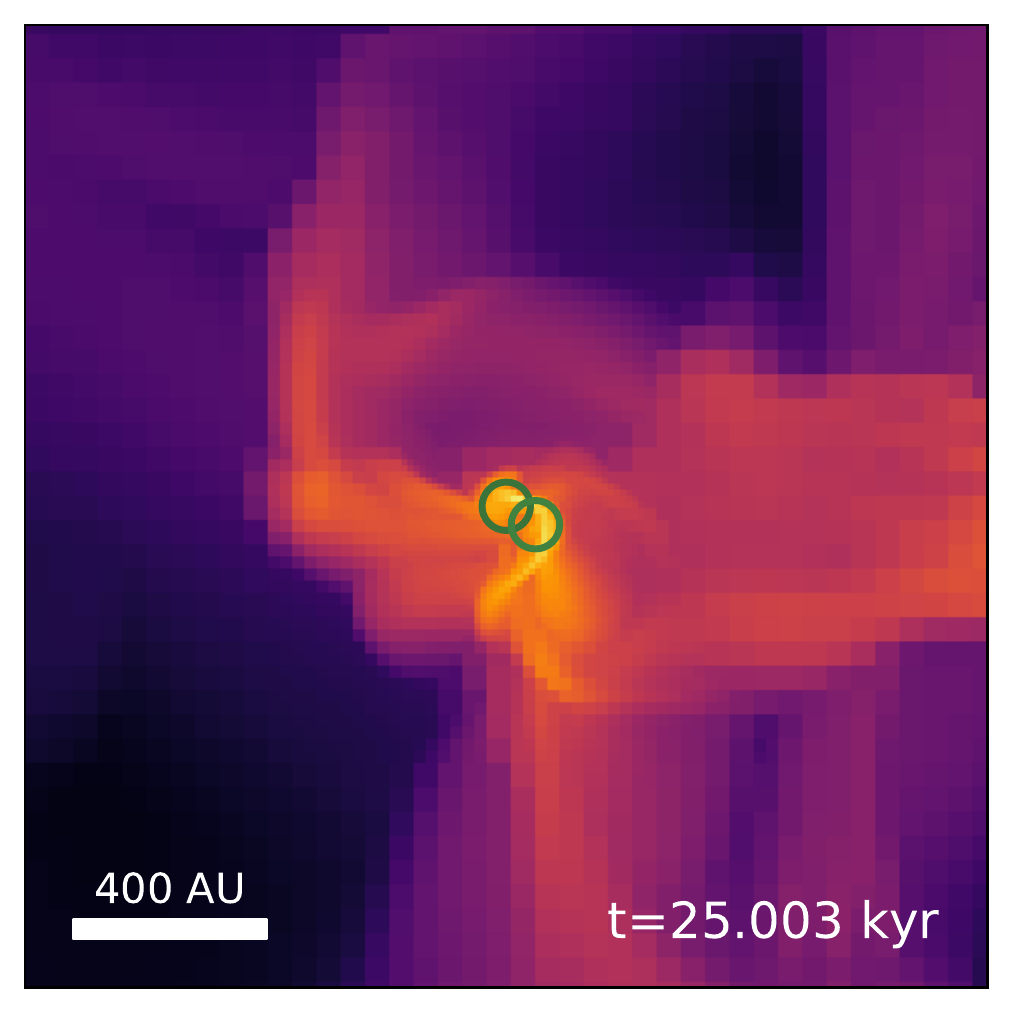}
\includegraphics[scale=0.5]{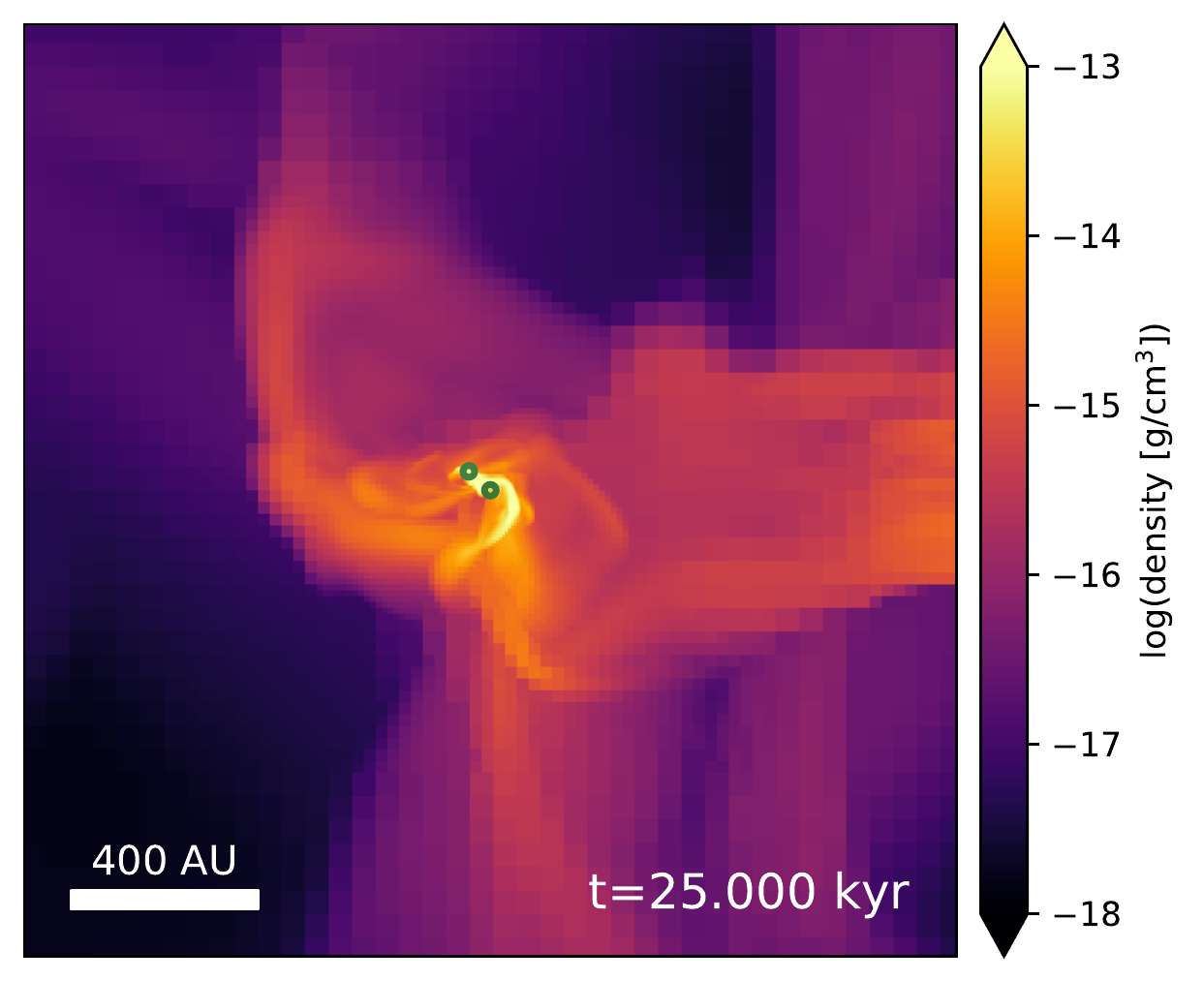}
\\
\includegraphics[scale=0.5]{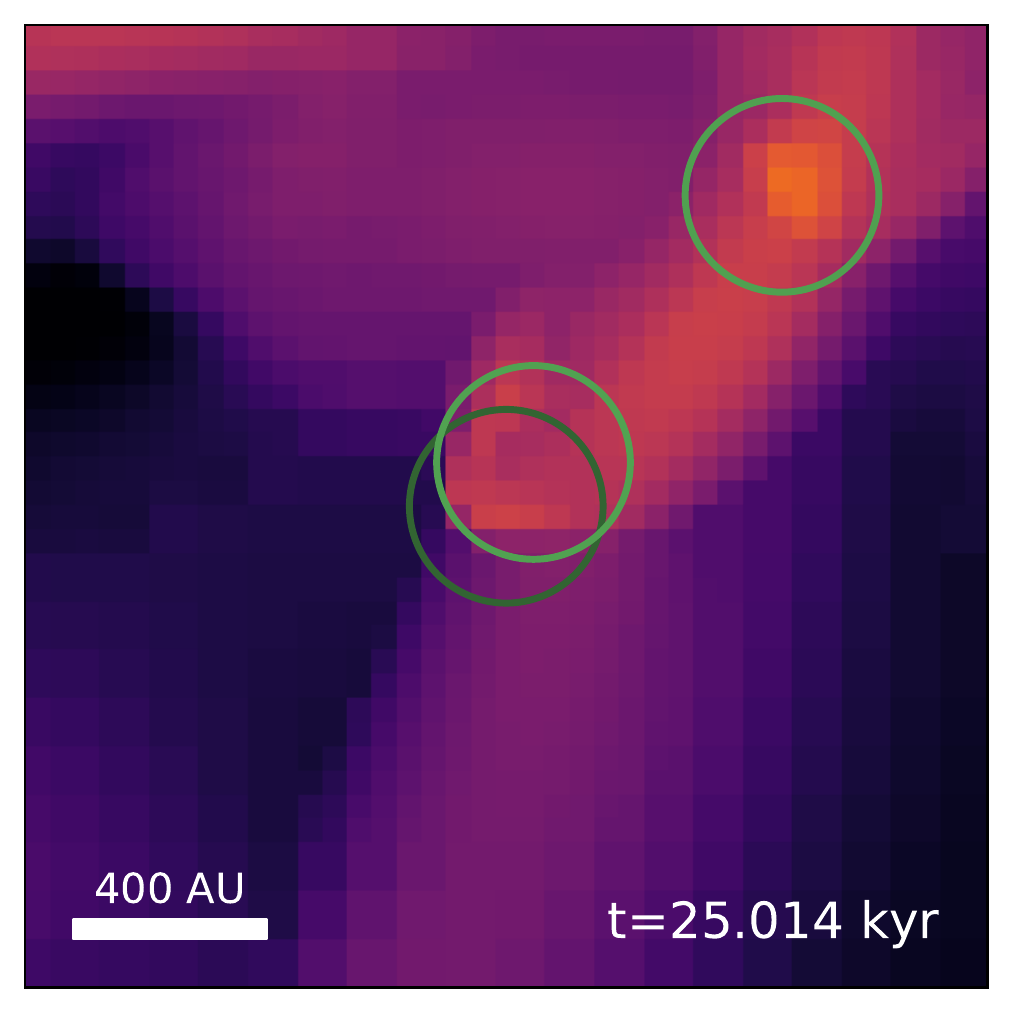}
\includegraphics[scale=0.5]{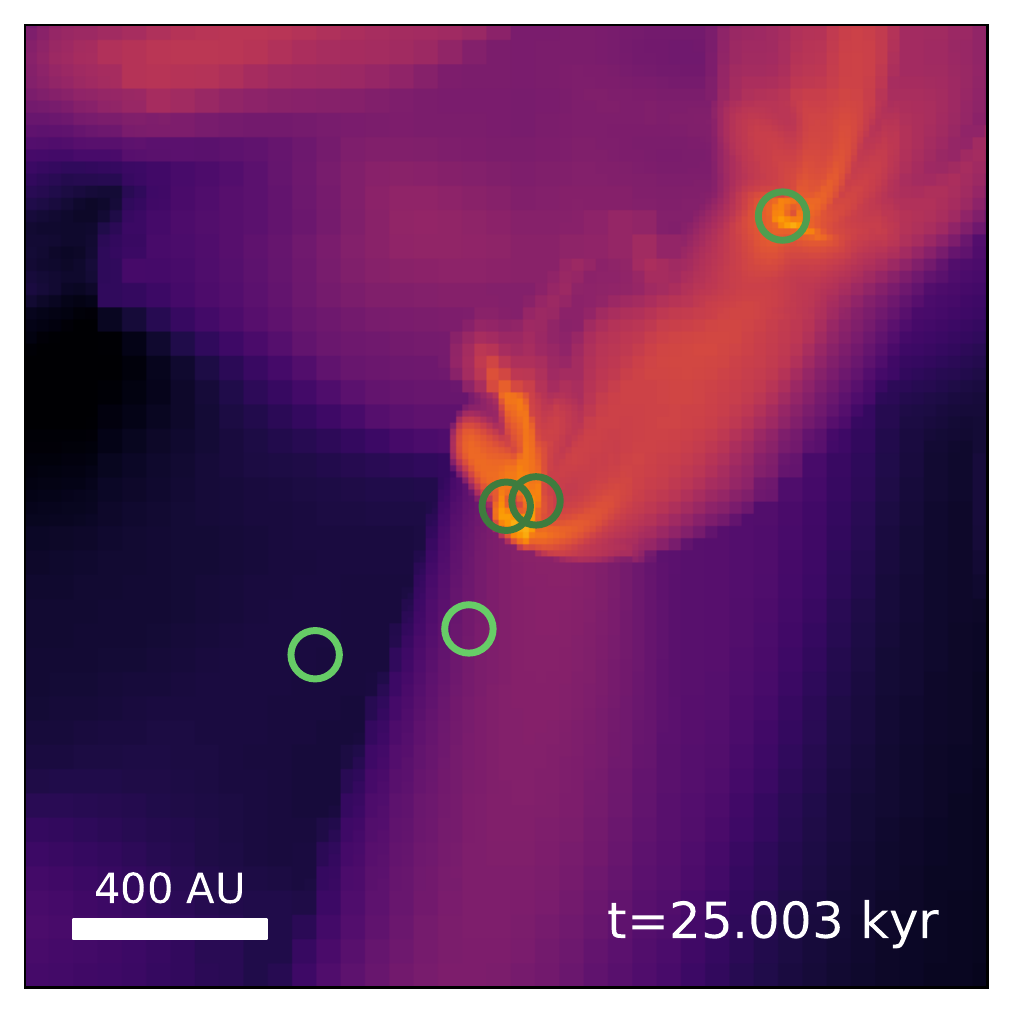}
\includegraphics[scale=0.5]{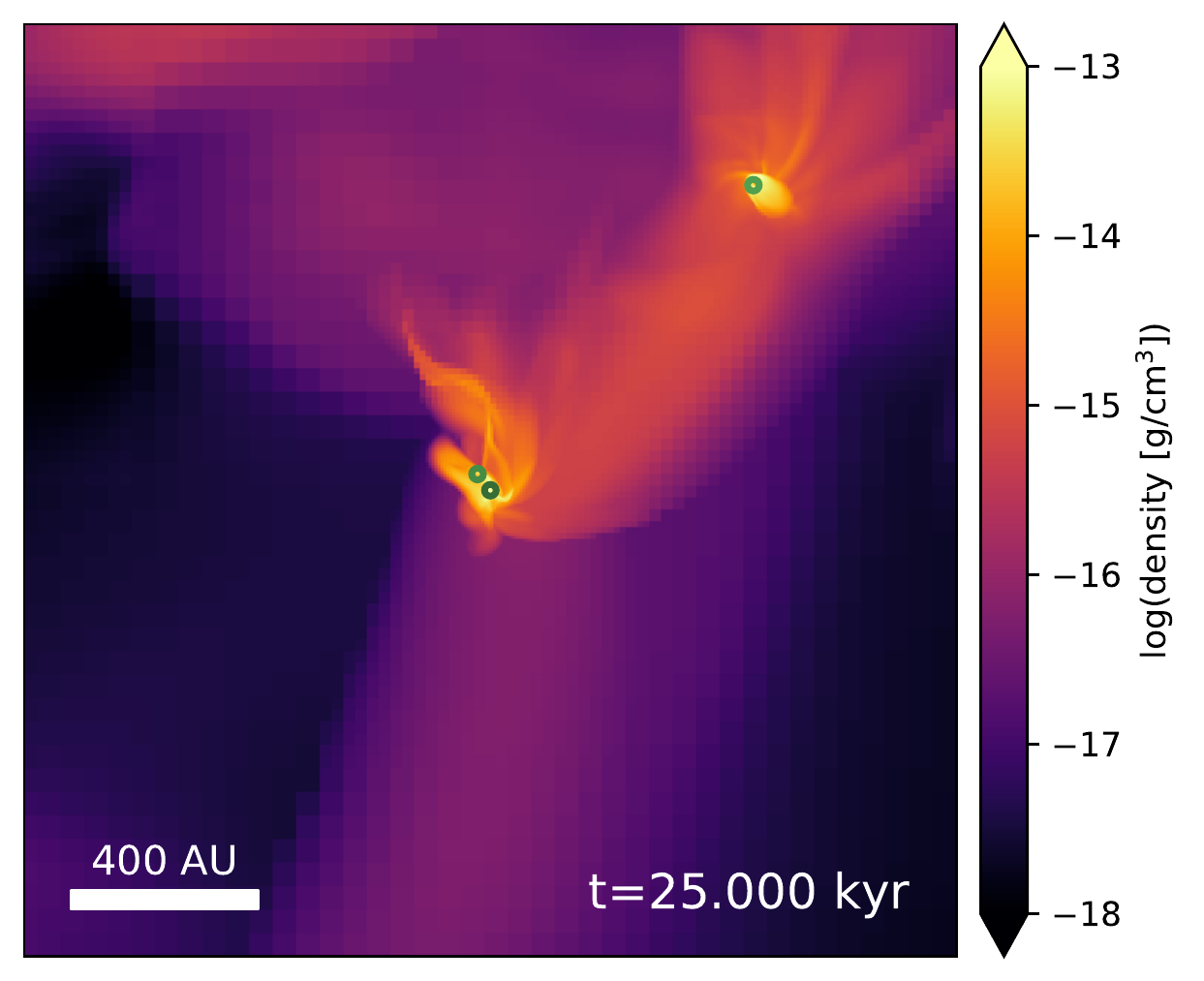}
\\
\includegraphics[scale=0.5]{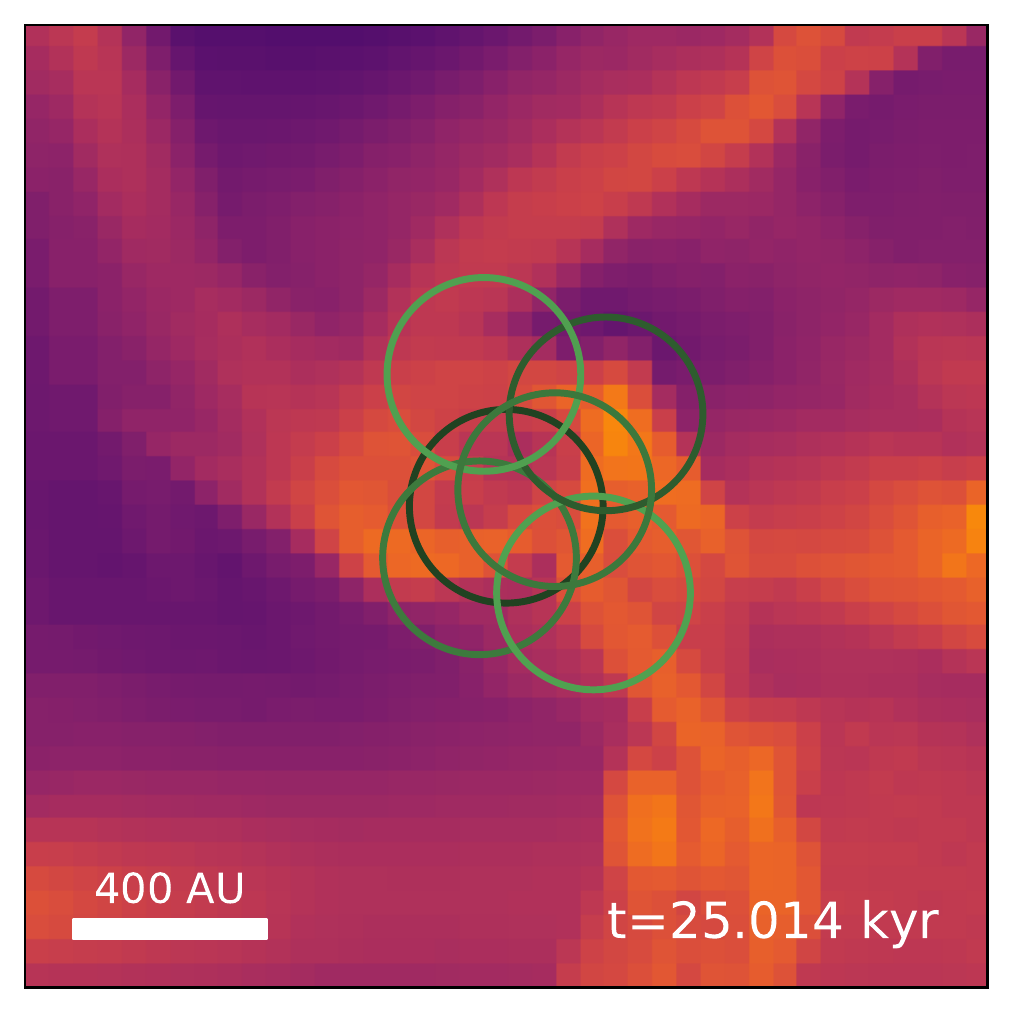}
\includegraphics[scale=0.5]{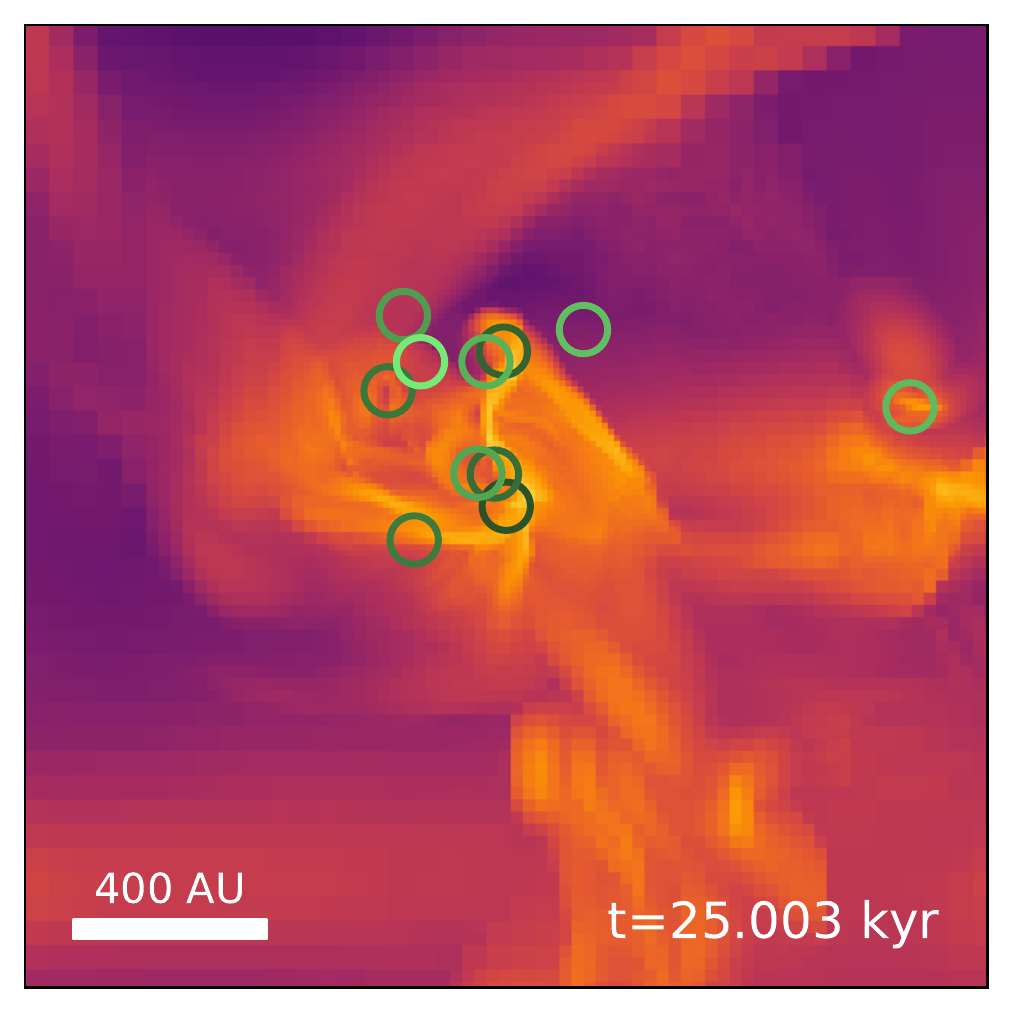}
\includegraphics[scale=0.5]{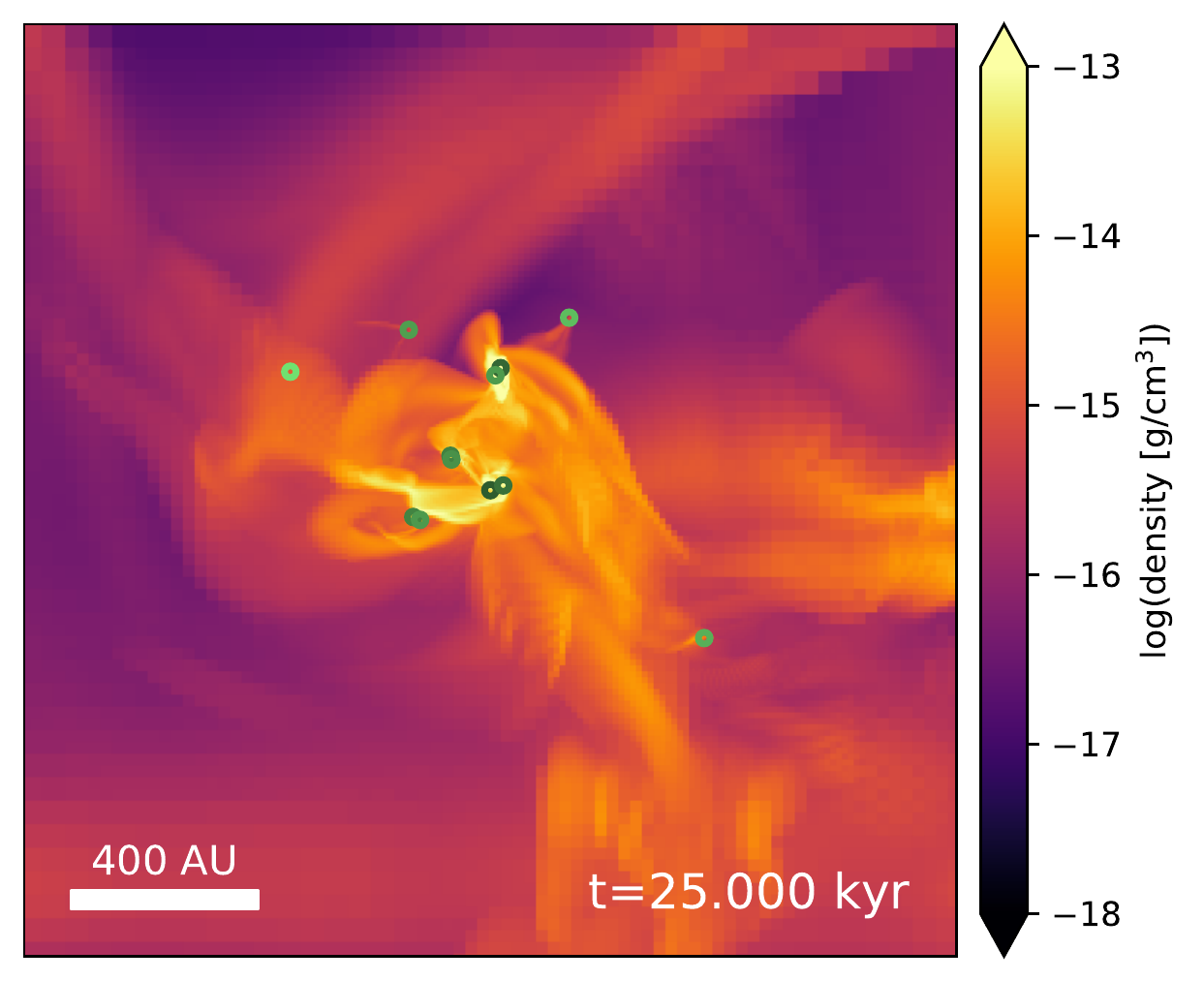}
\caption{Examples of different star formation regions in our simulations (the layout of the maps is the same as in Fig.~\ref{fig:unresolved})
From top to bottom there is an increasing geometric complexity: an isolated sink, a binary, two small regions merging and a complex star cluster. The more complex the star forming regions, the more difficult it is to reproduce the results when changing the resolution.}
\label{fig:example_regions_res}
\end{figure*}

\begin{figure*}
\center
\includegraphics[scale=0.60]{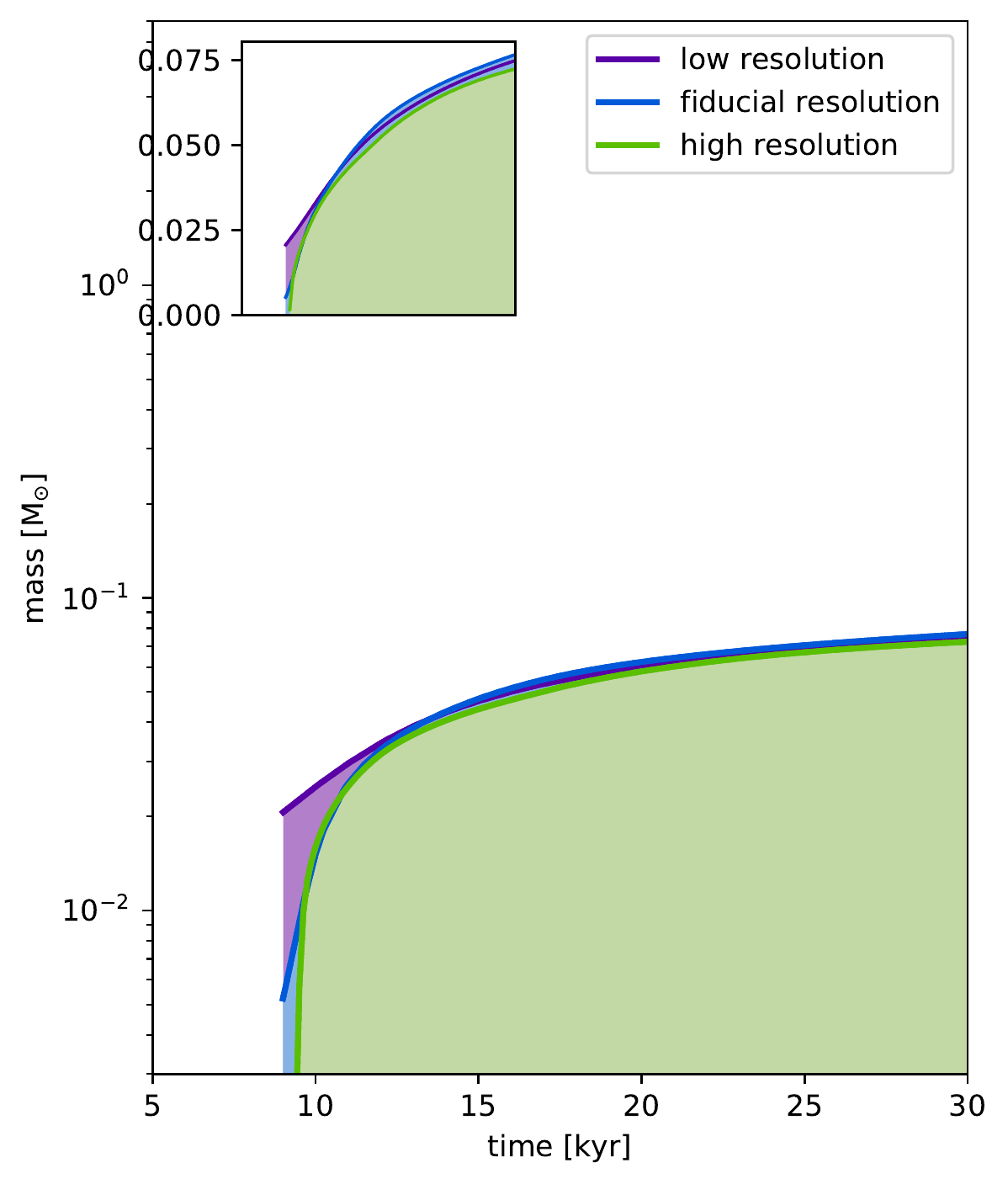}
\includegraphics[scale=0.60]{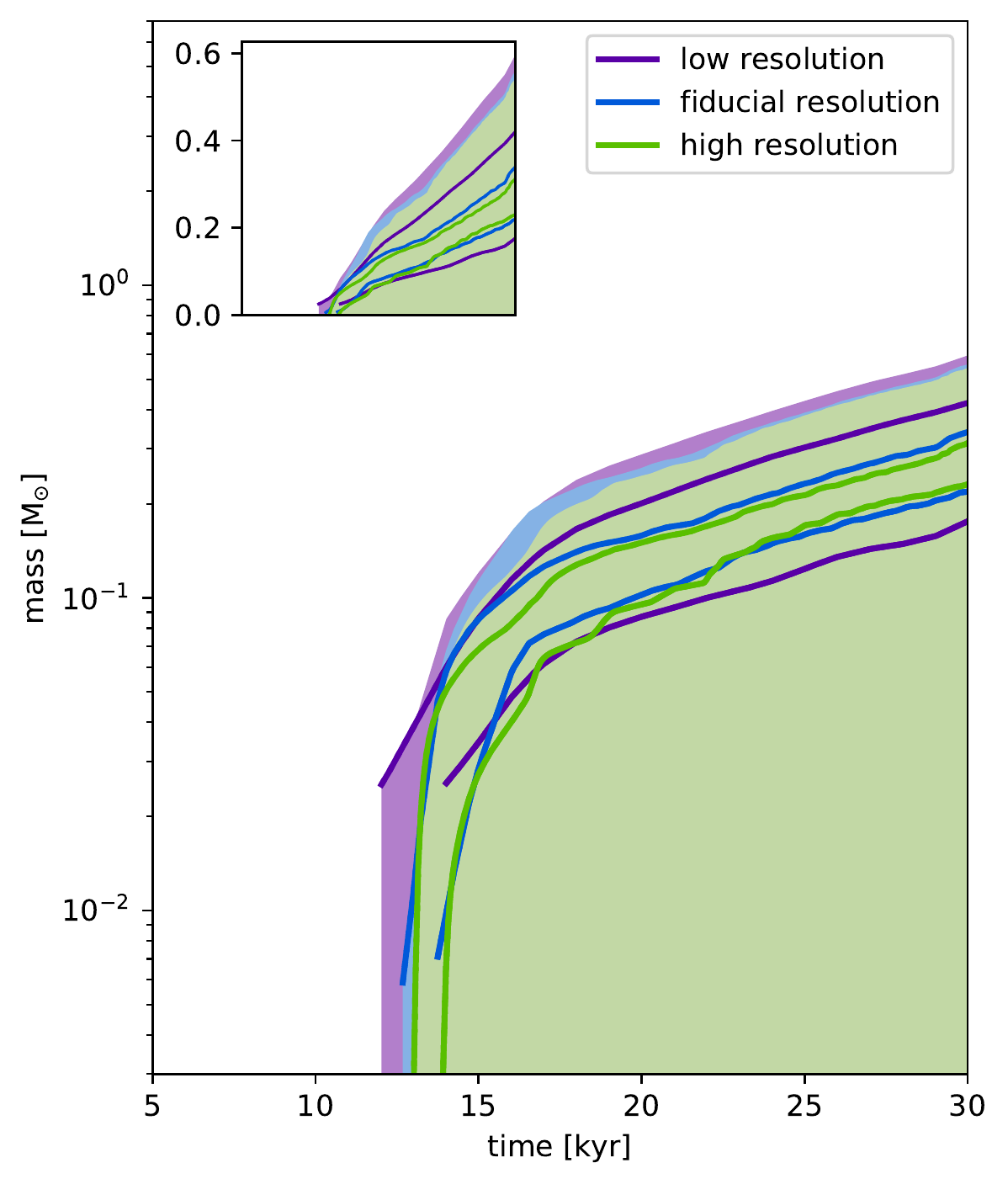}
\\
\includegraphics[scale=0.60]{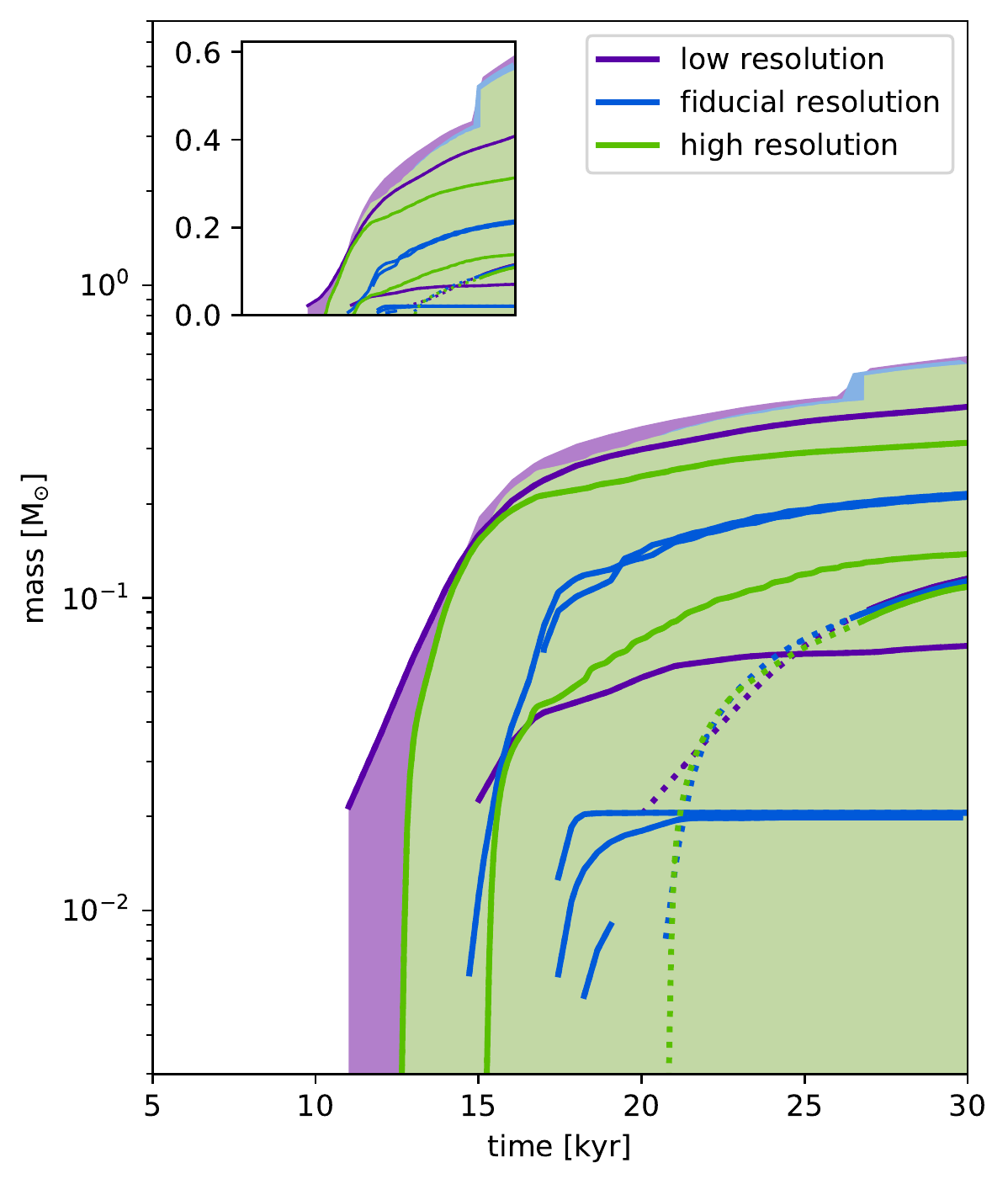}
\includegraphics[scale=0.60]{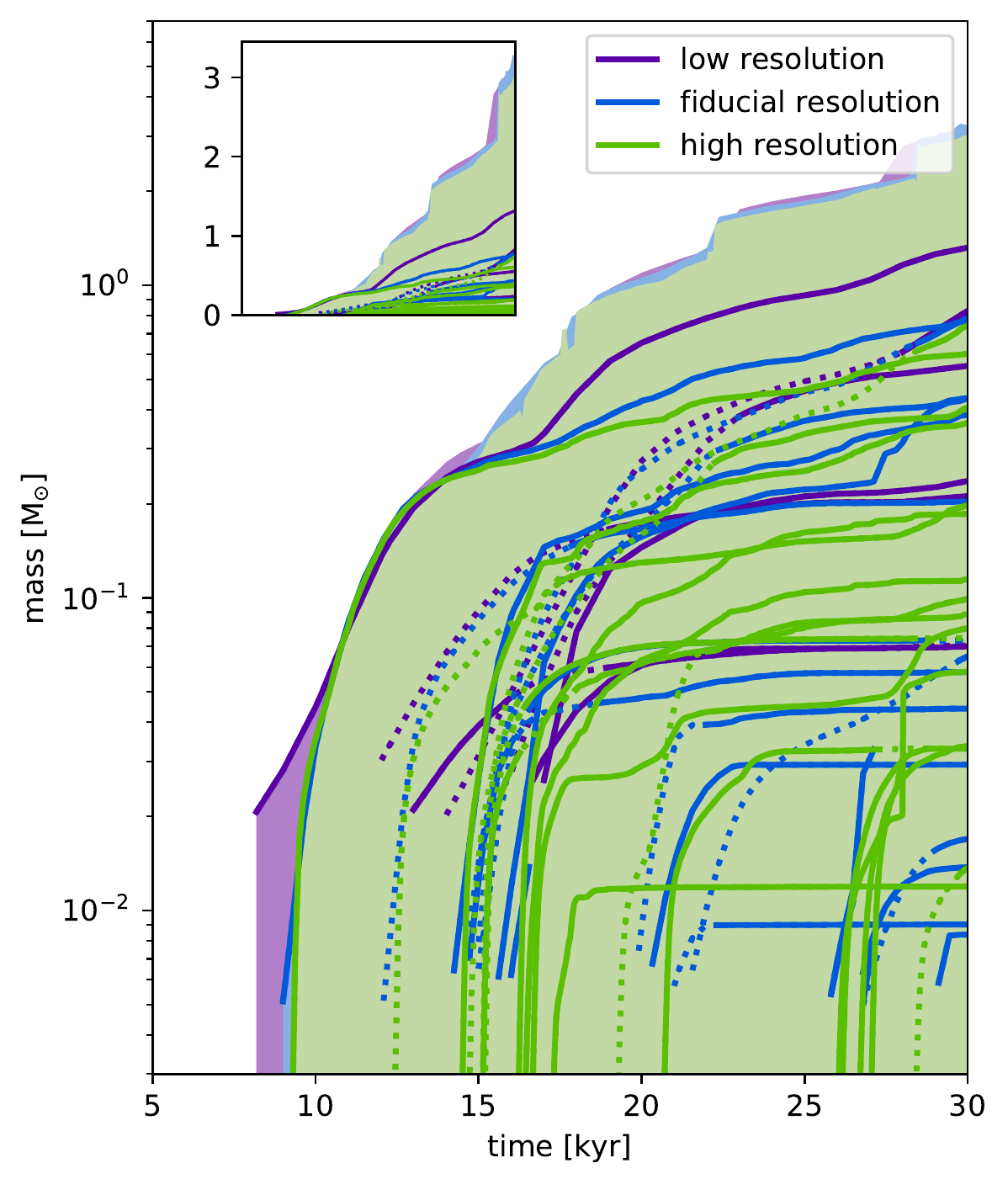}
\caption{Mass evolution diagrams of the regions shown in Fig.~ \ref{fig:example_regions_res}. The inset shows the mass on a linear scale. The dotted lines show the part of the track where a sink already exists but has not merged with the region yet. Top left: isolated sink. Top right: binary. Bottom left: binary merging with another sink region. Bottom right: a complex star cluster.}
\label{fig:example_regions_tracks}
\end{figure*}

\begin{figure*}
\center
\includegraphics[scale=0.70]{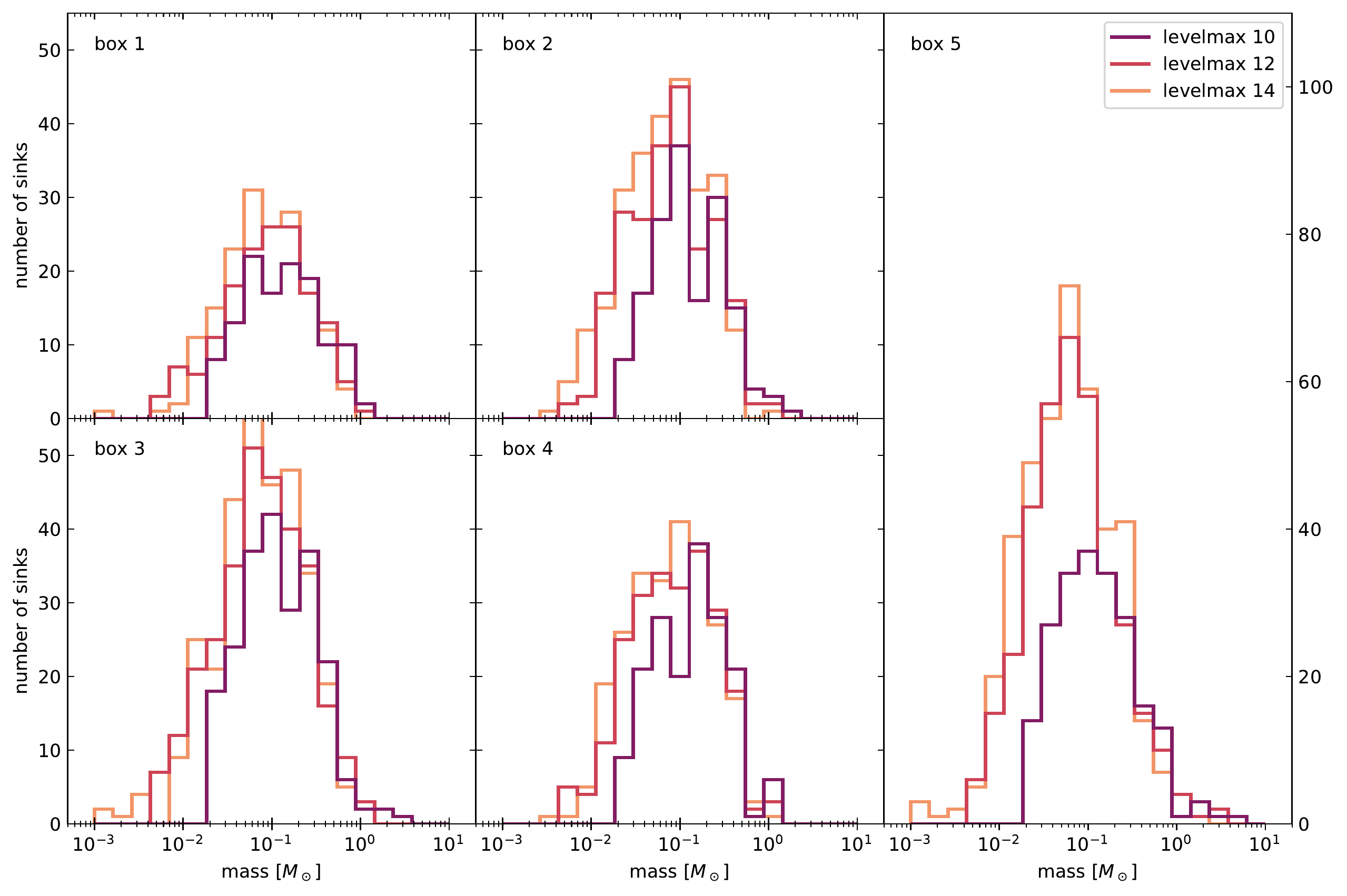}
\caption{IMF in each box for different resolutions at 30 kyr}
\label{fig:IMF_res_all}
\end{figure*}

\begin{figure*}
\center
\includegraphics[scale=0.60]{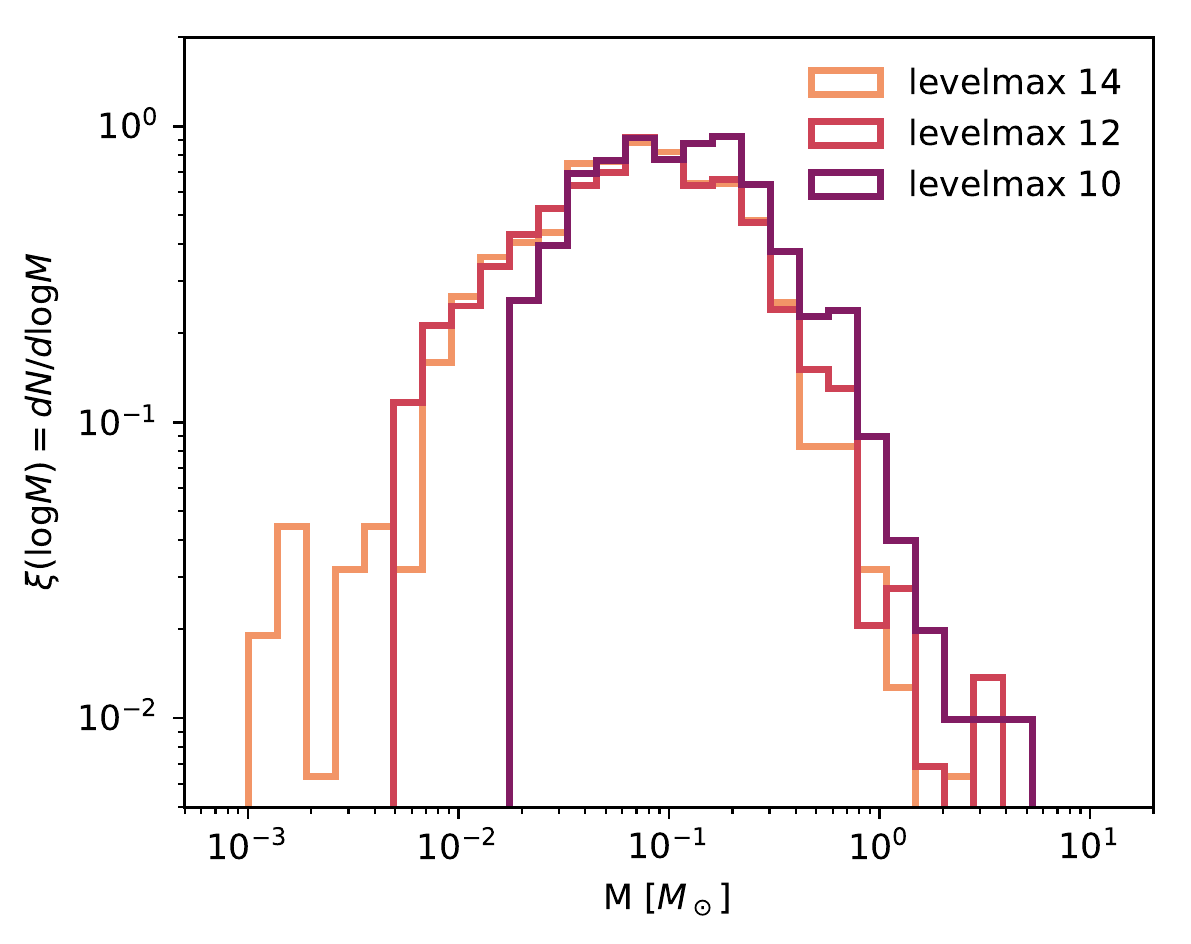}
\caption{Stacked IMF 30 kyr for the three different resolutions.}
\label{fig:IMF_res_stacked}
\end{figure*}

We tested the convergence of our results by running the same simulation with three different maximum refinement levels: 10 (low), 12 (fiducial) and 14 (high), 
corresponding to a spatial resolution of respectively 50.4, 12.6 and 3.15 AU.

\subsection{Sink formation and resolving individual stars}

The sink formation recipe has to be adapted for each resolution.
We choose the sink formation threshold to correspond to the density at which a next refinement would have been triggered if it would have been allowed,
\begin{equation}
\rho_{\mathrm{sink}} = \frac{15}{\pi} \frac{c_\mathrm{s}^2}{G (4 \Delta x_\mathrm{min})^2}
\end{equation}
The seed mass is chosen to be the isothermal Jeans mass at 10~K and the sink formation density,
\begin{equation}
m_\mathrm{seed} = m_\mathrm{J} = \frac{\pi}{6} \rho \left( c_s \sqrt{\frac{15}{\pi G \rho}_{\rm sink}} \right)^3 
\end{equation}
The exact values can be found in Table~\ref{table:res_params}.

It is important to mention that the polytropic equation of state is kept the same for all resolutions.
The properties of the first Larson core depend on the EOS but not on the resolution.
For the same reason, the merging timescale for young sinks is kept constant for all resolutions.
A list of the parameters that are kept fixed is given in Table~\ref{table:sim_params}.

Fig.~\ref{fig:unresolved} shows a time sequence of the same star forming region modelled at different resolutions.
The top row corresponds to the first time in the sequence and shows the formation of the same sink in all cases.
In the second row, the images corresponds to a later time when a companion star is formed quite close to the first protostar.
This second sink never formed in the low resolution case.
This is a typical example that shows that the low resolution simulation is not capable of resolving all binaries.

In Fig.~\ref{fig:unresolved_track} we have represented a diagram that shows the mass evolution of the various sinks in this particular star forming region.
After 16~kyr, we see that the fiducial and high resolution runs do contain a second sink with a similar mass, while the low resolution only has one sink and hence failed to form a binary.
Note however that the total mass in sinks at this time is very robust and is the same for all resolution.
Around time 22~kyr, a third member in the region appears for all resolutions. After time 25~kyr, another star forming region merges with this one and 
the dynamical state of the resulting start cluster becomes a lot more complex. Note again however that the total mass in the region is still always the same for all resolutions.

In Fig.~\ref{fig:example_regions_res}, we show 4 different star forming regions at the same time but at different resolutions.
They represent 3 typical configurations: an isolated sink, a small cluster with only 2 or 3 members and finally a large cluster with many sinks.
As one might expect, the simpler the geometry, the faster the convergence and the more robust the numerical solution.
Fig.~\ref{fig:example_regions_tracks} shows the sink mass evolution in these same 4 regions.
For the isolated sink case, the agreement between the different resolutions is nearly perfect.
For the other cases, we see that the convergence of the individual sink mass is slower, but the total mass in the region is very robust with increasing resolution.

In summary, our resolution study supports the claim that having a spatial resolution of 12.6 AU is enough to get converged results for the number of sinks and for the value of their individual masses..
This resolution corresponds to the size of the first Larson core. 

\subsection{Influence on the IMF}

Since we are mostly interested in the IMF and less in the exact mass of individual stars, we now study the effect of our different resolutions to the statistics of sinks.
Fig.~\ref{fig:IMF_res_all} shows the IMF of all initial conditions at 30~kyr and for different resolutions.
In all cases, we see that the high mass end agrees quite well.
The peak and the low mass end of the IMF for the fiducial and high resolution runs are also quite close, while the low resolution case is clearly not converged.
In box 2 and 5, we can also see a small excess of high mass stars in the low resolution runs.
These are unresolved star clusters with multiple members resolved only at higher resolution.
Fig.~\ref{fig:IMF_res_stacked} shows the stacked IMF (data from all initial conditions gathered together).
Also here we see that only the low resolution is incomplete, while the peak in the other two cases remains the same.
From these results, and our findings in the previous section, we conclude that the IMFs in our fiducial runs are already numerically converged.

\end{document}